\documentclass[aps,nofootinbib,notitlepage]{revtex4-1}

\usepackage{amsmath}
\usepackage{amssymb}
\usepackage{amsfonts}
\usepackage{mathrsfs}
\usepackage{mathtools}
\usepackage{multirow}

\usepackage{hyperref}

\usepackage[font=small]{caption}
\usepackage[section]{placeins}
\usepackage{accents}
\usepackage{color}

\DeclareSymbolFont{bbold}{U}{bbold}{m}{n}
\DeclareSymbolFontAlphabet{\mathbbold}{bbold}

\newcommand{\ud}{\mathrm{d}}

\newcommand{\pb}[1]{\,\mbox{}_{#1}}

\newcommand{\scri}{\mathscr{I}}
\newcommand{\lie}{\pounds}
\newcommand{\var}{\mathchar'26\mkern-9mu \delta}
\newcommand{\dual}{{}^*}
\newcommand{\npdelta}{\delta}
\newcommand{\npDelta}{\mathbbold{\Delta}}
\newcommand{\Chi}{\mathrm{X}}
\newcommand{\interchange}[2]{#1 \longleftrightarrow #2}
\newcommand{\bs}[1]{\boldsymbol{#1}}
\renewcommand{\Re}{\operatorname{Re}}
\renewcommand{\Im}{\operatorname{Im}}

\DeclareMathOperator{\sgn}{sgn}

\allowdisplaybreaks[3]

\numberwithin{equation}{section}
\numberwithin{table}{section}

\begin{document}

\title{A class of conserved currents for linearized gravity in the Kerr spacetime}
\author{Alexander M. Grant, \'{E}anna \'{E}. Flanagan}
\affiliation{Department of Physics, Cornell University, Ithaca, NY 14853, USA}

\begin{abstract}
  We construct a class of conserved currents for linearized gravity on a Kerr background.
  Our procedure, motivated by the current for scalar fields discovered by Carter (1977), is given by taking the symplectic product of solutions to the linearized Einstein equations that are defined by symmetry operators.
  We consider symmetry operators that are associated with separation of variables in the Teukolsky equation, as well as those arising due the self-adjoint nature of the Einstein equations.
  In the geometric optics limit, the charges associated with these currents reduce to sums over gravitons of positive powers of their Carter constants, much like the conserved current for scalar fields.
  We furthermore compute the fluxes of these conserved currents through null infinity and the horizon and identify which are finite.
\end{abstract}

\maketitle

\tableofcontents

\section{Introduction and Summary} \label{section:intro}

In the Kerr spacetime, freely falling point particles possess a constant of motion, distinct from the energy $E$ and the $z$ component of angular momentum $L_z$, known as the Carter constant $K$~\cite{PhysRev.174.1559}.
Much like $E$ and $L_z$, which are associated with Killing vectors, this constant of motion can be written in terms of a symmetric rank two \emph{Killing tensor} $K_{ab}$ as~\cite{1970CMaPh..18..265W}

\begin{equation}
  K = K_{ab} p^a p^b,
\end{equation}
where $p^a$ is the four-momentum of the particle and $K_{ab}$ satisfies

\begin{equation}
  \nabla_{(a} K_{bc)} = 0.
\end{equation}
This Killing tensor is not associated with any isometry of the Kerr spacetime, although the Carter constant reduces to the particle's total squared angular momentum (which is associated with spherical symmetry) in the Schwarzschild limit.
We fix our conventions for $K_{ab}$ in equation~\eqref{eqn:carter_tensor} below.

In addition to point particles, one can also consider test fields on the Kerr background, that is, fields whose magnitudes are small enough that their gravitational backreaction can be neglected.
In the Kerr spacetime, scalar, spin-$1/2$, and electromagnetic test fields possess conserved charges that generalize the Carter constant:
\begin{itemize}
\item For a sourceless complex scalar field $\Phi$, the conserved charge is the Klein-Gordon inner product of $\Phi$ with $\pb{0} \mathcal{D} \Phi$~\cite{PhysRevD.16.3395}:

  \begin{equation} \label{eqn:carter_charge}
    \pb{0} K \equiv \frac{1}{2i} \int_\Sigma \ud^3 \Sigma^a \left[(\pb{0} \mathcal{D} \Phi) \nabla_a \overline{\Phi} - \overline{\Phi} \nabla_a \pb{0} \mathcal{D} \Phi\right],
  \end{equation}
  where $\Sigma$ is any spacelike hypersurface, the differential operator $\pb{0} \mathcal{D}$ is defined by

  \begin{equation} \label{eqn:carter_operator}
    \pb{0} \mathcal{D} \Phi \equiv \nabla_a (K^{ab} \nabla_b \Phi),
  \end{equation}
  and bars denote complex conjugation.
  The operator $\pb{0} \mathcal{D}$ commutes with the d'Alembertian, and so maps the space of solutions into itself.
  The charge $\pb{0} K$ is associated with the Carter constant in the following sense: for a solution of the form $\Phi \propto e^{-i\vartheta/\epsilon}$, which represents a collection of scalar quanta with Carter constants $\{K_\alpha\}$, the charge is given by (in the geometric optics limit $\epsilon \to 0$)

  \begin{equation}
    \pb{0} K = \frac{1}{\hbar} \sum_\alpha K_\alpha.
  \end{equation}
  That is, the charge is proportional to the sum of the Carter constants of each scalar quantum.
In the case of real scalar fields, the charge vanishes in the geometric optics limit.
\item A similar result holds for any spin-$1/2$ field $\psi$ satisfying the Dirac equation~\cite{PhysRevD.19.1093}.
In Kerr, there exists an antisymmetric \emph{Killing-Yano tensor} $f_{ab}$, which satisfies $\nabla_{(a} f_{b)c} = 0$ and $K_{ab} = f_{ac} f^c{}_b$, with our particular choice of $K_{ab}$ in equation~\eqref{eqn:carter_tensor}.
An operator $\pb{1/2} \mathcal{D}$, which is defined in terms of $f_{ab}$ and commutes with the Dirac operator, is given by

  \begin{equation}
    \pb{1/2} \mathcal{D} = i\gamma_5 \gamma^a \left(f_a{}^b \nabla_b - \frac{1}{6} \gamma^b \gamma^c \nabla_c f_{ab}\right),
  \end{equation}
  where $\gamma^a$ is the usual gamma matrix and, in terms of the Levi-Civita tensor $\epsilon_{abcd}$, $\gamma_5 \equiv i \epsilon_{abcd} \gamma^a \gamma^b \gamma^c \gamma^d$.
  The charge which generalizes the charge in equation~\eqref{eqn:carter_charge} is proportional to the following integral over a spacelike hypersurface $\Sigma$:

  \begin{equation}
    \pb{1/2} K \propto \int_\Sigma \ud^3 \Sigma_a\; \overline{(\pb{1/2} \mathcal{D} \psi)} \gamma^a \pb{1/2} \mathcal{D} \psi.
  \end{equation}
  As in the scalar field case, this charge is proportional to the sum of the Carter constants of the individual quanta in the geometric optics limit.
This construction works for massive as well as massless spin-$1/2$ particles, and even charged spin-$1/2$ particles in the case of the Kerr-Newman spacetime~\cite{PhysRevD.19.1093}.
\item For electromagnetic fields, there are several conserved charges which satisfy the requirement of reducing, in the geometric optics limit, to a sum of (some power) of the Carter constants of the photons; some examples are given by~\cite{Andersson:2015xla}, which we have considered in~\cite{Grant:2019qyo} (along with additional examples).
\end{itemize}
It would be interesting to find similar conserved currents in the case of linearized gravity.

One application of such a conserved current would be to gravitational wave astronomy, in the form of further advances in the so-called \emph{extreme mass-ratio inspiral problem}.
The gravitational waves radiated during the inspiral of compact objects into supermassive black holes will be an important signal for LISA~\cite{AmaroSeoane:2012km}.
There is therefore a major effort currently underway to accurately compute gravitational waveforms that these sources would produce (see, for example,~\cite{Wardell:2015kea} and the references therein).
As there is a great separation of scales in the masses of the inspiralling object and the supermassive black hole, this is known as the extreme mass-ratio inspiral (EMRI) problem.
The compact object is treated as a point particle, and given an orbit, which on short timescales is geodesic, the radiation can be computed using black hole perturbation theory.
However, on long timescales, the orbital parameters change due to the effects of radiation reaction, and so on these timescales the computed radiation must be corrected.
Special classes of orbits, such as circular or equatorial orbits, can be evolved in the adiabatic limit by using the fluxes of energy and angular momentum to infinity and down the horizon to evolve the orbital energy and angular momentum, since for these orbits the Carter constant is completely determined by the energy and angular momentum (see, for example,~\cite{Hughes:1999bq}).

Generic orbits require a method of obtaining time-averaged rates of change of an orbit's Carter constant.
A formula for this quantity to leading adiabatic order has been derived directly from the self-force~\cite{Mino:2003yg} (see~\cite{Isoyama:2018sib} for recent efforts in this problem, including extensions of this result to the resonant case).
It is qualitatively similar to the formulae for energy and angular momentum fluxes, having terms corresponding to infinity and to the horizon~\cite{Sago01052006}.
There is, however, no known derivation of this formula from a conserved current.
Such a derivation would provide a unified framework with which to understand these results, and may be necessary to obtain results at higher order.
These higher-order results may be necessary for parameter estimation, or perhaps even simply detection, of signals from EMRIs.

Unfortunately, no conserved currents generalizing the Carter constant for general stress-energy tensors exist.
More precisely, we have shown that, given a general, conserved stress-energy tensor in Kerr, there is no functional of the stress-energy tensor and its derivatives on a spacelike hypersurface $\Sigma$ that a) reduces to the Carter constant for a point particle and b) is independent of the choice of hypersurface $\Sigma$ when the stress-energy tensor is of compact spatial support~\cite{Grant:2015xqa}.
This implies that there can be no generic derivation of a flux formula for a ``Carter constant'' that applies to arbitrary fields and sources.
It is still possible, however, that such derivations could exist for specific types of fields.
In particular, it may be possible to derive a flux formula for determining the evolution of an orbit's Carter contant in linearized gravity from an appropriate conserved current.

Motivated by this possibility, in this paper we construct four conserved currents, denoted $\pb{\pb{2} \mathring{\mathcal C}} j^a [\var \bs g]$, $\pb{\pb{2} \mathring{\mathcal D}} j^a [\var \bs g]$, and $\pb{\pb{\pm 2} \Omega} j^a [\var \bs g]$, that generalize the Carter constant in Kerr, in the sense that each of their charges reduce to the sum of some positive power of the Carter constants of the gravitons in the geometric optics limit.
Moreover, we show that these currents have the further property that their fluxes at null infinity and the horizon are finite for well-behaved solutions that describe radiation.
While these currents themselves are new, their construction involves symmetry operators which have been studied extensively in the literature (see, for example,~\cite{PhysRevLett.41.203, Chrzanowski:1975wv, Aksteiner:2016mol}).

The organization of this paper is as follows.
Section~\ref{section:kerr_perturbation} is a review of the theory of linearized gravity in Kerr, using both the spinor and Newman-Penrose formalisms, and fixes conventions which we use throughout.
It also reviews the Teukolsky formalism and separation of variables in the Kerr spacetime.
Section~\ref{section:symmetry} defines symmetry operators, which are the maps from the space of solutions into itself, such as the operator $\pb{0} \mathcal{D}$ in equation~\eqref{eqn:carter_operator} above.
We give particular examples of symmetry operators for linearized gravity in Kerr, and show how they act on expansions that arise in the Teukolsky formalism.
In section~\ref{section:currents}, we first define the symplectic product, a generalization of the Klein-Gordon inner product used in the scalar case, which we then use to generate the conserved currents that we consider in this paper.
In section~\ref{section:geometric_optics}, we review the geometric optics limit of solutions in linearized gravity on a curved background and use it to deduce the limits of currents defined in section~\ref{section:currents}.
In section~\ref{section:fluxes}, we compute fluxes of these currents through the horizon and null infinity.
We conclude in section~\ref{section:discussion} with general discussion and a summary of the properties of these currents in table~\ref{table:summary}.
Appendices~\ref{appendix:fluxes_integration} and~\ref{appendix:asymptotics} contain details of the calculations in section~\ref{section:fluxes}.

We use the following conventions in this paper: we follow most texts on spinors by using the $(+, -, -, -)$ sign convention for the metric and bars to denote complex conjugation.
We denote tensors with indices removed by bold face.
For any linear operator $T_{a_1 \cdots a_p}{}^{b_1 \cdots b_q}$ which maps tensors of rank $q$ to those of rank $p$, we write $T_{a_1 \cdots a_p}{}^{b_1 \cdots b_q} S_{b_1 \cdots b_q}$ as $\boldsymbol{T} \cdot \boldsymbol{S}$ when indices have been removed.
Furthermore, we will leave explicit the soldering forms $\sigma_a{}^{AA'}$ which form the isomorphism between the tangent vector space and the space of Hermitian spinors~\cite{penrose1987spinors}.

\section{Kerr perturbations: review and definitions} \label{section:kerr_perturbation}

\subsection{Spinor formalism} \label{section:spinors}

In this paper, we will be using a combination of the spinor and Newman-Penrose formalisms in order to describe linearized gravity about some arbitrary vacuum solution of the Einstein equations.
In general, we follow the notation of Penrose and Rindler~\cite{penrose1987spinors, penrose1988spinors}.
The spinor formalism is particularly convenient in Kerr, since not only is there a rank two Killing tensor $K_{ab}$ as discussed in section~\ref{section:intro}, but also a rank two symmetric spinor $\zeta_{AB}$ which satisfies the \emph{Killing spinor equation}~\cite{penrose1988spinors}:

\begin{equation} \label{eqn:killing_spinor}
  \nabla^{A'}{}_{(A} \zeta_{BC)} = 0.
\end{equation}
This Killing spinor generates the related \emph{conformal Killing tensor} $\Sigma_{ab}$ given by

\begin{equation} \label{eqn:carter_tensor}
  \Sigma_{ab} \equiv \sigma_a{}^{AA'} \sigma_b{}^{BB'} \zeta_{AB} \bar{\zeta}_{A'B'} \equiv \frac{1}{2} K_{ab} - \frac{1}{4} \Re\left[\zeta_{CD} \zeta^{CD}\right] g_{ab},
\end{equation}
which we use to define our Killing tensor $K_{ab}$~\cite{1970CMaPh..18..265W}.
Note that, given a Killing spinor $\zeta_{AB}$, equation~\eqref{eqn:carter_tensor} fixes the ambiguity in $K_{ab}$, which is otherwise only defined only up to terms of the form $\lambda g_{ab}$, for constant $\lambda$, or up to terms that are products of Killing vectors.

Petrov type D spacetimes possess a Killing spinor intimately connected with the Weyl spinor $\Psi_{ABCD}$~\cite{1970CMaPh..18..265W}, the symmetric spinor constructed from the Weyl tensor:

\begin{equation}
  C_{abcd} \equiv \sigma_a{}^{AA'} \sigma_b{}^{BB'} \sigma_c{}^{CC'} \sigma_d{}^{DD'} \left(\epsilon_{AB} \epsilon_{CD} \overline{\Psi}_{A'B'C'D'} + \epsilon_{A'B'} \epsilon_{C'D'} \Psi_{ABCD}\right).
\end{equation}
Since $\Psi_{ABCD}$ is symmetric, it can be written as a symmetric product of four spinors

\begin{equation}
  \Psi_{ABCD} = \alpha_{(A} \beta_B \gamma_C \delta_{D)}.
\end{equation}
For spacetimes of Petrov type D, there is a choice of these spinors such that $\alpha_A = \beta_A$ and $\gamma_A = \delta_A$ (this is one of many equivalent definitions of a type D spacetime).
Normalizing $\alpha_A$ and $\gamma_A$ to be a spin basis $(o, \iota)$ (that is, setting $o_A \iota^A = 1$), one finds

\begin{equation} \label{eqn:type_d}
  \Psi_{ABCD} = 6 \Psi_2 o_{(A} o_B \iota_C \iota_{D)}.
\end{equation}
We are using the following notation for contractions of spinors with a given spin basis~\cite{penrose1987spinors}: given a symmetric spinor field $S_{B_1 \cdots B_n}$ and a spin basis $(o, \iota)$, we define (for any integer $i$ with $0 \leq i \leq n$)

\begin{equation}
  S_i = S_{B_1 \cdots B_n} \iota^{B_1} \cdots \iota^{B_i} o^{B_{i + 1}} o^{B_n}.
\end{equation}

Thus, in equation~\eqref{eqn:type_d} $\Psi_2$ means the Weyl scalar $\Psi_{ABCD} \iota^A \iota^B o^C o^D$.
The spin basis $(o, \iota)$ is called a \emph{principal spin basis} for the Weyl spinor if it satisfies equation~\eqref{eqn:type_d}.
On such a basis, we define the Killing spinor $\zeta_{AB}$ by

\begin{equation}
  \zeta_{AB} \equiv \zeta o_{(A} \iota_{B)},
\end{equation}
where $\zeta \sqrt[3]{\Psi_2}$ is constant~\cite{1970CMaPh..18..265W}.
For the remainder of the paper, we will restrict ourselves (generally) to a principal spin basis of the background Weyl spinor.

With these definitions in hand, we turn to the construction of linearized gravity in Kerr.
We fix the background Kerr metric $g_{ab}$, and consider a one-parameter family of metrics $g_{ab} (\lambda)$, with $g_{ab} (0) = g_{ab}$.
In general, we will use a notational convention where, for any quantity $Q$, $Q(\lambda)$ will denote the quantity at an arbitrary value of $\lambda$, and $Q$ without an argument will denote $Q(0)$, the background value.
The \emph{linearization} $\var Q$ of $Q(\lambda)$ is defined by\footnote{We are using $\var$, instead of the more conventional $\delta$, in order to avoid confusion with the Newman-Penrose operator $\delta$.}

\begin{equation}
  \var Q = \left.\frac{\ud Q}{\ud \lambda}\right|_{\lambda = 0}.
\end{equation}
The linearized Einstein equations take the form

\begin{equation} \label{eqn:linearized_efe}
  \pb{2} \mathcal{E}^{abcd} \var g_{cd} = 8\pi \var T^{ab},
\end{equation}
where

\begin{equation} \label{eqn:einstein_operator}
  \pb{2} \mathcal{E}^{abcd} \equiv -\nabla^{(c} g^{d)(a} \nabla^{b)} + \frac{1}{2} (g^{cd} \nabla^{(a} \nabla^{b)} + g^{ac} g^{bd} \Box) - \frac{1}{2} g^{ab} (g^{cd} \Box - \nabla^{(c} \nabla^{d)}).
\end{equation}
is the linearized Einstein operator and $\var T^{ab}$ is the linearized stress-energy tensor.
Here the covariant derivative $\nabla_a$ is that associated with $g_{ab}$; the covariant derivative associated with $g_{ab} (\lambda)$ is denoted $\nabla_a (\lambda)$.
The prepended subscript 2 in $\pb{2} \mathcal{E}^{abcd}$ refers to the fact that linearized gravity is a spin-2 field.

To describe linearized perturbations using spinors, we consider the following quantity:

\begin{equation}
  (\var g)_{AA'BB'} \equiv \sigma^a{}_{AA'} \sigma^b{}_{BB'} \var g_{ab}.
\end{equation}
Note that this is \emph{not} the variation of a spinor; we are performing the variation \emph{first}, and then computing a spinor field using the soldering forms $\sigma^a{}_{AA'}$ that are associated with the background spacetime\footnote{We note that there have been recent developments on a variational formalism for spinors~\cite{Backdahl:2015yua} which we will not be using.
We instead follow the traditional approach of~\cite{penrose1987spinors}.}.
In general, the placement of parentheses around a quantity that we are varying implies that we take the variation first, and then perform the operation, such as raising or lowering indices: for example, $(\var g)^{ab} = g^{ac} g^{bd} \var g_{cd}$, whereas $\var g^{ab}$ would be the variation of the raised metric, and in fact $\var g^{ab} = -(\var g)^{ab}$.

In a similar manner, one can define a spinor $(\var \Psi)_{ABCD}$ that is frequently called the perturbed Weyl spinor~\cite{penrose1987spinors} (although it is also not the variation of a spinor), again using the background soldering forms:

\begin{equation} \label{eqn:perturbed_weyl_spinor}
  (\var \Psi)_{ABCD} \equiv \frac{1}{4} \sigma^a{}_{AE'} \sigma^b{}_B{}^{E'} \sigma^c{}_{CF'} \sigma^d{}_D{}^{F'} \var C_{abcd}.
\end{equation}
Using the form of the perturbed Riemann tensor, one finds that~\cite{penrose1987spinors}

\begin{equation} \label{eqn:perturbed_weyl_spinor_metric}
  (\var \Psi)_{ABCD} = \frac{1}{2} \nabla^{A'}{}_{(C} \nabla^{B'}{}_{D} (\var g)_{AB)A'B'} + \frac{1}{4} (\var g)_e{}^e \Psi_{ABCD}.
\end{equation}
The equations of motion for the perturbed Weyl spinor are derived from the Bianchi identity, and are~\cite{penrose1987spinors}

\begin{equation} \label{eqn:massless_spin_2}
  \nabla^{AA'} (\var \Psi)_{ABCD} = \frac{1}{2} (\var g)^{EFA'B'} \nabla_{BB'} \Psi_{EFCD} - \Psi_{EF(BC} \nabla_{D)}{}^{B'} (\var g)^{EFA'}{}_{B'} - \frac{1}{2} \Psi_{EF(BC} \nabla^{EB'} (\var g)_{D)}{}^{FA'}{}_{B'}.
\end{equation}
Thus, the equations of motion depend explicitly on the metric perturbation as well as the perturbed Weyl spinor.
Note further that equation~\eqref{eqn:massless_spin_2} reduces to the spin-2 massless spinor field equation $\nabla^{AA'} (\var \Psi)_{ABCD} = 0$ only when the manifold is conformally flat ($\Psi_{ABCD} = 0$).

The perturbed Weyl spinor, moreover, is not gauge invariant: under a gauge transformation $\var g_{ab} \to \var g_{ab} + 2 \nabla_{(a} \xi_{b)}$~\cite{penrose1987spinors},

\begin{equation}
  (\var \Psi)_{ABCD} \to (\var \Psi)_{ABCD} + \xi^{EE'} \nabla_{E'(A} \Psi_{BCD)E} + 2 \Psi_{E(ABC} \nabla_{D)E'} \xi^{EE'}.
\end{equation}
For type D spacetimes, however, $(\var \Psi)_0$ and $(\var \Psi)_4$ \emph{are} gauge invariant, and they are the pieces that correspond to gravitational radiation~\cite{Stewart:1974uz}.
Moreover, as is well known, the equations of motion for $(\var \Psi)_0$ and $(\var \Psi)_4$ can be ``decoupled'' from those for $(\var \Psi)_1$, $(\var \Psi)_2$, and $(\var \Psi)_3$, and each other~\cite{1973ApJ...185..635T}, as we will discuss in section~\ref{section:teukolsky}.
It suffices to use either $(\var \Psi)_0$ or $(\var \Psi)_4$ to describe a generic, well-behaved perturbation, up to $l = 0, 1$ modes~\cite{1973JMP....14.1453W}, and therefore we can describe such perturbations in terms of gauge invariant variables.

\subsection{Newman-Penrose formalism} \label{section:np}

We will also be using the Newman-Penrose notation: given a spin basis $(o, \iota)$, the null basis $\{l^a, n^a, m^a, \bar{m}^a\}$ is defined by

\begin{equation} \label{eqn:soldering}
  l^a = \sigma^a{}_{AA'} o^A \bar{o}^{A'}, \quad n^a = \sigma^a{}_{AA'} \iota^A \bar{\iota}^{A'}, \quad m^a = \sigma^a{}_{AA'} o^A \bar{\iota}^{A'},
\end{equation}
such that

\begin{equation}
  g_{ab} = 2(l_{(a} n_{b)} - m_{(a} \bar{m}_{b)}).
\end{equation}
Using these four vectors, one can define the Newman-Penrose operators by $D = l^a \nabla_a$, $\npDelta = n^a \nabla_a$, and $\npdelta = m^a \nabla_a$, as well as the twelve spin coefficients via the following eight equations:

\begin{equation} \label{eqn:spin_coefficients}
  \begin{aligned}
    D o_A &= \epsilon o_A - \kappa \iota_A, \qquad &D \iota_A &= \pi o_A - \epsilon \iota_A, \\
    \npDelta o_A &= \gamma o_A - \tau \iota_A, \qquad &\npDelta \iota_A &= \nu o_A - \gamma \iota_A, \\
    \npdelta o_A &= \beta o_A - \sigma \iota_A, \qquad &\npdelta \iota_A &= \mu o_A - \beta \iota_A, \\
    \bar{\npdelta} o_A &= \alpha o_A - \rho \iota_A, \qquad &\bar{\npdelta} \iota_A &= \lambda o_A - \alpha \iota_A.
  \end{aligned}
\end{equation}
The five Weyl scalars $\Psi_0$, $\Psi_1$, $\Psi_2$, $\Psi_3$, and $\Psi_4$, in Newman-Penrose notation, take the form~\cite{Newman:1961qr}

\begin{equation} \label{eqn:weyl_scalars}
  \Psi_i = -C_{abcd} \begin{cases}
    l^a m^b l^c m^d & i = 0 \\
    l^a n^b l^c m^d & i = 1 \\
    \frac{1}{2} l^a n^b (l^c n^d - m^c \bar{m}^d) & i = 2 \\
    l^a n^b \bar{m}^c n^d & i = 3 \\
    n^a \bar{m}^b n^c \bar{m}^d & i = 4
  \end{cases}.
\end{equation}
A null tetrad such that $\Psi_0 = \Psi_1 = \Psi_3 = \Psi_4 = 0$ and $\Psi_2 \neq 0$, for a Petrov type D spacetime, is called a \emph{principal tetrad} (as it is a tetrad associated with a principal spin basis).

Furthermore, at certain points throughout this paper, we will be using the notion of $'$ and $*$ transformations (reviewed in~\cite{Geroch:1973am}) to simplify the presentation.
These are defined by replacing, in some expression, the members of the spin basis via the following rules:

\begin{equation}
  \begin{aligned}
    ': &\; o_A \mapsto i\iota_A,& &\iota_A \mapsto io_A,& &\bar{o}_{A'} \mapsto -i\bar{\iota}_{A'},& &\bar{\iota}_{A'} \mapsto -i \bar{o}_{A'},& \\
    *: &\; o_A \mapsto o_A,& &\iota_A \mapsto \iota_A,& &\bar{o}_{A'} \mapsto -\bar{\iota}_{A'},& &\bar{\iota}_{A'} \mapsto -\bar{o}_{A'}.&
  \end{aligned}
\end{equation}
The $'$ and $*$ transformations elucidate certain symmetries that appear in Newman-Penrose notation.
The $'$ transformation, which merely switches $l^a \leftrightarrow n^a$ and $m^a \leftrightarrow \bar{m}^a$, is particularly important in Kerr, since it preserves $(o, \iota)$ as a principal spin basis.
As an example, applying the transformations to equation~\eqref{eqn:spin_coefficients} yields

\begin{equation}
  \begin{aligned}
    \epsilon' &= -\gamma, &\kappa' &= -\nu, &\pi' &= -\tau, \\
    \beta' &= -\alpha, &\sigma' &= -\lambda, &\mu' &= -\rho, \\
    \epsilon^* &= -\beta, &\kappa^* &= -\sigma, &\pi^* &= -\mu, \\
    \gamma^* &= -\alpha, &\tau^* &= -\rho, &\nu^* &= -\lambda.
  \end{aligned}
\end{equation}
As another example, consider the following equations, in Newman-Penrose notation, that the scalar $\zeta$ obeys in Kerr:

\begin{equation}
  D \zeta = -\zeta \rho, \qquad \Delta \zeta = \zeta \mu, \qquad \delta \zeta = -\zeta \tau, \qquad \bar{\delta} \zeta = \zeta \pi.
\end{equation}
The second equation can be derived from the first via a $'$ transformation, and likewise the fourth from the third, while the third follows from the first via a $*$ transformation.
In the future, we will only list one of the equations, and specify that the others can be obtained by the appropriate transformations.

\subsection{Teukolsky formalism} \label{section:teukolsky}

The Teukolsky formalism is a choice of variables for test fields in Kerr such that the equations of motion decouple, yielding equations that describe radiation, and furthermore, as we will discuss later in this section, separate in Boyer-Lindquist coordinates.
It builds off of the Newman-Penrose formalism: in the case of linearized gravity, the variables involve variations of the Weyl scalars.
Note that, taking variations of the Weyl scalars, we find that

\begin{equation} \label{eqn:np_vs_spinor}
  \var \Psi_0 = (\var \Psi)_0, \qquad \var \Psi_4 = (\var \Psi)_4.
\end{equation}
On the left-hand sides of these equations, there is a variation of the null tetrad as well as the Weyl tensor; on the right, only the Weyl tensor is varied, according to equation~\eqref{eqn:perturbed_weyl_spinor}.
Note that equation~\eqref{eqn:np_vs_spinor} only holds for $\var \Psi_0$ and $\var \Psi_4$, and only because the background is type D, as the tetrad is varied when varying equation~\eqref{eqn:weyl_scalars}.
This result is rather convenient, since we will have reason to use $\var \Psi_0$ and $(\var \Psi)_0$, for example, interchangeably.

The choice of variables that are employed here are the so-called ``master variables'' $\pb{s} \Omega$, defined by~\cite{1973ApJ...185..635T}

\begin{equation} \label{eqn:decoupled_variables}
  \pb{s} \Omega \equiv \begin{cases}
    \zeta^4 \var \Psi_4 & s = -2 \\
    \Phi & s = 0 \\
    \var \Psi_0 & s = 2
  \end{cases}.
\end{equation}
The value of $s$ is known as the \emph{spin-weight} of the particular variable.
Moreover, for $s > 0$, one can write these variables in terms of an operator $\pb{s} \boldsymbol{M}$, which maps from the space of gauge fields (such as the metric perturbation $\var g_{ab}$) to the corresponding master variable $\pb{s} \Omega$.
For example, for $|s| = 2$,

\begin{equation} \label{eqn:M_def}
  \pb{s} \Omega = \pb{s} M^{ab} \var g_{ab}.
\end{equation}
From equations~\eqref{eqn:perturbed_weyl_spinor},~\eqref{eqn:decoupled_variables}, and~\eqref{eqn:M_def} (see, for example,~\cite{Chrzanowski:1975wv}),

\begin{subequations} \label{eqn:M}
  \begin{align}
    & \phantom{_-} \begin{aligned}
      \pb{2} M^{ab} = - \frac{1}{2} \Big\{(\npdelta &+ \bar{\pi} - 3\beta - \bar{\alpha}) (\npdelta + \bar{\pi} - 2\beta - 2\bar{\alpha}) l^a l^b + (D - \bar{\rho} - 3\epsilon + \bar{\epsilon}) (D - \bar{\rho} - 2\epsilon + 2\bar{\epsilon}) m^a m^b \\
      &- \big[(D - \bar{\rho} - 3\epsilon + \bar{\epsilon}) (\npdelta + 2\bar{\pi} - 2\beta) + (\npdelta + \bar{\pi} - 3\beta - \bar{\alpha}) (D - 2\bar{\rho} - 2\epsilon)\big] l^a m^b\Big\}, \\
  \end{aligned} \\
    & \begin{aligned}
      \pb{-2} M^{ab} = - \frac{1}{2} \zeta^4 \Big\{(\bar{\npdelta} &- \bar{\tau} + 3\alpha + \bar{\beta}) (\bar{\npdelta} - \bar{\tau} + 2\alpha + 2\bar{\beta}) n^a n^b + (\npDelta + \bar{\mu} + 3\gamma - \bar{\gamma}) (\npDelta + \bar{\mu} + 2\gamma - 2\bar{\gamma}) \bar{m}^a \bar{m}^b \\
      &- \big[(\npDelta + \bar{\mu} + 3\gamma - \bar{\gamma}) (\bar{\npdelta} - 2\bar{\tau} + 2\alpha) + (\bar{\npdelta} - \bar{\tau} + 3\alpha + \bar{\beta}) (\npDelta + 2\bar{\mu} + 2\gamma)\big] n^a \bar{m}^b\Big\}.
    \end{aligned}
  \end{align}
\end{subequations}

In terms of these variables, and in a type D spacetime, the equations of motion for the scalar field $\Phi$ ($s = 0$) and linearized gravity ($s = \pm 2$) may be written in the form~\cite{1973ApJ...185..635T}

\begin{equation} \label{eqn:teukolsky}
  \pb{s} \Box \pb{s} \Omega = 8\pi \pb{s} \boldsymbol{\tau} \cdot \pb{|s|} \boldsymbol{T},
\end{equation}
known as the \emph{Teukolsky equation}.
Here, $\pb{s} \Box$ is a second-order differential operator (the \emph{Teukolsky operator}) that equals, for $s \geq 0$,

\begin{subequations}
  \begin{align} \label{eqn:teukolsky_operator}
    &\phantom{_-} \begin{aligned}
      \pb{s} \Box = 2 \{[&D - (2s - 1) \epsilon + \bar{\epsilon} - 2s\rho - \bar{\rho}] (\npDelta - 2s\gamma + \mu) - [\npdelta - \bar{\alpha} - (2s - 1) \beta - 2s\tau + \bar{\pi}] (\bar{\npdelta} - 2s\alpha + \pi) \\
      &- 2 (2s - 1) (s - 1) \Psi_2\},
    \end{aligned} \\
    &\begin{aligned}
      \pb{-s} \Box = 2 \{[&\npDelta + (2s - 1) \gamma - \bar{\gamma} + \bar{\mu}] [D + 2s\epsilon + (2s - 1) \rho] - [\bar{\npdelta} + (2s - 1) \alpha + \bar{\beta} - \bar{\tau}] [\npdelta + 2s\beta + (2s - 1) \tau] \\
      &- 2 (2s - 1) (s - 1) \Psi_2\}.
    \end{aligned}
  \end{align}
\end{subequations}
On the right-hand side of equation~\eqref{eqn:teukolsky}, $\pb{s} \boldsymbol{\tau}$ is an operator which converts $\pb{s} \boldsymbol{T}$, the source term for the equations of motion (for example, $\pb{2} T^{ab}$ is the stress-energy tensor $\var T^{ab}$), into the source term for the Teukolsky equation~\eqref{eqn:teukolsky}.
For example, one choice of $\pb{\pm 2} \tau_{ab}$ is given by inspection of equations~(2.13) and~(2.15) of~\cite{1973ApJ...185..635T}:

\begin{subequations} \label{eqn:tau}
  \begin{align}
    &\phantom{_-} \begin{aligned}
      \pb{2} \tau_{ab} &= \left[(\npdelta + \bar{\pi} - \bar{\alpha} - 3\beta - 4\tau) l_{(a|} - (D - 3\epsilon + \bar{\epsilon} - 4\rho - \bar{\rho}) m_{(a|}\right] \\
      &\qquad\qquad\qquad\times \left[(D - \epsilon + \bar{\epsilon} - \bar{\rho}) m_{|b)} - (\npdelta + \bar{\pi} - \bar{\alpha} - \beta) l_{|b)}\right],
    \end{aligned} \\
    &\begin{aligned}
      \pb{-2} \tau_{ab} &= \zeta^4 \left[(\npDelta + 3\gamma - \bar{\gamma} + 4\mu + \bar{\mu}) \bar{m}_{(a|} - (\bar{\npdelta} - \bar{\tau} + \bar{\beta} + 3\alpha + 4\pi) n_{(a|}\right] \\
      &\qquad\qquad\qquad\times \left[(\bar{\npdelta} - \bar{\tau} + \bar{\beta} + \alpha) n_{|b)} - (\npDelta + \gamma - \bar{\gamma} + \bar{\mu}) \bar{m}_{|b)}\right].
    \end{aligned}
  \end{align}
\end{subequations}
A freedom in $\pb{\pm 2} \tau_{ab}$ is discussed in section~\ref{section:gauge} below.
One can also rewrite Teukolsky's original result as an operator equation~\cite{PhysRevLett.41.203}, as we will find useful in section~\ref{section:tsi}.
In terms of $\pb{s} \boldsymbol{M}$,

\begin{equation} \label{eqn:decoupling}
  \pb{s} \boldsymbol{\tau} \cdot \pb{|s|} \boldsymbol{\mathcal{E}} = \pb{s} \Box \pb{s} \boldsymbol{M},
\end{equation}
where, for $|s| = 2$, $\pb{|s|} \boldsymbol{\mathcal{E}}$ is the linearized Einstein operator~\eqref{eqn:einstein_operator}.
Applying equation~\eqref{eqn:decoupling} to a metric perturbation and using equation~\eqref{eqn:M_def} and the linearized Einstein equation~\eqref{eqn:linearized_efe} yields the Teukolsky equation~\eqref{eqn:teukolsky} for $|s| = 2$.
Since all of the operations just described are $\mathbb{C}$-linear, equation~\eqref{eqn:decoupling} holds for complexified metric perturbations as well.

So far, we have not tied our discussion to a particular coordinate system, nor a particular tetrad (other than enforcing that we use a principal null tetrad), since we have only required the background metric to be Petrov type D.
We now work in Kerr, and in Boyer-Lindquist coordinates $(t, r, \theta, \phi)$, where the metric takes the form

\begin{equation}
  \ud s^2 = \ud t^2 - \Sigma \left(\frac{\ud r^2}{\Delta} + \ud \theta^2\right) - (r^2 + a^2) \sin^2 \theta \ud \phi^2 - \frac{2Mr}{\Sigma} \left(a \sin^2 \theta \ud \phi - \ud t\right)^2,
\end{equation}
where $\Delta = r^2 - 2Mr + a^2$ and $\Sigma = r^2 + a^2 \cos^2 \theta = |\zeta|^2$, and where we have chosen

\begin{equation} \label{eqn:zeta_bl}
  \zeta = r - ia\cos \theta.
\end{equation}
This choice of $\zeta$ has the property that $\bs{t} \equiv \partial_t$ can be defined in terms of $\zeta_{AB}$~\cite{penrose1988spinors}:

\begin{equation} \label{eqn:time}
  t^{AA'} = -\frac{2}{3} \nabla_B{}^{A'} \zeta^{AB}.
\end{equation}
Using the Kinnersley tetrad (a principal tetrad of the background Weyl tensor), which is given by

\begin{equation} \label{eqn:kinnersley}
  \begin{gathered}
    \bs{l} = \frac{(r^2 + a^2) \partial_t + a \partial_\phi}{\Delta} + \partial_r, \quad \bs{n} = \frac{(r^2 + a^2) \partial_t + a \partial_\phi}{2\Sigma} - \frac{\Delta}{2\Sigma} \partial_r, \\
    \bs{m} = \frac{1}{\sqrt{2} \bar{\zeta}} \left(ia \sin \theta \partial_t + \partial_\theta + \frac{i}{\sin \theta} \partial_\phi\right),
  \end{gathered}
\end{equation}
we find that $\Psi_2 = -M/\zeta^3$.
Furthermore, the non-zero spin coefficients are given by

\begin{equation} \label{eqn:kinnersley_spin_coefficients}
  \begin{gathered}
    \rho = -\frac{1}{\zeta}, \quad \mu = -\frac{\Delta}{2\Sigma \zeta}, \quad \gamma = \mu + \frac{r - M}{2\Sigma}, \\
    \beta = \frac{\cot \theta}{2\sqrt{2} \bar{\zeta}}, \quad \pi = \alpha + \bar{\beta} = \frac{ia}{\sqrt{2} \zeta^2} \sin \theta, \quad \tau = -\frac{ia}{\sqrt{2} \Sigma} \sin \theta.
  \end{gathered}
\end{equation}

We now review how the source-free version of the Teukolsky equation~\eqref{eqn:teukolsky} separates in these coordinates.
Consider, for integers $s$ and $n$, the operators~\cite{1974ApJ...193..443T, chandrasekhar1983mathematical}

\begin{equation} \label{eqn:D_L_operators}
  \mathscr{D}_n = \partial_r + \frac{r^2 + a^2}{\Delta} \partial_t + \frac{a}{\Delta} \partial_\phi + 2n \frac{r - M}{\Delta}, \quad \mathscr{L}_s = \partial_\theta - i\left(a \sin \theta \partial_t + \frac{1}{\sin \theta} \partial_\phi\right) + s \cot \theta.
\end{equation}
Note that these operators satisfy

\begin{equation} \label{eqn:DL_raising}
  \Delta^{-m} \mathscr{D}_n \Delta^m = \mathscr{D}_{n + m}, \qquad \sin^{-r} \theta \mathscr{L}_s \sin^r \theta = \mathscr{L}_{r + s}.
\end{equation}
We also define the operators $\mathscr{D}_n^+$ and $\mathscr{L}_s^+$, by taking $\mathscr{D}_n$ and $\mathscr{L}_s$ and setting $\partial_t \to -\partial_t$ and $\partial_\phi \to -\partial_\phi$; note that $\mathscr{L}_s^+ = \overline{\mathscr{L}_s}\;$~\footnote{Note that here, and below, our definition of the complex conjugate $\overline{\mathcal O}$ of an operator $\mathcal{O}$ is $\overline{\mathcal O} (f) = \overline{\mathcal{O} (\bar f)}$, where $f$ is the argument of this operator.
  This is consistent with the standard notation for the Newman-Penrose operator $\bar{\delta}$.}.
Equations analogous to equations~\eqref{eqn:DL_raising} hold for $\mathscr{D}_n^+$ and $\mathscr{L}_s^+$.
We will also need a way to express these operators in terms of Newman-Penrose operators; using equations~\eqref{eqn:kinnersley} and~\eqref{eqn:kinnersley_spin_coefficients}, we find

\begin{equation} \label{eqn:np_reverse_engineer}
  \mathscr{L}_s = \sqrt{2} \zeta \left(\bar{\npdelta} + 2s \bar{\beta}\right), \qquad \mathscr{D}_n = D + 2n \rho \mu^{-1} (\gamma - \mu), \qquad \mathscr{D}_n^+ = -\rho \mu^{-1} [\npDelta - 2n (\gamma - \mu)].
\end{equation}
Note that these formulae are only valid for the Kinnersley tetrad.
For real frequencies $\omega$ and integers $m$, we further define operators $\mathscr{D}_{nm\omega}$ and $\mathscr{L}_{sm\omega}$ by the requirement that, for any function $f(r, \theta)$,

\begin{equation} \label{eqn:DL_fourier}
  \mathscr{D}_n \left[e^{i(m\phi - \omega t)} f(r, \theta)\right] \equiv e^{i(m\phi - \omega t)} \mathscr{D}_{nm\omega} f(r, \theta), \quad \mathscr{L}_s \left[e^{i(m\phi - \omega t)} f(r, \theta)\right] \equiv e^{i(m\phi - \omega t)} \mathscr{L}_{sm\omega} f(r, \theta).
\end{equation}
This equation yields the formulae

\begin{equation}
  \mathscr{D}_{nm\omega} \equiv \partial_r + \frac{iK_{m\omega}}{\Delta} + 2n \frac{r - M}{\Delta}, \quad \mathscr{L}_{sm\omega} \equiv \partial_\theta + Q_{m\omega} + s \cot \theta,
\end{equation}
where

\begin{equation}
  K_{m\omega} \equiv am - \omega (r^2 + a^2), \quad Q_{m\omega} \equiv m \csc \theta - a \omega \sin \theta
\end{equation}
(note that the conventions for $K_{m\omega}$ in~\cite{chandrasekhar1983mathematical} and~\cite{1973ApJ...185..635T} differ by a sign; here, we use the convention of~\cite{chandrasekhar1983mathematical}).

The operator on the left-hand side of the Teukolsky equation~\eqref{eqn:teukolsky} takes the following simple form:

\begin{equation} \label{eqn:RS_decomp}
  \pb{s} \Box = \pb{s} \mathcal{R} + \pb{s} \mathcal{S},
\end{equation}
where

\begin{subequations} \label{eqn:RS}
  \begin{align}
    \pb{s} \mathcal{R} &\equiv \begin{cases}
      \Delta \mathscr{D}_1 \mathscr{D}^+_s - 2 (2s - 1) r \partial_t & s \geq 0 \\
      \Delta \mathscr{D}_{1 + s}^+ \mathscr{D}_0 - 2 (2s + 1) r \partial_t & s \leq 0
    \end{cases}, \label{eqn:R} \\
    \pb{s} \mathcal{S} &\equiv \begin{cases}
      \mathscr{L}^+_{1 - s} \mathscr{L}_s + 2i (2s - 1) a \cos \theta \partial_t & s \geq 0 \\
      \mathscr{L}_{1 + s} \mathscr{L}^+_{-s} + 2i (2s + 1) a \cos \theta \partial_t & s \leq 0
    \end{cases}, \label{eqn:S}
  \end{align}
\end{subequations}
where it can be readily shown that either the top or bottom lines of equations~\eqref{eqn:R} and~\eqref{eqn:S} yield equal results for $s = 0$; that is, $\pb{+0} \mathcal{R} = \pb{-0} \mathcal{R}$ and $\pb{+0} \mathcal{S} = \pb{-0} \mathcal{S}$.
Note that $\pb{s} \mathcal{R}$ is a differential operator that only depends on $r$, $t$, and $\phi$, while $\pb{s} \mathcal{S}$ only depends on $\theta$, $t$, and $\phi$.
As such, it is clear that the \emph{sourceless} Teukolsky equation~\eqref{eqn:teukolsky} separates in $r$ and $\theta$, and so one can write~\cite{1973ApJ...185..635T}

\begin{equation} \label{eqn:mode_expansion}
  \pb{s} \Omega (t, r, \theta, \phi) = \int_{-\infty}^\infty \ud \omega \sum_{l = |s|}^\infty \sum_{|m| \leq l} \pb{s} \widehat{\Omega}_{lm\omega} (r) \pb{s} \Theta_{lm\omega} (\theta) e^{i(m\phi - \omega t)}.
\end{equation}
Inserting this expansion into the sourceless Teukolsky equation~\eqref{eqn:teukolsky}, followed by using equations~\eqref{eqn:RS_decomp}, \eqref{eqn:RS}, \eqref{eqn:DL_raising}, and~\eqref{eqn:DL_fourier}, one finds that (for $s \geq 0$), the functions $\pb{\pm s} \widehat{\Omega}_{lm\omega}$ and $\pb{\pm s} \Theta_{lm\omega}$ satisfy~\cite{chandrasekhar1983mathematical}

\begin{subequations} \label{eqn:separated_teukolsky}
  \begin{align}
    \left[\mathscr{L}_{(1 - s)(\mp m)(\mp \omega)} \mathscr{L}_{s(\pm m)(\pm \omega)} \pm 2(2s - 1) \omega a \cos \theta\right] \pb{\pm s} \Theta_{lm\omega} &= -\pb{\pm s} \lambda_{lm\omega} \pb{\pm s} \Theta_{lm\omega}, \label{eqn:angular_teukolsky} \\
    \left[\Delta \mathscr{D}_{(1 - s)(\pm m)(\pm \omega)} \mathscr{D}_{0(\mp m)(\mp \omega)} \pm 2i (2s - 1) \omega r\right] \Delta^{(s \pm s)/2} \pb{\pm s} \widehat{\Omega}_{lm\omega} &= \Delta^{(s \pm s)/2} \pb{\pm s} \lambda_{lm\omega} \pb{\pm s} \widehat{\Omega}_{lm\omega}, \label{eqn:radial_teukolsky}
  \end{align}
\end{subequations}
where $\pb{\pm s} \lambda_{lm\omega}$ is a separation constant.
This constant reduces to $(l + s)(l - s + 1) = l(l + 1) - s(s - 1)$ in the Schwarzschild limit~\cite{1973ApJ...185..649P, chandrasekhar1983mathematical}.

The functions $\pb{s} \Theta_{lm\omega}$ are regular solutions to a Sturm-Liouville problem on $[0, \pi]$ with eigenvalues $\pb{s} \lambda_{lm\omega}$.
Thus, there is only one solution for each value of $l$, $m$, and $\omega$, up to scaling.
Note, moreover, that the differential operator on the left-hand side of equation~\eqref{eqn:angular_teukolsky} commutes with the following three operations: complex conjugation, $(s, m, \omega) \to (-s, -m, -\omega)$, and $(s, \theta) \to (-s, \pi - \theta)$.
As such, we can simultaneously diagonalize this operator with each of these operations, choosing $\pb{s} \lambda_{lm\omega}$ and $\pb{s} \Theta_{lm\omega}$ to be real, as well as choosing

\begin{equation} \label{eqn:Theta_flips}
  \pb{s} \Theta_{lm\omega} (\theta) = (-1)^{m + s} \pb{-s} \Theta_{l(-m)(-\omega)} (\theta), \qquad \pb{s} \Theta_{lm\omega} (\pi - \theta) = (-1)^{l + m} \pb{-s} \Theta_{lm\omega} (\theta)
\end{equation}
(a convention which is used by~\cite{1982JPhA...15.3737G}), as well as

\begin{equation} \label{eqn:lambda_flips}
  \pb{s} \lambda_{lm\omega} = \pb{-s} \lambda_{lm\omega} = \pb{s} \lambda_{l(-m)(-\omega)}.
\end{equation}
Finally, the scaling freedom in $\pb{s} \Theta_{lm\omega}$ is fixed by imposing the following normalization condition~\cite{1973ApJ...185..635T}

\begin{equation} \label{eqn:Theta_normalization}
  \int_0^\pi \pb{s} \Theta_{lm\omega} (\theta) \pb{s} \Theta_{l'm\omega} (\theta) \sin \theta \ud \theta = \delta_{ll'}.
\end{equation}
The functions

\begin{equation}
  \pb{s} Y_{lm\omega} (\theta, t, \phi) \equiv e^{i(m\phi - \omega t)} \pb{s} \Theta_{lm\omega} (\theta)
\end{equation}
are the so-called \emph{spin-weighted spheroidal harmonics}, and are orthogonal for different $l$, $m$, and $\omega$.

We now define another expansion for $\pb{s} \Omega$, subtly different from that in equation~\eqref{eqn:mode_expansion}, which results in a convenient way of expanding $\overline{\pb{s} \Omega}$ as well.
To do so, note that the differential operator on the right-hand side of equation~\eqref{eqn:radial_teukolsky} commutes with taking $(m, \omega) \to (-m, -\omega)$ followed by complex conjugation.
As such, we can construct two linearly independent solutions labelled by $p = \pm 1$ [their eigenvalue under this operation, multiplied by a conventional factor of $(-1)^{m + s}$]:

\begin{equation} \label{eqn:p_def}
  \pb{s} \widehat{\Omega}_{lm\omega p} (r) \equiv \frac{1}{2} \left[\pb{s} \widehat{\Omega}_{lm\omega} (r) + p(-1)^{m + s} \overline{\pb{s} \widehat{\Omega}_{l(-m)(-\omega)} (r)}\right],
\end{equation}
and so
\begin{equation}
  \pb{s} \widehat{\Omega}_{lm\omega} (r) = \sum_{p = \pm 1} \pb{s} \widehat{\Omega}_{lm\omega p} (r).
\end{equation}
It is occasionally more convenient to re-express the expansion~\eqref{eqn:mode_expansion} in terms of $\pb{s} \widehat{\Omega}_{lm\omega p} (r)$, instead of $\pb{s} \widehat{\Omega}_{lm\omega} (r)$:

\begin{equation} \label{eqn:decoupled_expansion}
  \pb{s} \Omega (t, r, \theta, \phi) = \int_{-\infty}^\infty \ud \omega \sum_{l = |s|}^\infty \sum_{|m| \leq l} \sum_{p = \pm 1} e^{i(m\phi - \omega t)} \pb{s} \Theta_{lm\omega} (\theta) \pb{s} \widehat{\Omega}_{lm\omega p} (r).
\end{equation}
A simple consequence of equations~\eqref{eqn:Theta_flips} and~\eqref{eqn:p_def} is that

\begin{equation} \label{eqn:decoupled_bar_expansion}
  \overline{\pb{s} \Omega (t, r, \theta, \phi)} = \int_{-\infty}^\infty \ud \omega \sum_{l = |s|}^\infty \sum_{|m| \leq l} \sum_{p = \pm 1} p e^{i(m\phi - \omega t)} \pb{-s} \Theta_{lm\omega} (\theta) \pb{s} \widehat{\Omega}_{lm\omega p} (r),
\end{equation}
and so this is a convenient expansion of the complex conjugate of the master variables.
Note, however, that these expansions are different in status from the expansion~\eqref{eqn:mode_expansion}, as the coefficients in this expansion must satisfy
\begin{equation} \label{eqn:eigenvalue_condition}
  \overline{\pb{s} \widehat{\Omega}_{l(-m)(-\omega)p} (r)} = p (-1)^{m + s} \pb{s} \widehat{\Omega}_{lm\omega p} (r).
\end{equation}

\section{Symmetry operators} \label{section:symmetry}

As defined by Kalnins, McLenaghan, and Williams~\cite{1992RSPSA.439..103K}, a \emph{symmetry operator} is an $\mathbb{R}$-linear operator that maps the space of solutions to the equations of motion, which must be linear, into itself.
For the space of complexified solutions to real equations of motion, there exists a trivial symmetry operator mapping solutions to their complex conjugates.
In his original paper, Carter constructed the symmetry operator for scalar fields in equation~\eqref{eqn:carter_operator}, which commutes with the d'Alembertian~\cite{PhysRevD.16.3395}.
If an operator commutes with the operators in the sourceless equations of motion, then it must be a symmetry operator: if a field $\phi$ satisfies $\mathcal{L} \phi = 0$, and $[\mathcal{D}, \mathcal{L}] = 0$, then

\begin{equation}
  \mathcal{L} \mathcal{D} \phi = \mathcal{D} \mathcal{L} \phi = 0,
\end{equation}
and so $\mathcal{D} \phi$ is a solution.
Lie derivatives with respect to Killing vectors are examples of symmetry operators which commute with the equations of motion.
Further examples of symmetry operators can be created by composing symmetry operators associated with Killing vectors, but these are, in a sense, ``reducible''.

In this section we review two classes of \emph{irreducible} symmetry operators that appear in the Kerr spacetime: those that derive from separation of variables, and those that arise from taking the adjoint of the Teukolsky equation.
Note that, recently, additional symmetry operators have been discussed in the Kerr spacetime~\cite{Aksteiner:2016mol}, which we do not discuss in this paper.

\subsection{Separation of variables} \label{section:separation}

The first class of symmetry operators we consider is associated with the separability of the underlying equations of motion.
To see that there is always a symmetry operator associated with separability, consider as an example the following partial differential equation (in two variables $x, y$):

\begin{equation} \label{eqn:example}
  \mathcal{L} \phi \equiv \left[\mathcal{X} (x, \partial_x, \ldots) + \mathcal{Y} (y, \partial_y, \partial_y^2, \ldots)\right] \phi = 0,
\end{equation}
for some differential operators $\mathcal{X}$ and $\mathcal{Y}$.
Since $\mathcal{X}$ only depends upon $x$ and $\mathcal{Y}$ only depends upon $y$, $\mathcal{X}$ and $\mathcal{Y}$ must commute.
Moreover, $\mathcal{L} = \mathcal{X} + \mathcal{Y}$, and so $\mathcal{X}$ and $\mathcal{Y}$ must both commute with $\mathcal{L}$, and so $\mathcal{X}$ and $\mathcal{Y}$ are symmetry operators.
In addition, if there are additional variables $z_1, \ldots, z_n$, and $\mathcal{X}$ and $\mathcal{Y}$ only depend on derivatives with respect to these variables, then this argument still holds.

Irreducible symmetry operators arise in Kerr, similarly, via a separation of variables argument.
As discussed in section~\ref{section:teukolsky}, the Teukolsky equation separates, yielding the two operators $\pb{s} \mathcal{R}$ and $\pb{s} \mathcal{S}$ in equations~\eqref{eqn:R} and~\eqref{eqn:S} (respectively).
These operators are analogous to the operators $\mathcal{X}$ and $\mathcal{Y}$ in equation~\eqref{eqn:example} above, and depend on derivatives with respect to additional variables $t$ and $\phi$.
One combination of $\pb{s} \mathcal{R}$ and $\pb{s} \mathcal{S}$ is particularly interesting, namely

\begin{equation} \label{eqn:D}
  \pb{s} \mathcal{D} \equiv \frac{1}{2} \left(\pb{s} \mathcal{R} - \pb{s} \mathcal{S}\right).
\end{equation}
One can show that, for $s = 0$, this is in fact the scalar symmetry operator~\eqref{eqn:carter_operator} discussed by Carter~\cite{PhysRevD.16.3395}.

In the case of linearized gravity, $\pb{s} \mathcal{D}$ is a map from the space of solutions of the homogeneous Teukolsky equation~\eqref{eqn:teukolsky} of spin weight $s$ into itself.
In section~\ref{section:symmetry_separated}, we will review a procedure (a version of Chrzanowski metric reconstruction~\cite{Chrzanowski:1975wv}) which will allow us to construct another operator $\pb{s} \mathcal{D}_{ab}{}^{cd}$ from $\pb{s} \mathcal{D}$ that maps the space of complexified metric perturbations into itself.
The symmetry operator $\pb{s} \mathcal{D}_{ab}{}^{cd}$ will be more useful than $\pb{s} \mathcal{D}$, since the symplectic product for linearized gravity naturally acts on the space of metric perturbations.

\subsection{Adjoint symmetry operators} \label{section:tsi}

In Kerr, for spins higher than 0, there is a second set of irreducible symmetry operators that can be constructed, following an argument due to Wald~\cite{PhysRevLett.41.203}.
This argument holds, as do many of our equations, for all $|s| \leq 2$; however, we will only explicitly use $|s| = 2$ in this paper.

The argument is as follows.
We first define the adjoint of a linear differential operator.
Consider a linear differential operator $\bs{\mathcal L}$ that takes tensor fields of rank $p$ to tensor fields of rank $q$.
We say that an operator which takes tensor fields of rank $q$ to tensor fields of rank $p$ is the adjoint $\bs{\mathcal L}^\dagger$ of $\bs{\mathcal L}$ if, for all tensor fields $\bs{\phi}$ of rank $p$ and tensor fields $\bs{\psi}$ of rank $q$, there exists a vector field $j^a [\bs{\phi}, \bs{\psi}]$ such that

\begin{equation} \label{eqn:adjoint}
   \boldsymbol{\psi} \cdot (\bs{\mathcal L} \cdot \boldsymbol{\phi}) - \boldsymbol{\phi} \cdot (\bs{\mathcal L}^\dagger \cdot \boldsymbol{\psi}) = \nabla_a j^a [\boldsymbol{\phi}, \boldsymbol{\psi}].
\end{equation}
Note that this is not the usual definition of adjoint, which has a complex conjugate acting on $\boldsymbol{\psi}$ in the first term and on $(\mathcal{L}^\dagger \boldsymbol{\psi})$ in the second.
Chrzanowski~\cite{Chrzanowski:1975wv} and Gal'tsov~\cite{1982JPhA...15.3737G} use the usual definition, whereas Wald uses the definition~\eqref{eqn:adjoint}.

We now give some examples of adjoints of the operators considered in section~\ref{section:teukolsky}.
First, we note that one can easily show that, for two operators $\bs{\mathcal L}_1$ and $\bs{\mathcal L}_2$,

\begin{equation}
  (\bs{\mathcal L}_1 \bs{\mathcal L}_2)^\dagger = \bs{\mathcal L}_2^\dagger \bs{\mathcal L}_1^\dagger.
\end{equation}
Moreover, the adjoints of the various Newman-Penrose operators, using equations~\eqref{eqn:soldering}, \eqref{eqn:spin_coefficients}, and~\eqref{eqn:adjoint}, are given by

\begin{equation} \label{eqn:adjoint_np}
  D^\dagger = -D - (\epsilon + \bar{\epsilon}) + \rho + \bar{\rho},
\end{equation}
together with the corresponding expressions obtained via $'$ and $*$ transformations.
Using equations~\eqref{eqn:adjoint} and~\eqref{eqn:einstein_operator}, one finds that $\pb{2} \bs{\mathcal E}$ is self-adjoint:

\begin{equation} \label{eqn:einstein_dagger}
  \pb{2} \bs{\mathcal E}^\dagger = \pb{2} \bs{\mathcal E}.
\end{equation}
Similarly, one can show from equations~\eqref{eqn:adjoint_np} and~\eqref{eqn:teukolsky_operator} that

\begin{equation} \label{eqn:teukolsky_dagger}
  \pb{s} \Box^\dagger = \pb{-s} \Box,
\end{equation}
as was first noted by Cohen and Kegeles~\cite{PhysRevD.10.1070}.
Finally, the adjoint of the operator $\pb{s} \boldsymbol{\tau}$ [equation~\eqref{eqn:tau}] that enters into the Teukolsky equation~\eqref{eqn:teukolsky}, for $|s| = 2$, is given by

\begin{equation} \label{eqn:tau_dagger}
  \pb{s} \tau_{ab}^{\dagger} = \begin{cases}
    [m_{(a|} (D + 2\epsilon - \rho) - l_{(a|} (\npdelta + 2\beta - \tau)] [l_{|b)} (\npdelta + 4\beta + 3\tau) - m_{|b)} (D + 4\epsilon + 3\rho)] & s = 2 \\
    [\bar{m}_{(a|} (\npDelta - 2\gamma + \mu) - n_{(a|} (\bar{\npdelta} - 2\alpha + \pi)] [n_{|b)} (\bar{\npdelta} - 4\alpha - 3\pi) - \bar{m}_{|b)} (\npDelta - 4\gamma - 3\mu)] \zeta^4 & s = -2 \\
  \end{cases}.
\end{equation}

We now take the adjoint of equation~\eqref{eqn:decoupling}, yielding [from equations~\eqref{eqn:teukolsky_dagger} and~\eqref{eqn:einstein_dagger}]

\begin{equation} \label{eqn:adjoint_teukolsky}
  \pb{|s|} \boldsymbol{\mathcal{E}} \cdot \pb{s} \boldsymbol{\tau}^\dagger = \pb{s} \boldsymbol{M}^\dagger \pb{-s} \Box.
\end{equation}
Suppose that we have a solution $\pb{-s} \psi$ to the vacuum Teukolsky equation $\pb{-s} \Box \pb{-s} \psi = 0$; note that $\pb{-s} \psi$ is not necessarily the master variable $\pb{-s} \Omega$ associated with $\var g_{ab}$ via equation~\eqref{eqn:M_def}.
Then, from equations~\eqref{eqn:adjoint_teukolsky},

\begin{equation}
  \pb{|s|} \boldsymbol{\mathcal{E}} \cdot \pb{s} \boldsymbol{\tau}^\dagger \pb{-s} \psi = 0.
\end{equation}
Thus, $\pb{s} \boldsymbol{\tau}^\dagger \pb{-s} \psi$ is a \emph{complex} metric perturbation that solves the vacuum linearized Einstein equations.

Thus, the operator $\pb{s} \boldsymbol{\tau}^\dagger$ allows the construction of complex vacuum metric perturbations from vacuum solutions to the Teukolsky equation.
From a single solution $\pb{-s} \psi$ to the vacuum Teukolsky equation~\eqref{eqn:teukolsky} of spin weight $-s$, one can therefore apply $\pb{s'} \bs{M}$ (for some other $s'$, where $|s'| = |s|$) to either $\pb{s} \bs{\tau}^\dagger \pb{-s} \psi$ or $\overline{\pb{s} \bs{\tau}^\dagger \pb{-s} \psi}$, both of which yield solutions to the vacuum Teukolsky equation:

\begin{equation} \label{eqn:tsi}
  \pb{s'} \Box \pb{s'} \boldsymbol{M} \cdot \pb{s} \boldsymbol{\tau}^\dagger \pb{-s} \psi = 0, \qquad \pb{s'} \Box \pb{s'} \boldsymbol{M} \cdot \overline{\pb{s} \boldsymbol{\tau}^\dagger \pb{-s} \psi} = 0.
\end{equation}
That is, there exist two symmetry operators of the form

\begin{equation} \label{eqn:C}
  \pb{s', s} \mathcal{C} \equiv \pb{s'} \boldsymbol{M} \cdot \pb{s} \boldsymbol{\tau}^\dagger, \qquad \pb{s', s} \widetilde{\mathcal C} \equiv \pb{s'} \boldsymbol{M} \cdot \overline{\pb{s} \boldsymbol{\tau}^\dagger}.
\end{equation}
The operator $\pb{s', s} \mathcal{C}$ maps from the space of solutions to the vacuum Teukolsky equation~\eqref{eqn:teukolsky} of spin weight $-s$ to the space of solutions to the vacuum Teukolsky equation of spin weight $s'$.
Similarly, $\pb{s', s} \widetilde{\mathcal C}$ maps from the space of solutions to the complex conjugate of the vacuum Teukolsky equation~\eqref{eqn:teukolsky} of spin weight $-s$ into the space of solutions to the vacuum Teukolsky equation of spin weight $s'$.

As in section~\ref{section:separation}, these operators act on the master variables, rather than metric perturbations.
However, one can also construct the operators (for $|s| = 2$)

\begin{equation} \label{eqn:C_tensor}
  \pb{s} \mathcal{C}_{ab}{}^{cd} \equiv \pb{s} \tau_{ab}^\dagger \pb{-s} M^{cd},
\end{equation}
which are symmetry operators for metric perturbations.
That is, they are $\mathbb{R}$-linear maps from the space of complexified solutions to the vacuum linearized Einstein equations into itself.
This follows from the operator identity (derived from equations~\eqref{eqn:adjoint_teukolsky} and~\eqref{eqn:C_tensor})

\begin{equation}
  \pb{|s|} \boldsymbol{\mathcal{E}} \cdot \pb{s} \boldsymbol{\mathcal{C}} = \pb{s} \boldsymbol{M}^\dagger \pb{-s} \Box \pb{-s} \boldsymbol{M} = \pb{s} \boldsymbol{M}^\dagger \pb{-s} \boldsymbol{\tau} \cdot \pb{|s|} \boldsymbol{\mathcal{E}},
\end{equation}
where the second equality from equation~\eqref{eqn:decoupling}.
Applying this operator identity to (in general) a complex vacuum metric perturbation, the right-hand side yields zero.
Note that the two cases $s = \pm 2$ in equations~\eqref{eqn:tau_dagger} and~\eqref{eqn:M} differ by a $'$ transformation, along with a factor of $\zeta^4$, and so $\pb{2} \mathcal{C}_{ab}{}^{cd}$ and $\pb{-2} \mathcal{C}_{ab}{}^{cd}$ are related by a $'$ transformation.
Furthermore, the metric perturbations generated by $\pb{\pm 2} \mathcal{C}_{ab}{}^{cd}$ are in a trace-free gauge by construction.

Finally, we note that this argument has been used in a fully tetrad-invariant form, using a spinor form of the Teukolsky equations, to generate symmetry operators for metric perturbations of the sort that we review in this section~\cite{Aksteiner:2016mol}.
For simplicity, we use the Newman-Penrose form of the Teukolsky equations instead.

\subsection{Issues of gauge} \label{section:gauge}

Since the operators $\pb{\pm 2} \tau_{ab}^\dagger$ map into the space of metric perturbations which are solutions to the linearized Einstein equation, the solutions which these operators generate will be in a particular gauge.
This gauge freedom can be understood in the following way: the operators $\pb{\pm 2} \tau_{ab}$ in equation~\eqref{eqn:teukolsky} are only defined up to transformations of the form

\begin{equation} \label{eqn:tau_freedom}
  \pb{\pm 2} \tau_{ab} \to \pb{\pm 2} \tau_{ab} + 2 \xi_{(a} \nabla_{b)},
\end{equation}
as they act upon the stress-energy tensor, for which $\nabla_a T^{ab} = 0$.
As such, we find that $\pb{\pm 2} \tau_{ab}^\dagger$ have the corresponding freedom

\begin{equation}
  \pb{\pm 2} \tau_{ab}^\dagger \to \pb{\pm 2} \tau_{ab}^\dagger + 2 \nabla_{(a} \xi_{b)}.
\end{equation}
Note here that, in the second term, the covariant derivative acts upon the argument of these operators in addition to acting on $\xi_b$.
The particular choice~\eqref{eqn:tau} of $\pb{\pm 2} \tau_{ab}$ fixes this freedom, and so the metric perturbations generated by $\pb{\pm 2} \mathcal{C}_{ab}{}^{cd}$ are in a particular gauge.
The gauge conditions which they satisfy are~\cite{Chrzanowski:1975wv}

\begin{equation} \label{eqn:radn_gauges}
  g^{ab} \pb{\pm 2} \tau_{ab}^\dagger = 0, \qquad l^a \pb{2} \tau_{ab}^\dagger = 0, \qquad n^a \pb{-2} \tau_{ab}^\dagger = 0.
\end{equation}
For $\pb{2} \tau_{ab}^\dagger$, this is the \emph{ingoing radiation gauge condition}, whereas for $\pb{-2} \tau_{ab}^\dagger$, this is the \emph{outgoing radiation gauge condition}.

We now show that the solutions $\pb{2} \boldsymbol{\mathcal{C}} \cdot \var \boldsymbol{g}$ and $\pb{-2} \boldsymbol{\mathcal{C}} \cdot \var \boldsymbol{g}$ do not differ by a gauge transformation, in the case where $\var g_{ab}$ is real.
This is in contrast to the case in electromagnetism~\cite{Grant:2019qyo}, where the analogous solutions do, in fact, differ by a gauge transformation.
While the total solutions $\pb{2} \boldsymbol{\mathcal{C}} \cdot \var \boldsymbol{g}$ and $\pb{-2} \boldsymbol{\mathcal{C}} \cdot \var \boldsymbol{g}$ do not differ by a gauge transformation, we will also show that the imaginary parts of each of these two solutions \emph{are} related by a gauge transformation, and so they represent the same physical solution.

To proceed, we first note the following identities [note a conventional factor of two difference with~\cite{PhysRevD.19.1641}, which comes from the difference between their equation (2.21) and our equation~\eqref{eqn:perturbed_weyl_spinor_metric}]

\begin{subequations} \label{eqn:nonzero_operators}
  \begin{align}
    \overline{\pb{2} \boldsymbol{M}} \cdot \pb{2} \boldsymbol{\mathcal{C}} &\circeq \frac{1}{2} (D + \epsilon - 3\bar{\epsilon}) (D + 2\epsilon - 2\bar{\epsilon}) (D + 3\epsilon - \bar{\epsilon}) (D + 4\epsilon) \pb{-2} \boldsymbol{M}, \\
    \overline{\pb{-2} \boldsymbol{M}} \cdot \pb{2} \boldsymbol{\mathcal{C}} &\circeq \frac{1}{2} \bar{\zeta}^4 (\npdelta + 3\bar{\alpha} + \beta) (\npdelta + 2\bar{\alpha} + 2\beta) (\npdelta + \bar{\alpha} + 3\beta) (\npdelta + 4\beta) \pb{-2} \boldsymbol{M}, \\
    \overline{\pb{-2} \boldsymbol{M}} \cdot \overline{\pb{2} \boldsymbol{\mathcal{C}}} &\circeq \frac{3}{2} \overline{\zeta^4 \Psi_2} \left[\bar{\tau} (\npdelta + 4\bar{\alpha}) - \bar{\rho} (\npDelta + 4\bar{\gamma}) - \bar{\mu} (D + 4\bar{\epsilon}) + \bar{\pi} (\bar{\npdelta} + 4\bar{\beta}) + 2\overline{\Psi}_2\right] \overline{\pb{-2} \boldsymbol{M}} \nonumber \\
    &= \frac{3}{2} \overline{\zeta^3 \Psi_2} t^a [\nabla_a + 4 (\iota_B \nabla_a o^B)] \overline{\pb{-2} \boldsymbol{M}}, \label{eqn:nonzero_operators_t}
  \end{align}
\end{subequations}
where ``$\circeq$'' means ``equality modulo equations of motion''.
Moreover, apart from those that occur in this equation, all other combinations of $\pb{\pm 2} \boldsymbol{M}$ and $\overline{\pb{\pm 2} \boldsymbol{M}}$ acting on $\pb{2} \boldsymbol{\mathcal{C}}$ and $\overline{\pb{2} \boldsymbol{\mathcal{C}}}$ are zero for vacuum solutions.
Here we have used the equation

\begin{equation}
  D \rho = (\rho + \epsilon + \bar{\epsilon}) \rho
\end{equation}
(along with its $'$- and $*$-transformed versions) in order to simplify, as well as equation~\eqref{eqn:time}.
One can furthermore use a $'$-transformation to write down versions of equation~\eqref{eqn:nonzero_operators} involving $\pb{-2} \bs{\mathcal C}$, noting that $\Psi_2 \to \Psi_2$ under a $'$-transformation, and $\zeta$ must flip sign (note that $t^a$ keeps the same sign).

To determine whether certain linear combinations of $\pb{\pm 2} \mathcal{C}_{ab}{}^{cd} \var g_{cd}$ (and their complex conjugates) differ by gauge transformations, we need the following relation, which only holds for $\var \Psi_4$ and $\var \Psi_0$ coming from the same real vacuum metric perturbations:

\begin{equation} \label{eqn:tsr}
  \begin{split}
    (D + \epsilon - 3\bar{\epsilon}) (D + 2\epsilon - 2\bar{\epsilon}) (D + 3\epsilon - \bar{\epsilon}) (D + 4\epsilon) \zeta^4 \var \Psi_4 &= (\bar{\npdelta} - \alpha - 3\bar{\beta}) (\bar{\npdelta} - 2\alpha - 2\bar{\beta}) (\bar{\npdelta} - 3\alpha - \bar{\beta}) (\bar{\npdelta} - 4\alpha) \zeta^4 \var \Psi_0 \\
    &\hspace{1em}+ 3\overline{\zeta^3 \Psi_2} t^a [\nabla_a - 4 (\iota_B \nabla_a o^B)] \overline{\var \Psi_0};
  \end{split}
\end{equation}
we will also need this equation's $'$-transform.
This relation can be derived using the perturbed Bianchi identities and Newman-Penrose equations, as mentioned in~\cite{PhysRevD.14.317}; for a more modern derivation, see for example~\cite{Aksteiner:2016pjt}.
Using equations~\eqref{eqn:nonzero_operators} and~\eqref{eqn:tsr}, along with their $'$-transforms, we find that (applied to a real, vacuum metric perturbation),

\begin{equation} \label{eqn:tsr_operators}
  \overline{\pb{2} \bs{M}} \cdot \pb{2} \bs{\mathcal C} \circeq \overline{\pb{2} \bs{M}} \cdot \pb{-2} \bs{\mathcal C} - \overline{\pb{2} \bs{M}} \cdot \overline{\pb{-2} \bs{\mathcal C}}.
\end{equation}
The $'$-transform of this equation merely switches $2 \to -2$.
As remarked below equation~\eqref{eqn:nonzero_operators}, one has that

\begin{equation}
  \overline{\pb{2} \bs{M}} \cdot \overline{\pb{2} \bs{\mathcal C}} \circeq 0
\end{equation}
(along with its $'$-transform), and so one therefore has that

\begin{equation}
  \overline{\pb{2} \boldsymbol{M}} \cdot \Im\left[\pb{+2} \boldsymbol{\mathcal{C}} - \pb{-2} \boldsymbol{\mathcal{C}}\right] \cdot \var \bs{g} = 0, \qquad \overline{\pb{-2} \boldsymbol{M}} \cdot \Im\left[\pb{+2} \boldsymbol{\mathcal{C}} - \pb{-2} \boldsymbol{\mathcal{C}}\right] \cdot \var \bs{g} = 0. \\
\end{equation}

This equation does not, as it stands, guarantee that $\Im[\pb{2} \mathcal{\bs C} \cdot \var \bs{g}]$ and $\Im[\pb{-2} \mathcal{\bs C} \cdot \var \bs{g}]$ are related by a gauge transformation, just that the master variables associated with these two metric perturbations are equal.
This implies that their difference is a metric perturbation that contributes to $\var M$ and $\var a$; that is, it only has monopole and dipole terms~\cite{1973JMP....14.1453W}.
One would expect that $\Im[\pb{\pm 2} \mathcal{C}_{ab}{}^{cd} \var g_{cd}]$, as they are constructed wholly from the radiative Weyl scalars $\var \Psi_0$ and $\var \Psi_4$ (which do not have monopole or dipole pieces), would not have non-radiating pieces.
This statement is in fact correct due to arguments in~\cite{Stewart:1978tm}.
In conclusion, we find that $\Im[\pb{2} \mathcal{\bs C} \cdot \var \bs{g}]$ and $\Im[\pb{-2} \mathcal{\bs C} \cdot \var \bs{g}]$ differ by a gauge transformation:
\begin{equation}
  \Im[\pb{2} \mathcal{C}_{ab}{}^{cd} \var g_{cd}] = \Im[\pb{-2} \mathcal{C}_{ab}{}^{cd} \var g_{cd}] + 2 \nabla_{(a} \xi_{b)},
\end{equation}
for some vector field $\xi^a$.
The main theorem of~\cite{Aksteiner:2016pjt} provides an alternative proof of this result, as does the discussion in section 4.3 of~\cite{Aksteiner:2016mol}.

\subsection{Action of symmetry operators on expansions} \label{section:symmetry_separated}

In section~\ref{section:teukolsky}, we showed that the master variables (and their complex conjugates) have convenient expansions [equations~\eqref{eqn:decoupled_expansion} and~\eqref{eqn:decoupled_bar_expansion}] in terms of spin-weighted spheroidal harmonics.
We show in this section that the symmetry operators considered in this paper which act on the master variables are ``diagonal'', in the sense that they act upon each term in these expansions by simply multiplying each term by an overall constant.
We then construct a similar expansion for vacuum metric perturbations, and show that the action of the symmetry operators that we have defined for metric perturbations are also diagonal on this expansion.

First, let us consider the action of the symmetry operator $\pb{s} \mathcal{D}$ defined in equation~\eqref{eqn:D}.
From equations~\eqref{eqn:RS}, \eqref{eqn:DL_raising},~\eqref{eqn:DL_fourier}, and~\eqref{eqn:separated_teukolsky}, it follows that

\begin{equation} \label{eqn:D_action}
  \pb{s} \mathcal{D} \pb{s} \Omega = \int_{-\infty}^\infty \sum_{l = |s|}^\infty \sum_{|m| \leq l} \sum_{p = \pm 1} \pb{|s|} \lambda_{lm\omega} e^{i(m\phi - \omega t)} \pb{s} \Theta_{lm\omega} (\theta) \pb{s} \widehat{\Omega}_{lm\omega p} (r).
\end{equation}
Later in this section, we will also show that a similar diagonalization occurs for a tensor version of this operator, which we will define in equation~\eqref{eqn:D_tensor}.

Next, we consider the symmetry operators $\pb{s', s} \widetilde{\mathcal C}$ defined in equation~\eqref{eqn:C}.
We begin by noting that these symmetry operators simplify with the choice of Boyer-Lindquist coordinates and the Kinnersley tetrad, yielding the so-called ``spin-inversion'' operators~\cite{Chrzanowski:1975wv, 1982JPhA...15.3737G}:

\begin{subequations} \label{eqn:spinversion}
  \begin{align}
    \pb{2, 2} \widetilde{\mathcal C} &= \frac{1}{2} \mathscr{D}_0^4, \qquad &\pb{-2, -2} \widetilde{\mathcal C} &= \frac{1}{32} \Delta^2 \left(\mathscr{D}_0^+\right)^4 \Delta^2, \label{eqn:radial_spinversion} \\
    \pb{2, -2} \widetilde{\mathcal C} &= \frac{1}{8} \mathscr{L}_{-1}^+ \mathscr{L}_0^+ \mathscr{L}_1^+ \mathscr{L}^+_2, \qquad &\pb{-2, 2} \widetilde{\mathcal C} &= \frac{1}{8} \mathscr{L}_{-1} \mathscr{L}_0 \mathscr{L}_1 \mathscr{L}_2. \label{eqn:angular_spinversion}
  \end{align}
\end{subequations}
The constant numerical factors here are consistent with those of Wald~\cite{PhysRevLett.41.203} and Chrzanowski~\cite{Chrzanowski:1975wv}, but disagree with those of other authors (such as~\cite{chandrasekhar1983mathematical, 1982JPhA...15.3737G}) due to normalization conventions.

These operators are referred to as spin-inversion operators for the following reason.
Considering their action on the terms in the expansion~\eqref{eqn:decoupled_bar_expansion} of $\overline{\pb{s} \Omega}$, they are either purely radial [equation~\eqref{eqn:radial_spinversion}] or purely angular [equation~\eqref{eqn:angular_spinversion}].
Due to this fact, along with the expansions in equations~\eqref{eqn:decoupled_expansion} and~\eqref{eqn:decoupled_bar_expansion}, it is apparent that, when acting on the terms in these expansions, the operator $\pb{2, 2} \widetilde{\mathcal C}$ maps from the space of solutions to the radial Teukolsky equation~\eqref{eqn:radial_teukolsky} with $s = -2$ to $s = 2$, and similarly $\pb{-2, -2} \widetilde{\mathcal C}$ maps from solutions with $s = 2$ to $s = -2$.
Similarly, for the angular operators, due to the fact that the expansion for $\overline{\pb{s} \Omega}$ is in terms of $\pb{-s} \Theta_{lm\omega}$, $\pb{2, -2} \widetilde{\mathcal C}$ maps from the space of solutions to angular Teukolsky equation~\eqref{eqn:angular_teukolsky} with $s = 2$ to $s = -2$, and similarly $\pb{-2, 2} \widetilde{\mathcal C}$ maps from $s = -2$ to $s = 2$.

We now show that the spin-inversion operators merely multiply each term in the expansion~\eqref{eqn:decoupled_bar_expansion} by some constant, starting with the angular spin-inversion operators.
The angular Teukolsky equation~\eqref{eqn:angular_teukolsky} is a Sturm-Liouville problem, which only has one solution for a given value of $l$, $m$, and $\omega$ (up to normalization).
If the angular spin-inversion operators, when acting upon individual terms in the expansion~\eqref{eqn:decoupled_bar_expansion}, map between the two spaces of solutions with $s = \pm 2$, then these maps can be entirely characterized by two overall constants, which we denote by $\pb{\pm 2} C_{lm\omega}$:

\begin{equation} \label{eqn:angular_starobinsky}
  \mathscr{L}_{-1(\pm m)(\pm \omega)} \mathscr{L}_{0(\pm m)(\pm \omega)} \mathscr{L}_{1(\pm m)(\pm \omega)} \mathscr{L}_{2(\pm m)(\pm \omega)} \pb{\pm 2} \Theta_{lm\omega} \equiv \pb{\pm 2} C_{lm\omega} \pb{\mp 2} \Theta_{lm\omega}.
\end{equation}
This equation is known as the \emph{angular Teukolsky-Starobinsky identity}.
Since these operators are entirely real, this constant $\pb{\pm 2} C_{lm\omega}$ is also real.
Moreover, the normalization condition for $\pb{s} \Theta_{lm\omega}$ implies that~\cite{chandrasekhar1983mathematical}

\begin{equation}
  \pb{2} C_{lm\omega} = \pb{-2} C_{lm\omega} \equiv C_{lm\omega},
\end{equation}
where

\begin{equation} \label{eqn:starobinsky}
  C_{lm\omega}^2 = \pb{2} \lambda_{lm\omega}^2 (\pb{2} \lambda_{lm\omega} + 2)^2 - 8 \omega^2 \pb{2} \lambda_{lm\omega} [\alpha_{m\omega}^2 (5 \pb{2} \lambda_{lm\omega} + 6) - 12a^2] + 144 \omega^4 \alpha_{m\omega}^4,
\end{equation}
and

\begin{equation}
  \alpha_{m\omega}^2 = a^2 - am/\omega.
\end{equation}

We now turn to the case of the radial operators in equation~\eqref{eqn:radial_spinversion}, which are somewhat more complicated.
This is because there are two solutions to the radial equation~\eqref{eqn:radial_teukolsky}, as it is second-order, and not a Sturm-Liouville problem.
However, as noted in section~\ref{section:teukolsky}, the two solutions can be characterized by their eigenvalues under the transformation $(m, \omega) \to (-m, -\omega)$, followed by complex conjugation.
Since the radial spin-inversion operator is also invariant under this transformation, we must therefore have that

\begin{equation} \label{eqn:radial_starobinsky}
  \Delta^2 \mathscr{D}_{0(\mp m)(\mp \omega)}^4 \Delta^{(s \pm s)/2} \pb{\pm 2} \widehat{\Omega}_{lm\omega p} \equiv 2^{\pm 2} \pb{\pm 2} C_{lm\omega p} \Delta^{(s \mp s)/2} \pb{\mp 2} \widehat{\Omega}_{lm\omega p}
\end{equation}
(the factor of $2^{\pm 2}$ is purely conventional, and is present only to make our final expressions simpler).
This equation is known as the \emph{radial Teukolsky-Starobinsky identity}.

To determine the values of the constants $\pb{\pm 2} C_{lm\omega p}$, we need to use the fact that $\pb{\pm 2} \Omega$ come from the same real metric perturbation.
The values of these constants given by Teukolsky and Press in their original paper~\cite{1974ApJ...193..443T} only hold for the $p = 1$ case (as pointed out by Bardeen~\cite{bardeen}\footnote{That~\cite{1974ApJ...193..443T} only considers $p = 1$ can be seen from their equation~(3.21), along with the remark below their equation~(3.22) that the quantities $S_2$ and $S_2^\dagger$ that appear in this equation are given by $\pb{2} S_{lm}$ and $\pb{-2} S_{lm}$ (in this chapter, these are denoted $\pb{2} \Theta_{lm\omega}$ and $\pb{-2} \Theta_{lm\omega}$).
  These two statements imply that the radial functions $R_s$ discussed in~\cite{1974ApJ...193..443T} obey

  \[
     \overline{R_s (-m, -\omega)} = R_s (m, \omega).
  \]
  In this paper, due to differences in notation and the conventions in equation~\eqref{eqn:Theta_flips}, this is equivalent to the statement that $\pb{s} \widehat{\Omega}_{l(-m)(-\omega)} = (-1)^{m + s} \overline{\pb{s} \widehat{\Omega}_{lm\omega}}$, which by equation~\eqref{eqn:eigenvalue_condition} implies that $p = 1$.}).
The values of $\pb{\pm 2} C_{lm\omega p}$ are found using equation~\eqref{eqn:tsr_operators}, since (in terms of $\pb{s} \Omega$) the complex conjugate of this equation (and its $'$-transform) can be written as

\begin{equation} \label{eqn:tsi_scalar}
  \pb{-s, -s} \widetilde{\mathcal C}\; \overline{\pb{s} \Omega} = \pb{-s, s} \widetilde{\mathcal C}\; \overline{\pb{-s} \Omega} - \pb{-s, s} \mathcal{C} \pb{-s} \Omega.
\end{equation}
Using equations~\eqref{eqn:spinversion},~\eqref{eqn:angular_starobinsky}, and~\eqref{eqn:radial_starobinsky}, as well as~\eqref{eqn:nonzero_operators_t}, we find that

\begin{subequations} \label{eqn:C_action}
  \begin{align}
    \pb{-s, s} \widetilde{\mathcal C}\; \overline{\pb{-s} \Omega} &= \frac{1}{8} \int_{-\infty}^\infty \ud \omega \sum_{l = 2}^\infty \sum_{|m| \leq l} \sum_{p = \pm 1} p C_{lm\omega} e^{i(m\phi - \omega t)} \pb{-s} \Theta_{lm\omega} \pb{-s} \widehat{\Omega}_{lm\omega p}, \\
    \pb{-s, -s} \widetilde{\mathcal C}\; \overline{\pb{s} \Omega} &= \frac{1}{8} \int_{-\infty}^\infty \ud \omega \sum_{l = 2}^\infty \sum_{|m| \leq l} \sum_{p = \pm 1} p \pb{s} C_{lm\omega p} e^{i(m\phi - \omega t)} \pb{-s} \Theta_{lm\omega} \pb{-s} \widehat{\Omega}_{lm\omega p}, \\
    \pb{-s, s} \mathcal{C} \pb{-s} \Omega &= \frac{3iM}{2} \sgn(s) \int_{-\infty}^\infty \ud \omega \sum_{l = 2}^\infty \sum_{|m| \leq l} \sum_{p = \pm 1} \omega e^{i(m\phi - \omega t)} \pb{-s} \Theta_{lm\omega} \pb{-s} \widehat{\Omega}_{lm\omega p},
  \end{align}
\end{subequations}
and so equation~\eqref{eqn:tsi_scalar} implies that

\begin{equation} \label{eqn:complex_starobinsky}
  \pb{\pm 2} C_{lm\omega p} = C_{lm\omega} \mp 12i pM\omega.
\end{equation}

At this point, we have shown how symmetry operators on the space of master variables act diagonally on the expansions~\eqref{eqn:decoupled_expansion} and~\eqref{eqn:decoupled_bar_expansion}.
We would like a similar diagonalization for the operator $\pb{s} \bs{\mathcal C}$, but (\emph{a priori}) there does not exist an analogous expansion for the metric perturbation.
We now construct such an expansion.
To begin, if a) $\pb{s} \psi$ is a solution to the vacuum Teukolsky equation~\eqref{eqn:teukolsky}, b) it is the master variable associated with some real solution to the linearized Einstein equations, and c)

\begin{equation}
  \pb{s} \Omega = \pb{s} M^{ab} \Im[\pb{s} \tau_{ab}^\dagger \pb{-s} \psi],
\end{equation}
then we call $\pb{s} \psi$ a \emph{Debye potential} for $\var g_{ab}$ (for the origin of this terminology, see~\cite{PhysRevD.10.1070}).
The first of these conditions ensures that $\pb{2} \psi$ and $\zeta^{-4} \pb{-2} \psi$ satisfy the same relation as (respectively) $\var \Psi_0$ and $\var \Psi_4$ in equation~\eqref{eqn:tsr}.
The second of these conditions ensures that $\Im[\pb{s} \tau_{ab}^\dagger \pb{-s} \psi]$ and (by the first condition) $\Im[\pb{-s} \tau_{ab}^\dagger \pb{s} \psi]$ are the same as $\var g_{ab}$, up to gauge and $l = 0, 1$ terms.

The easiest way to satisfy these conditions is as follows.
First, note that, by equations~\eqref{eqn:C_tensor} and~\eqref{eqn:C_action},

\begin{equation} \label{eqn:psi_motivation}
  \begin{split}
    \pb{s} M^{ab} &\Im\left\{\pb{s} \mathcal{C}_{ab}{}^{cd} \Im[\pb{-s} \tau_{cd}^\dagger \pb{s} \Omega]\right\} \\
    &= \frac{1}{16} \pb{s} M^{ab} \Re\left[\pb{s} \tau_{ab}^{\dagger} \int_{-\infty}^\infty \ud \omega \sum_{l = 2}^\infty \sum_{|m| \leq l} \sum_{p = \pm 1} p \pb{s} C_{lm\omega p} e^{i(m\phi - \omega t)} \pb{-s} \Theta_{lm\omega} \pb{-s} \widehat{\Omega}_{lm\omega p}\right] \\
    &= \frac{1}{256} \int_{-\infty}^\infty \ud \omega \sum_{l = 2}^\infty \sum_{|m| \leq l} \sum_{p = \pm 1} (C_{lm\omega}^2 + 144M^2 \omega^2) e^{i(m\phi - \omega t)} \pb{s} \Theta_{lm\omega} \pb{s} \widehat{\Omega}_{lm\omega p}. \\
  \end{split}
\end{equation}
We now define $\pb{s} \psi$, for a given $\pb{s} \Omega$, by

\begin{equation} \label{eqn:psi_def}
  \begin{split}
    \pb{s} \psi &\equiv 256 \pb{s} M^{ab} \Im\left[\pb{s} \tau_{ab}^\dagger \int_{-\infty}^\infty \ud \omega \sum_{l = 2}^\infty \sum_{|m| \leq l} \sum_{p = \pm 1} \frac{e^{i(m\phi - \omega t)} \pb{-s} \Theta_{lm\omega} (\theta) \pb{-s} \widehat{\Omega}_{lm\omega p} (r)}{C_{lm\omega}^2 + 144 M^2 \omega^2}\right] \\
    &= 16i \int_{-\infty}^\infty \ud \omega \sum_{l = 2}^\infty \sum_{|m| \leq l} \sum_{p = \pm 1} \frac{pe^{i(m\phi - \omega t)} \pb{s} \Theta_{lm\omega} (\theta) \pb{s} \widehat{\Omega}_{lm\omega p} (r)}{\pb{s} C_{lm\omega p}},
  \end{split}
\end{equation}
where the second line comes from equation~\eqref{eqn:C_action}, and $\pb{s} \widehat{\Omega}_{lm\omega p}$ is given in terms of $\pb{s} \Omega$ by equations~\eqref{eqn:mode_expansion} and~\eqref{eqn:p_def}.
Since $C_{lm\omega}^2 + 144M^2 \omega^2$ is real, $\pb{s} \psi$ satisfies the first of the above requirements, and by equation~\eqref{eqn:psi_motivation} it also satisfies the second.
Moreover, the second line implies that

\begin{equation}
  \pb{s} \widehat{\psi}_{lm\omega(-p)} = \frac{16 ip}{\pb{s} C_{lm\omega p}} \pb{s} \widehat{\Omega}_{lm\omega p}.
\end{equation}
where the expansion coefficients $\pb{s} \widehat{\psi}_{lm\omega p}$ are defined by an expansion analogous to equation~\eqref{eqn:decoupled_expansion}, together with the behavior under complex conjugation given by equation~\eqref{eqn:eigenvalue_condition}.
This condition is satisfied, due to the fact that
\begin{equation}
  \overline{\pb{s} C_{l(-m)(-\omega) p}} = \pb{s} C_{lm\omega p},
\end{equation}
by equations~\eqref{eqn:lambda_flips}, \eqref{eqn:starobinsky} and~\eqref{eqn:complex_starobinsky}, as well as by using equation~\eqref{eqn:eigenvalue_condition} for $\pb{s} \widehat{\Omega}_{lm\omega p}$.
While this would also be a perfectly reasonable definition of $\pb{s} \psi$, it is not apparent in this form that $\pb{s} \psi$ is generated by a real metric perturbation, which is crucial, and is explicit in equation~\eqref{eqn:psi_def}.
Finally, note that equations analogous to equation~\eqref{eqn:C_action} also hold for $\pb{s} \psi$ in terms of $\pb{s} \psi_{lm\omega p}$.

We can now define an expansion for the metric perturbation.
First, we define

\begin{equation} \label{eqn:hertz}
  \var_\pm g_{ab} \equiv \pb{\pm 2} \tau_{ab}^\dagger \pb{\mp 2} \psi,
\end{equation}
which (as remarked above) satisfy

\begin{equation} \label{eqn:hertz_condition}
  \pb{s} M^{ab} \Im[\var_+ g_{ab}] = \pb{s} M^{ab} \Im[\var_- g_{ab}] = \pb{s} \Omega.
\end{equation}
These metric perturbations have convenient expansions of the form

\begin{equation} \label{eqn:hertz_expansion}
  \var_\pm g_{ab} = \int_{-\infty}^\infty \ud \omega \sum_{l = 2}^\infty \sum_{|m| \leq l} \sum_{p = \pm 1} (\var_\pm g_{lm\omega p})_{ab},
\end{equation}
where

\begin{equation} \label{eqn:hertz_coefficients}
  (\var_\pm g_{lm\omega p})_{ab} \equiv \pb{\pm 2} \tau_{ab}^\dagger \left[e^{i(m\phi - \omega t)} \pb{\mp 2} \Theta_{lm\omega} (\theta) \pb{\mp 2} \widehat{\psi}_{lm\omega p} (r)\right].
\end{equation}
Note that the relationship between $\var_\pm g_{ab}$ and their coefficients is not $\mathbb{C}$-linear, due to the transformation properties of these coefficients under complex conjugation resulting from equation~\eqref{eqn:eigenvalue_condition}.

This procedure, which allowed us to construct a metric perturbation $\Im[\var_\pm g_{ab}]$ from $\pb{\mp 2} \Omega$ such that the master variables associated with this metric perturbation are $\pb{\pm 2} \Omega$, is similar to the one laid out in~\cite{Chrzanowski:1975wv}, which is referred to in the literature as \emph{Chrzanowski metric reconstruction}.
We now provide an operator form of this procedure: define

\begin{equation} \label{eqn:Pi}
  \begin{split}
    \pb{s} \Pi_{ab} \pb{s} \Omega &\equiv 256 \pb{s} \mathcal{C}_{ab}{}^{cd} \Im\left[\pb{-s} \tau_{cd}^\dagger \int_{-\infty}^\infty \ud \omega \sum_{l = 2}^\infty \sum_{|m| \leq l} \sum_{p = \pm 1} \frac{e^{i(m\phi - \omega t)} \pb{s} \Theta_{lm\omega} \pb{s} \widehat{\Omega}_{lm\omega p}}{C^2_{lm\omega} + 144 M^2 \omega^2}\right] \\
    &= 16i \pb{s} \tau_{ab}^\dagger \int_{-\infty}^\infty \ud \omega \sum_{l = 2}^\infty \sum_{|m| \leq l} \sum_{p = \pm 1} \frac{pe^{i(m\phi - \omega t)} \pb{-s} \Theta_{lm\omega} \pb{-s} \widehat{\Omega}_{lm\omega p}}{\pb{-s} C_{lm\omega p}},
\end{split}
\end{equation}
which satisfies

\begin{equation}
  \pb{s} M^{ab} \Im[\pb{s} \Pi_{ab} \pb{s} \Omega] = \pb{s} M^{ab} \Im[\pb{-s} \Pi_{ab} \pb{-s} \Omega] = \pb{s} \Omega.
\end{equation}
Note that the operator $\pb{s} \Pi_{ab}$ is non-local, since it requires an expansion in spin-weighted spheroidal harmonics for its definition.
This operator allows us to define a version of the operator $\pb{s} \mathcal{D}$ defined in section~\ref{section:separation} that maps to the space of complexified solutions of the linearized Einstein equations, much like $\pb{s} \mathcal{C}_{ab}{}^{cd}$:

\begin{equation} \label{eqn:D_tensor}
  \pb{s} \mathcal{D}_{ab}{}^{cd} \equiv \pb{s} \Pi_{ab} \pb{s} \mathcal{D} \pb{s} M^{cd}.
\end{equation}
We also define a version of this operator \emph{without} the intermediate factor of $\pb{s} \mathcal{D}$:

\begin{equation} \label{eqn:Chi}
  \pb{s} \Chi_{ab}{}^{cd} \equiv \pb{s} \Pi_{ab} \pb{s} M^{cd}.
\end{equation}

Now that we have both a definition of an expansion for the metric perturbation, along with a variety of symmetry operators defined which map the space of metric perturbations into itself, we can proceed to show that these symmetry operators act diagonally on these expansions.
Note, again, that there is no convenient notion of an expansion of the form~\eqref{eqn:hertz_expansion} for a general $\var g_{ab}$, and so we only compute the action of our various symmetry operators on $\var_\pm g_{ab}$.
The simplest case is $\pb{s} \mathcal{C}_{ab}{}^{cd}$, which satisfies [by equation~\eqref{eqn:C_action}]\footnote{Note that, as mentioned above below equation~\eqref{eqn:hertz_coefficients}, the relationship between $\var_\pm g_{ab}$ and their coefficients is not $\mathbb{C}$-linear.
  This explains the apparent contradiction of the left-hand side of equations~\eqref{eqn:C_tensor_action_angular} and~\eqref{eqn:C_tensor_action_radial} being $\mathbb{C}$-antilinear, but the right-hand sides appearing to be $\mathbb{C}$-linear.}

\begin{subequations} \label{eqn:C_tensor_action}
  \begin{align}
    \pb{\pm 2} \mathcal{C}_{ab}{}^{cd} \overline{\var_\pm g_{cd}} &= \pb{\pm 2} \tau_{ab}^\dagger \pb{\mp 2, \pm 2} \widetilde{\mathcal C}\; \overline{\pb{\mp 2} \psi} \nonumber \\
    &= \frac{1}{8} \int_{-\infty}^\infty \ud \omega \sum_{l = 2}^{\infty} \sum_{|m| \leq l} \sum_{p = \pm 1} pC_{lm\omega} (\var_\pm g_{lm\omega p})_{ab}, \label{eqn:C_tensor_action_angular} \\
    \pb{\pm 2} \mathcal{C}_{ab}{}^{cd} \overline{\var_\mp g_{cd}} &= \pb{\pm 2} \tau_{ab}^\dagger \pb{\mp 2, \mp 2} \widetilde{\mathcal C}\; \overline{\pb{\pm 2} \psi} \nonumber \\
    &= \frac{1}{8} \int_{-\infty}^\infty \ud \omega \sum_{l = 2}^{\infty} \sum_{|m| \leq l} \sum_{p = \pm 1} p \pb{\pm 2} C_{lm\omega p} (\var_\pm g_{lm\omega p})_{ab}, \label{eqn:C_tensor_action_radial} \\
    \pb{\pm 2} \mathcal{C}_{ab}{}^{cd} \var_\pm g_{cd} &= \pb{\pm 2} \tau_{ab}^\dagger \pb{\mp 2, \pm 2} \mathcal{C} \pb{\mp 2} \psi \nonumber \\
    &= \pm \frac{3iM}{2} \int_{-\infty}^\infty \ud \omega \sum_{l = 2}^{\infty} \sum_{|m| \leq l} \sum_{p = \pm 1} \omega (\var_\pm g_{lm\omega p})_{ab}.
  \end{align}
\end{subequations}
These equations demonstrate that the action on the expansion~\eqref{eqn:hertz_expansion} is diagonal, up to mappings from $\overline{(\var_\pm g_{lm\omega p})_{ab}} \to (\var_\pm g_{lm\omega p})_{ab}$ and $(\var_\mp g_{lm\omega p})_{ab}$, as well as mappings from $(\var_\pm g_{lm\omega p})_{ab} \to (\var_\mp g_{lm\omega p})_{ab}$.
More useful later in this paper will be the action of $\pb{s} \mathcal{C}_{ab}{}^{cd}$ on $\Im[\var_\pm g_{ab}]$:

\begin{equation}
  \begin{split}
    \pb{\pm 2} \mathcal{C}_{ab}{}^{cd} \Im[\var_+ g_{cd}] &= \pb{\pm 2} \mathcal{C}_{ab}{}^{cd} \Im[\var_- g_{cd}] \\
    &= \frac{i}{16} \int_{-\infty}^\infty \ud \omega \sum_{l = 2}^{\infty} \sum_{|m| \leq l} \sum_{p = \pm 1} p \pb{\pm 2} C_{lm\omega p} (\var_\pm g_{lm\omega p})_{ab}.
  \end{split}
\end{equation}

Similarly, we will consider the action of $\pb{s} \mathcal{D}_{ab}{}^{cd}$ and $\pb{s} \Chi_{ab}{}^{cd}$ on $\Im[\var_\pm g_{ab}]$.
We have that [by equation~\eqref{eqn:hertz_condition}]

\begin{equation}
  \pb{s} \Pi_{ab} \pb{s} \Omega = \pb{s} \Chi_{ab}{}^{cd} \Im[\var_\pm g_{cd}],
\end{equation}
along with [by equations~\eqref{eqn:hertz} and~\eqref{eqn:Pi}]

\begin{equation} \label{eqn:Pi_hertz}
  \pb{\pm 2} \Pi_{ab} \pb{\pm 2} \Omega = \var_\pm g_{ab},
\end{equation}
and so we find that

\begin{equation} \label{eqn:Chi_tensor_action}
  \pb{\pm 2} \Chi_{ab}{}^{cd} \Im[\var_+ g_{cd}] = \pb{\pm 2} \Chi_{ab}{}^{cd} \Im[\var_- g_{cd}] = \var_\pm g_{ab},
\end{equation}
Similarly, by the $\mathbb{R}$-linearity of equation~\eqref{eqn:Pi_hertz}, we find that [from equation~\eqref{eqn:D_action}]

\begin{equation} \label{eqn:D_tensor_action}
  \pb{\pm 2} \mathcal{D}_{ab}{}^{cd} \Im[\var_+ g_{cd}] = \pb{\pm 2} \mathcal{D}_{ab}{}^{cd} \Im[\var_- g_{cd}] = \int_{-\infty}^\infty \ud \omega \sum_{l = 2}^{\infty} \sum_{|m| \leq l} \sum_{p = \pm 1} \pb{2} \lambda_{lm\omega} (\var_\pm g_{lm\omega p})_{ab}.
\end{equation}

\subsection{Projection operators} \label{section:projection}

The final set of symmetry operators that we introduce are projection operators acting on the space of master variables $\pb{s} \Omega$.
Before we introduce these operators, however, it is relevant to discuss the asymptotic properties of the master variables.
First, define the tortoise coordinate $r^*$ by

\begin{equation} \label{eqn:tortoise}
  \frac{\ud r^*}{\ud r} \equiv \frac{r^2 + a^2}{\Delta}.
\end{equation}
This coordinate satisfies $r^* \to \infty$ as $r \to \infty$ and $r^* \to -\infty$ as $r \to r_+$, where $r_+$ is the location of the horizon, satisfying $\Delta|_{r = r_+} = 0$.

Now, the vacuum Teukolsky radial equation~\eqref{eqn:radial_teukolsky} is a second-order ordinary differential equation in $r$, and so its solution space is spanned by two solutions (for given values of $s$, $l$, $m$, and $\omega$) that are characterized by their asymptotic behavior at either $r = r_+$ or $r = \infty$.
One can show, from the asymptotic form of the vacuum Teukolsky radial equation~\eqref{eqn:radial_teukolsky}, that one can choose two independent solutions $\pb{s} R_{lm\omega}^{\textrm{in}} (r)$ and $\pb{s} R_{lm\omega}^{\textrm{out}} (r)$ with the following asymptotic forms as $r^* \to -\infty$~\cite{1974ApJ...193..443T}:

\begin{equation} \label{eqn:radial_falloffs_in_out}
  \pb{s} R_{lm\omega}^{\textrm{in}} (r) \to e^{-ik_{m\omega} r^*}/\Delta^s, \quad \pb{s} R_{lm\omega}^{\textrm{out}} (r) \to e^{ik_{m\omega} r^*},
\end{equation}
where

\begin{equation}
  k_{m\omega} \equiv \omega - am/(2Mr_+).
\end{equation}
Similarly, at $r^* \to \infty$, one can choose two independent solutions $\pb{s} R_{lm\omega}^{\textrm{down}} (r)$ and $\pb{s} R_{lm\omega}^{\textrm{up}} (r)$, which have the following asymptotic forms:

\begin{equation} \label{eqn:radial_falloffs_down_up}
  \pb{s} R_{lm\omega}^{\textrm{down}} (r) \to e^{-i\omega r^*}/r, \quad \pb{s} R_{lm\omega}^{\textrm{up}} (r) \to e^{i\omega r^*}/r^{2s + 1}.
\end{equation}
A general solution can therefore be expanded in terms of these solutions as

\begin{equation}
  \begin{split}
    \pb{s} \widehat{\Omega}_{lm\omega} (r) &= \pb{s} \widehat{\Omega}_{lm\omega}^{\textrm{down}} \pb{s} R_{lm\omega}^{\textrm{down}} (r) + \pb{s} \widehat{\Omega}_{lm\omega}^{\textrm{up}} \pb{s} R_{lm\omega}^{\textrm{up}} (r) \\
    &= \pb{s} \widehat{\Omega}_{lm\omega}^{\textrm{in}} \pb{s} R_{lm\omega}^{\textrm{in}} (r) + \pb{s} \widehat{\Omega}_{lm\omega}^{\textrm{out}} \pb{s} R_{lm\omega}^{\textrm{out}} (r).
  \end{split}
\end{equation}
Moreover, from the asymptotic behavior in equations~\eqref{eqn:radial_falloffs_in_out} and~\eqref{eqn:radial_falloffs_down_up}, we have

\begin{equation}
  \overline{\pb{s} R_{l(-m)(-\omega)}^{\textrm{in/out/down/up}} (r)} = \pb{s} R_{lm\omega}^{\textrm{in/out/down/up}} (r),
\end{equation}
and so, from the definition~\eqref{eqn:p_def},

\begin{equation}
  \begin{split}
    \pb{s} \widehat{\Omega}_{lm\omega p} (r) &= \pb{s} \widehat{\Omega}_{lm\omega p}^{\textrm{down}} \pb{s} R_{lm\omega}^{\textrm{down}} (r) + \pb{s} \widehat{\Omega}_{lm\omega p}^{\textrm{up}} \pb{s} R_{lm\omega}^{\textrm{up}} (r) \\
    &= \pb{s} \widehat{\Omega}_{lm\omega p}^{\textrm{in}} \pb{s} R_{lm\omega}^{\textrm{in}} (r) + \pb{s} \widehat{\Omega}_{lm\omega p}^{\textrm{out}} \pb{s} R_{lm\omega}^{\textrm{out}} (r),
  \end{split}
\end{equation}
where

\begin{equation}
  \pb{s} \widehat{\Omega}_{lm\omega p}^{\textrm{in/out/down/up}} \equiv \frac{1}{2} \left[\pb{s} \widehat{\Omega}_{lm\omega}^{\textrm{in/out/down/up}} + p(-1)^{m + s} \overline{\pb{s} \widehat{\Omega}_{l(-m)(-\omega)}^{\textrm{in/out/down/up}}}\right].
\end{equation}

We now define projection operators associated with this expansion as follows: for example, define $\pb{s} \mathcal{P}^{\textrm{in}}$ by

\begin{equation}
  \begin{split}
    \pb{s} \mathcal{P}^{\textrm{in}} \pb{s} \Omega &= \pb{s} \mathcal{P}^{\textrm{in}} \int_{-\infty}^\infty \ud \omega \sum_{l = |s|}^\infty \sum_{|m| \leq l} e^{i(m\phi - \omega t)} \pb{s} \Theta_{lm\omega} (\theta) \left[\pb{s} \widehat{\Omega}_{lm\omega}^{\textrm{in}} \pb{s} R_{lm\omega}^{\textrm{in}} (r) + \pb{s} \widehat{\Omega}_{lm\omega}^{\textrm{out}} \pb{s} R_{lm\omega}^{\textrm{out}} (r)\right] \\
    &\equiv \int_{-\infty}^\infty \ud \omega \sum_{l = |s|}^\infty \sum_{|m| \leq l} e^{i(m\phi - \omega t)} \pb{s} \Theta_{lm\omega} (\theta) \pb{s} \widehat{\Omega}_{lm\omega}^{\textrm{in}} \pb{s} R_{lm\omega}^{\textrm{in}} (r).
  \end{split}
\end{equation}
Analogous definitions can be given for $\pb{s} \mathcal{P}^{\textrm{out}}$, $\pb{s} \mathcal{P}^{\textrm{down}}$, and $\pb{s} \mathcal{P}^{\textrm{up}}$.
Since these operators require an expansion in spin-weighted spheroidal harmonics, they are necessarily non-local.

The reason we introduce these projection operators is that, as we show in appendix~\ref{appendix:asymptotics}, whether $\pb{s} \tau_{ab}^\dagger \pb{-s} \Omega$ falls off as $1/r$ (that is, whether it is an asymptotically flat metric perturbation) depends on the values $\pb{-s} \Omega_{lm\omega}^{\textrm{down/out}}$.
This was first remarked by Chrzanowski in~\cite{Chrzanowski:1975wv}.
As such, we define a projected version of $\pb{s} \tau_{ab}^\dagger$, which we call $\pb{s} \mathring{\tau}_{ab}^\dagger$, such that $\pb{s} \mathring{\tau}_{ab}^\dagger \pb{-s} \Omega$ is always well-behaved as $r \to \infty$:

\begin{equation}
  \pb{2} \mathring{\tau}_{ab}^\dagger \equiv \pb{2} \tau_{ab}^\dagger \pb{-2} \mathcal{P}^{\textrm{down}}, \qquad \pb{-2} \mathring{\tau}_{ab}^\dagger \equiv \pb{-2} \tau_{ab}^\dagger \pb{2} \mathcal{P}^{\textrm{up}}.
\end{equation}
Using this operator, we can define

\begin{equation} \label{eqn:C_tensor_proj}
  \pb{s} \mathring{\mathcal C}_{ab}{}^{cd} \equiv \pb{s} \mathring{\tau}_{ab}^\dagger \pb{-s} M^{cd},
\end{equation}
which allows for the definition of

\begin{equation}
  \pb{s} \mathring{\Pi}_{ab}{}^{cd} \pb{s} \Omega \equiv 256 \pb{s} \mathring{\mathcal C}_{ab}{}^{cd} \Im\left[\pb{-s} \tau_{cd}^\dagger \int_{-\infty}^\infty \ud \omega \sum_{l = 2}^\infty \sum_{|m| \leq l} \sum_{p = \pm 1} \frac{e^{i(m\phi - \omega t)} \pb{s} \Theta_{lm\omega} \pb{s} \widehat{\Omega}_{lm\omega p}}{C^2_{lm\omega} + 144 M^2 \omega^2}\right].
\end{equation}
Finally, this last operator allows for the definitions

\begin{equation} \label{eqn:D_Chi_tensor_proj}
  \pb{s} \mathring{\mathcal D}_{ab}{}^{cd} \equiv \pb{s} \mathring{\Pi}_{ab} \pb{s} \mathcal{D} \pb{s} M^{cd}, \qquad \pb{s} \ring{\Chi}_{ab}{}^{cd} \equiv \pb{s} \mathring{\Pi}_{ab} \pb{s} M^{cd}.
\end{equation}

\section{Conserved Currents} \label{section:currents}

We next turn to conserved currents that can be constructed using these symmetry operators.
First, we review the general theory of symplectic products, which are bilinear currents constructed from the Lagrangian formulation of a given classical field theory.
We then select a handful of conserved currents that can be constructed using symplectic products and symmetry operators, whose properties we discuss throughout the rest of this paper.

\subsection{Symplectic product} \label{section:symplectic}

Given a theory that possesses a Lagrangian formulation with Lagrangian density $\mathcal{L}$, one method of generating conserved quantities is to use the symplectic product defined in this section.
Following Burnett and Wald~\cite{Burnett57}, we start with a general Lagrangian four-form $\boldsymbol{L} [\boldsymbol{\phi}] \equiv \dual \mathcal{L} [\boldsymbol{\phi}]$ that is locally constructed from dynamical fields $\boldsymbol{\phi}$, where ${}^*$ denotes the Hodge dual.
It then follows that

\begin{equation}
  \var \boldsymbol{L} [\boldsymbol{\phi}] \equiv \boldsymbol{E} [\boldsymbol{\phi}] \cdot \var \boldsymbol{\phi} - \ud \boldsymbol{\Theta} [\boldsymbol{\phi}; \var \boldsymbol{\phi}],
\end{equation}
where the three-form $\boldsymbol{\Theta} [\bs \phi; \var \bs \phi]$ is the \emph{symplectic potential}, and $\boldsymbol{E} [\bs \phi]$ is a tensor-valued differential form\footnote{Some of the indices of $\bs E[\bs \phi]$ are contracted with those of $\var \bs \phi$, yielding a four-form $\bs E[\bs \phi] \cdot \var \bs \phi$.} that encodes the equations of motion; that is, $\boldsymbol{E} [\bs \phi] = 0$ on shell.
Thus, on shell, the integral of $\var \boldsymbol{L} [\bs \phi]$ is just a boundary term, which we use to define $\boldsymbol{\Theta} [\boldsymbol{\phi}; \var \boldsymbol{\phi}]$.
We can then define the \emph{symplectic product} by taking a second, independent variation:

\begin{equation}
  \boldsymbol{\omega} [\bs \phi; \var_1 \boldsymbol{\phi}, \var_2 \boldsymbol{\phi}] \equiv \var_1 \boldsymbol{\Theta} [\boldsymbol{\phi}; \var_2 \boldsymbol{\phi}] - \var_2 \boldsymbol{\Theta} [\boldsymbol{\phi}; \var_1 \boldsymbol{\phi}].
\end{equation}
Thus, we have that

\begin{equation}
  \begin{split}
    \ud \boldsymbol{\omega} [\bs \phi; \var_1 \boldsymbol{\phi}, \var_2 \boldsymbol{\phi}] &= \var_1 \boldsymbol{E} [\bs \phi] \cdot \var_2 \boldsymbol{\phi} - \var_2 \boldsymbol{E} [\bs \phi] \cdot \var_1 \boldsymbol{\phi},
  \end{split}
\end{equation}
which vanishes if $\var_1 \boldsymbol{\phi}$ and $\var_2 \boldsymbol{\phi}$ are both solutions to the linearized equations of motion.
We define the corresponding vector current by

\begin{equation}
  \pb{S} j^a \left[\bs \phi; \var_1 \boldsymbol{\phi}, \var_2 \boldsymbol{\phi}\right] \equiv \left(\dual \boldsymbol{\omega} \left[\bs \phi; \var_1 \boldsymbol{\phi}, \var_2 \boldsymbol{\phi}\right]\right)^a.
\end{equation}

We now turn to two different Lagrangians whose symplectic products are particularly interesting.
First, we consider the symplectic product for the Einstein-Hilbert Lagrangian four-form by $\boldsymbol{L}_{\textrm{EH}} [\bs g] = R \boldsymbol{\epsilon}/(16\pi)$.
For this Lagrangian, we find (following~\cite{Burnett57}, for example; note the difference in sign due to using a different sign convention for $R^a{}_{bcd}$)

\begin{equation}
  (\Theta_{\textrm{EH}})_{abc} [\boldsymbol{g}; \var \boldsymbol{g}] = -\frac{1}{8\pi} \epsilon_{abcd}  g^{fg} \delta^d{}_{[e} \var C^e{}_{f]g},
\end{equation}
where $\var C^a{}_{bc}$ is the variation of the connection coefficients for $\nabla_a (\lambda)$:

\begin{equation}
  \var C^a{}_{bc} = \frac{1}{2} g^{ad} (\nabla_b \var g_{cd} + \nabla_c \var g_{bd} - \nabla_d \var g_{bc}).
\end{equation}
Thus, the symplectic (vector) current is given by

\begin{equation} \label{eqn:symp_form_linearized_gr}
  \begin{split}
    \pb{S} j_{\textrm{EH}}^a [\var_1 \boldsymbol{g}, \var_2 \boldsymbol{g}] &= \frac{1}{8\pi} \delta^a{}_{[b} \var_1 C^b{}_{c]d} \left[(\var_2 g)^{cd} - \frac{1}{2} (\var_2 g)^e{}_e g^{cd}\right] - \interchange{1}{2} \\
    &= \frac{1}{16\pi} \var_1 C^a{}_{bc} (\var_2 g)^{bc} + v^a [\var_1 \bs{g}] (\var_2 g)^b{}_b + w^{ab} [\var_1 \bs{g}] \nabla_b (\var_2 g)^c{}_c - \interchange{1}{2},
  \end{split}
\end{equation}
for some tensor fields $v^a [\var \bs{g}]$ and $w^{ab} [\var \bs{g}]$ which are unimportant for the discussion of this paper, as we only consider metric perturbations which are trace-free.
Here, for simplicity, the dependence on the background metric $g_{ab}$ is implicit.
This symplectic product provides a bilinear current on the space of metric perturbations which is conserved for vacuum solutions to the linearized Einstein equations.

Somewhat unexpectedly, one can also define a symplectic product for the master variables themselves.
In order to do so, we need a Lagrangian formulation for the Teukolsky equation.
Such a Lagrangian formulation was recently used to generate Noether currents for the master variables in~\cite{Toth:2018qrx}.
As noted by Bini, Cherubini, Jantzen, and Ruffini~\cite{Bini:2002jx}, the Teukolsky operator can be rewritten as a modified wave operator:

\begin{equation}
  \pb{s} \Box = (\nabla^a + s\Gamma^a) (\nabla_a + s\Gamma_a) - 4s^2 \Psi_2,
\end{equation}
where

\begin{equation}
  \Gamma^a = -2\left[\gamma l^a + (\epsilon + \rho) n^a - \alpha m^a - (\beta + \tau) \bar{m}^a\right].
\end{equation}
Since the equations of motion are now in the form of a modified wave equation, one can write down a Lagrangian four-form of the form (for $s \geq 0$)

\begin{equation}
  \boldsymbol{L}_{\textrm{BCJR}} [\pb{s} \Omega, \pb{-s} \Omega] = \dual (\ud + s\boldsymbol{\Gamma}) \pb{s} \Omega \wedge (\ud - s\boldsymbol{\Gamma}) \pb{-s} \Omega - 96s^2 \Psi_2 \pb{s} \Omega \pb{-s} \Omega \boldsymbol{\epsilon}.
\end{equation}
Note that, in this expression, the metric and $\Gamma^a$ are non-dynamical fields, and therefore do not get varied.
Varying this Lagrangian four-form results in the Teukolsky equations for spins $s$ and $-s$.
One can easily show that

\begin{equation}
  \boldsymbol{\Theta}_{\textrm{BCJR}} [\pb{s} \Omega, \pb{-s} \Omega; \var \pb{s} \Omega, \var \pb{-s} \Omega] = \var \pb{s} \Omega \dual (\ud - s\boldsymbol{\Gamma}) \pb{-s} \Omega + \var \pb{-s} \Omega \dual (\ud + s\boldsymbol{\Gamma}) \pb{s} \Omega,
\end{equation}
and so

\begin{equation} \label{eqn:symp_form_BCJR}
  \pb{S} j_{\textrm{BCJR}}^a \left[\var_1 \pb{s} \Omega, \var_1 \pb{-s} \Omega; \var_2 \pb{s} \Omega, \var_2 \pb{-s} \Omega\right] = \var_1 \pb{s} \Omega (\nabla^a - s\Gamma^a) \var_2 \pb{-s} \Omega + \var_1 \pb{-s} \Omega (\nabla^a + s\Gamma^a) \var_2 \pb{s} \Omega - \interchange{1}{2}.
\end{equation}
Here, we are dropping any dependence on the background values of $\pb{s} \Omega$ and $\pb{-s} \Omega$, since they do not appear on the right-hand side.

Although this current is bilinear on the space of variations of the master variables, it can be regarded as a bilinear current on the space of master variables themselves, since their equations of motion are linear.
Note further that this symplectic product is not the physical symplectic product for linearized gravity.

\subsection{Currents of interest} \label{section:definitions}

Using the results of sections~\ref{section:symmetry} and~\ref{section:symplectic}, we now define the following currents, for which we will be computing the geometric optics limit and the fluxes at the horizon and null infinity.
The first of these currents is a rescaled version of the symplectic product of $\pb{s} \bs{\mathcal C} \cdot \var \bs{g}$ and its complex conjugate:

\begin{equation} \label{eqn:C_current}
  \pb{\pb{s} \mathcal{C}} j^a [\var \bs{g}] \equiv 8i \pb{S} j_{\textrm{EH}}^a \Big[\pb{s} \bs{\mathcal C} \cdot \var \bs{g}, \overline{\pb{s} \bs{\mathcal C} \cdot \var \bs{g}}\Big],
\end{equation}
in terms of the symplectic product~\eqref{eqn:symp_form_linearized_gr} and the symmetry operator~\eqref{eqn:C_tensor}.
The normalization here is chosen to give a nicer limit in geometric optics; similarly, this current is simpler in the limit of geometric optics than other currents that can be constructed from $\pb{s} \bs{\mathcal C}$.
The currents defined in equation~\eqref{eqn:C_current} are entirely local, but they generally diverge at null infinity, as we will show in section~\ref{section:fluxes}.
The divergences can be removed by using $\pb{s} \mathring{\bs{\mathcal C}}$ instead of $\pb{s} \bs{\mathcal C}$.
We therefore define

\begin{equation} \label{eqn:C_current_proj}
  \pb{\pb{2} \mathring{\mathcal C}} j^a [\var \bs{g}] \equiv 8i \sum_{s = \pm 2} \pb{S} j_{\textrm{EH}}^a \bigg[\pb{s} \mathring{\bs{\mathcal C}} \cdot \var \bs{g}, \overline{\pb{s} \mathring{\bs{\mathcal C}} \cdot \var \bs{g}}\bigg],
\end{equation}
where $\pb{2} \mathring{\bs{\mathcal C}}$ is defined in equation~\eqref{eqn:C_tensor_proj}.
The motivation for including the sum over $s$ in this definition is due to the fact that $\pb{2} \mathring{\bs{\mathcal C}}$ and $\pb{-2} \mathring{\bs{\mathcal C}}$ are only nonzero for ingoing and outgoing solutions at null infinity, respectively.
The sum therefore ensures that the total current is nonzero for both types of solutions.

We next define similar currents involving $\pb{s} \bs{\Chi}$ and $\pb{s} \bs{\mathcal D}$:

\begin{align}
  \pb{\pb{s} \mathcal{D}} j^a [\var \bs{g}] &\equiv \frac{i}{16} \pb{S} j_{\textrm{EH}}^a \Big[\pb{s} \bs{\Chi} \cdot \var \bs{g}, \overline{\pb{s} \bs{\mathcal D} \cdot \var \bs{g}}\Big], \label{eqn:D_current} \\
  \pb{\pb{2} \mathring{\mathcal D}} j^a [\var \bs{g}] &\equiv \frac{i}{16} \sum_{s = \pm 2} \pb{S} j_{\textrm{EH}}^a \bigg[\pb{s} \mathring{\bs \Chi} \cdot \var \bs{g}, \overline{\pb{s} \mathring{\bs{\mathcal D}} \cdot \var \bs{g}}\bigg]. \label{eqn:D_current_proj}
\end{align}
Unlike the currents~\eqref{eqn:C_current} and~\eqref{eqn:C_current_proj}, both of these currents are nonlocal.
We will see below that the geometric optics limits of these currents are proportional to the Carter constants $K$ of the gravitons, as opposed to $K^4$ for the currents~\eqref{eqn:C_current} and~\eqref{eqn:C_current_proj}.

Finally, we define the currents

\begin{equation} \label{eqn:Omega_current}
  \pb{\pb{s} \Omega} j^a [\var \boldsymbol{g}] \equiv \frac{1}{4\pi i} \pb{S} j_{\textrm{BCJR}}^a \left[\pb{s} \Omega, \pb{-s} \Omega; \pb{s, s} \widetilde{\mathcal C}\; \overline{\pb{-s} \Omega}, \pb{-s, s} \widetilde{\mathcal C}\; \overline{\pb{-s} \Omega}\right],
\end{equation}
in terms of the symplectic product for the master variables in equation~\eqref{eqn:symp_form_BCJR} and the symmetry operator~\eqref{eqn:C}.
Note that $\pb{\pm 2} \Omega$ are functions of $\var g_{ab}$, by equation~\eqref{eqn:M_def}.
These currents are very similar to the currents $\pb{\pb{\pm 2} \mathcal{C}} j^a [\var \bs g]$, having the same geometric optics limit, and also being local; however, these currents have the advantage of also having finite fluxes at null infinity.

We now derive various properties of these currents in sections~\ref{section:geometric_optics} and~\ref{section:fluxes}.
For convenience, these properties are summarized at the end of this paper in table~\ref{table:summary}.

\section{Geometric Optics} \label{section:geometric_optics}

Using the symmetry operators in section~\ref{section:symmetry} and the symplectic products in section~\ref{section:symplectic}, one could define a multitude of currents that are conserved for vacuum solutions to the linearized Einstein equations.
In this section, we provide the motivation for the particular currents highlighted in section~\ref{section:definitions}.
This is accomplished by taking the geometric optics limit, in which solutions to the linearized Einstein equations represent null fluids of gravitons.
We express the associated currents in terms of the gravitons' constants of motion.

\subsection{Geometric optics on general backgrounds} \label{section:geometric_optics_gr}

The starting point for geometric optics is a harmonic ansatz for the metric perturbation:

\begin{equation}
  \var g_{ab} = \Re\left\{\left[a \varpi_{ab} + O(\epsilon)\right] e^{-i\vartheta/\epsilon}\right\},
\end{equation}
where $a$ and $\vartheta$ are real, $\varpi_{ab}$, the \emph{polarization tensor}, is a complex, symmetric tensor that is normalized to satisfy $\varpi_{ab} \bar{\varpi}^{ab} = 1$, and $\epsilon$ is a dimensionless parameter whose limit is taken to zero.
Inserting this ansatz into the linearized Einstein equations and the Lorenz gauge condition and equating coefficients of powers of $\epsilon$ yields the following results (see, for example, Misner, Thorne, and Wheeler~\cite{mtw}):

\begin{enumerate}

\item[i.] The wavevector $k^a$ defined by

  \begin{equation}
    k_a \equiv \nabla_a \vartheta
  \end{equation}
  is tangent to a congruence of null geodesics:

  \begin{equation}
    k^b \nabla_b k^a = 0, \qquad k_a k^a = 0.
  \end{equation}

\item[ii.] The polarization tensor $\varpi_{ab}$ is orthogonal to $k^a$ and parallel-transported along these geodesics:

  \begin{equation} \label{eqn:varpi_conditions}
    k^a \varpi_{ab} = 0, \qquad k^c \nabla_c \varpi_{ab} = 0.
  \end{equation}

\item[iii.] The amplitude $a$ evolves along these geodesics according to

  \begin{equation} \label{eqn:etendue}
    \nabla_a (a^2 k^a) = 0.
  \end{equation}

\end{enumerate}

We now consider this formalism in terms of spinors.
First, as $k^a$ is null, we can write

\begin{equation}
  k^{AA'} = \kappa^A \bar{\kappa}^{A'},
\end{equation}
for some spinor $\kappa^A$.
We choose a second spinor $\lambda^A$ such that $(\kappa, \lambda)$ form a spin basis.
The conditions~\eqref{eqn:varpi_conditions} and the normalization of $\varpi_{ab}$ imply that

\begin{equation} \label{eqn:phase_tensor}
  \varpi_{ab} = k_{(a} \alpha_{b)} + e_R q_a q_b + e_L \bar{q}_a \bar{q}_b,
\end{equation}
where $q_a \equiv \kappa_A \bar{\lambda}_{A'}$ and $\alpha^a$ is an arbitrary vector satisfying $\alpha^a k_a = 0$.
Because of the gauge freedom $\var g_{ab} \to \var g_{ab} + 2 \nabla_{(a} \xi_{b)}$, the first term can be removed by a gauge transformation (which preserves the Lorenz gauge condition), and so we can safely set $\alpha^a = 0$.

The last two terms in equation~\eqref{eqn:phase_tensor} are physically measurable.
The complex coefficients $e_R$ and $e_L$ correspond to right and left circular polarization.
By the normalization of $\varpi_{ab}$, we have that $|e_R|^2 + |e_L|^2 = 1$.
Moreover, these factors of $e_R$ and $e_L$ appear in the expansion for $(\var \Psi)_{ABCD}$:

\begin{equation} \label{eqn:Psi_go}
  (\var \Psi)_{ABCD} = -\frac{1}{\epsilon^2} a \kappa_A \kappa_B \kappa_C \kappa_D \left(e_R e^{-i\vartheta/\epsilon} + \bar{e}_L e^{i\vartheta/\epsilon}\right) + O(1/\epsilon).
\end{equation}

\subsection{Conserved currents} \label{section:geometric_optics_currents}

When considering nonlinear quantities in geometric optics, such as conserved currents, we will discard rapidly oscillating terms.
This effectively takes a spacetime average of these quantities over a scale that is large compared to $\epsilon$, but small compared to the radius of curvature of the background spacetime (see, for example,~\cite{Isaacson:1968zza}, or~\cite{Burnett:1989gp} for rigorous treatments of this averaging procedure via weak limits).
Such an average we will denote by $\langle \cdot\rangle$.

We start with a few simple results.
First, if a conserved current reduces in the limit of geometric optics to

\begin{equation} \label{eqn:conserved_quantity}
  \langle j^a\rangle = \frac{1}{\epsilon^n} [a^2 Q k^a + O(\epsilon)],
\end{equation}
for some quantity $Q$ and integer $n$, then $Q$ is a conserved quantity along the integral curves of $k^a$.
To see this, note that the leading order term in the conservation equation $\nabla_a \langle j^a\rangle = 0$ yields

\begin{equation} \label{eqn:optical_conservation}
  0 = a^2 k^a \nabla_a Q + Q \nabla_a (a^2 k^a) = a^2 k^a \nabla_a Q,
\end{equation}
from equation~\eqref{eqn:etendue}.
All currents that we consider in this paper will be of the form~\eqref{eqn:conserved_quantity} in the geometric optics limit..

The second result is that, under the assumption~\eqref{eqn:conserved_quantity}, the conserved charge associated with the current $j^a$ reduces to a sum over all gravitons of the conserved quantity $Q$ for each graviton.
This result means that equation~\eqref{eqn:conserved_quantity} is a physically appealing assumption.
The proof proceeds as follows~\cite{mtw}: first, we note that the effective stress-energy tensor appropriate to gravitational radiation in the geometric optics regime is given by~\cite{Isaacson:1968zza}

\begin{equation}
  \langle T_{ab}^{\textrm{eff}}\rangle = \frac{1}{32\pi} \left\langle (\nabla_a \var g_{cd}) [\nabla_b (\var g)^{cd}]\right\rangle + O(1/\epsilon) = \frac{a^2}{32\pi \epsilon^2} \left[k_a k_b + O(\epsilon)\right].
\end{equation}
On the other hand, the stress-energy tensor for a collection of gravitons with number-flux $\mathcal N_a$ and momentum $p_a = \hbar k_a/\epsilon$ is given by~\cite{mtw}

\begin{equation}
  T_{ab}^{\textrm{eff}} = p_{(a} \mathcal N_{b)},
\end{equation}
and so we find that

\begin{equation}
  a^2 k_a = 32 \pi \hbar \epsilon \mathcal N_a [1 + O(\epsilon)].
\end{equation}
Upon integrating a current $j^a$ given by equation~\eqref{eqn:conserved_quantity} over a hypersurface $\Sigma$, one finds the charge

\begin{equation} \label{eqn:conserved_charge}
  \int_\Sigma \langle j^a\rangle \ud^3 \Sigma_a = \frac{32\pi \hbar}{\epsilon^{n - 1}} \sum_\alpha Q_\alpha [1 + O(\epsilon)],
\end{equation}
where $\alpha$ labels the gravitons passing through the hypersurface.
That is, the charge is proportional to the sum of the conserved quantities over all of the gravitons passing through the surface.

\subsection{Computations} \label{section:geometric_optics_examples}

We now turn to computations of geometric optics limits for the conserved currents discussed in this paper.
For these calculations, we first define the quantities $\kappa_0$, $\kappa_1$, $r_a$, and $s_a$:

\begin{equation}
  \kappa_0 \equiv o_A \kappa^A, \qquad \kappa_1 \equiv \iota_A \kappa^A, \qquad r^a \equiv  \sigma^a{}_{AA'} o^A \bar{\kappa}^{A'}, \qquad s^a \equiv \sigma^a{}_{AA'} \iota^A \bar{\kappa}^{A'}.
\end{equation}
These quantities are constructed from the spinor $\kappa_A$ (which is related to the wavevector $k^a$) and the principal spin basis $(o, \iota)$.
They satisfy

\begin{equation}
  \begin{gathered}
    |\zeta \kappa_0 \kappa_1|^2 = \frac{\epsilon^2}{2 \hbar^2} K, \qquad r_a r^a = s_a s^a = r_a k^a = s_a k^a = 0, \\
    r_a \bar{r}^a = |\kappa_0|^2, \qquad s_a \bar{s}^a = |\kappa_1|^2, \qquad r_a \bar{s}^a = -\kappa_0 \bar{\kappa}_1,
  \end{gathered}
\end{equation}
where $K = \hbar^2 K_{ab} k^a k^b/\epsilon^2$ is the Carter constant for the gravitons.
The factors of $\hbar$ arise in this classical computation as part of converting from the wavevectors of the gravitons to their momenta, and hence their conserved quantities.

We now begin calculating the conserved currents defined in section~\ref{section:definitions}.
Since, to leading order in geometric optics, the differential operators present in this paper become c-numbers, a straightforward calculation starting from equations~\eqref{eqn:M} and~\eqref{eqn:tau_dagger} shows that

\begin{subequations}
  \begin{align}
    \pb{s} \tau_{ab}^\dagger &= \frac{1}{\epsilon^2} \begin{cases}
      \kappa_0^2 r_a r_b + O(\epsilon) & s = 2 \\
      \zeta^4 \kappa_1^2 s_a s_b + O(\epsilon) & s = -2
    \end{cases}, \label{eqn:tau_dagger_go} \\
    \pb{s} M^{ab} &= \frac{1}{2\epsilon^2} \begin{cases}
      \kappa_0^2 r^a r^b + O(\epsilon) & s = 2 \\
      \zeta^4 \kappa_1^2 s^a s^b + O(\epsilon) & s = -2
    \end{cases}, \label{eqn:M_go}
  \end{align}
\end{subequations}
and [starting from equation~\eqref{eqn:Psi_go}] that

\begin{equation} \label{eqn:Omega_go}
  \pb{s} \Omega = -\frac{a}{\epsilon^2} (e_R e^{-i\vartheta/\epsilon} + \bar{e}_L e^{i\vartheta/\epsilon})\begin{cases}
    \kappa_0^4 + O(\epsilon) & s = 2 \\
    (\zeta \kappa_1)^4 + O(\epsilon) & s = -2
  \end{cases}.
\end{equation}
As such, we find that

\begin{equation} \label{eqn:C_geometric_optics}
  \pb{s} \mathcal{C}_{ab}{}^{cd} \var g_{cd} = -\frac{a}{\epsilon^4} \zeta^4 (\kappa_1 \kappa_0)^2 (e_R e^{-i\vartheta/\epsilon} + \bar{e}_L e^{i\vartheta/\epsilon}) \begin{cases}
    r_a r_b \kappa_1^2 + O(\epsilon) & s = 2 \\
    s_a s_b \kappa_0^2 + O(\epsilon) & s = -2
  \end{cases}.
\end{equation}
This implies that

\begin{equation}
  \left\langle(\pb{s} \mathcal{C}_{bc}{}^{de} \var g_{de}) \nabla^a \overline{\pb{s} \mathcal{C}^{bc}{}_{de} \var g^{de}}\right\rangle = -\frac{2\pi i}{\hbar^7} K^4 (|e_R|^2 - |e_L|^2) \mathcal N^a [1 + O(\epsilon)].
\end{equation}
Thus, we find that the current $\pb{\pb{s} \mathcal{C}} j^a [\var \bs g]$ is given in this limit by

\begin{equation} \label{eqn:C_current_go}
  \begin{split}
    \left\langle\pb{\pb{s} \mathcal{C}} j^a [\var \boldsymbol{g}]\right\rangle &= \frac{1}{2\pi} \left\langle\Im\left[(\pb{s} \mathcal{C}_{bc}{}^{de} \var g_{de}) \nabla^a \overline{\pb{s} \mathcal{C}^{bc}{}_{de} \var g^{de}}\right]\right\rangle [1 + O(\epsilon)] \\
    &= \frac{1}{\hbar^7} K^4 \left(|e_R|^2 - |e_L|^2\right) \mathcal N^a [1 + O(\epsilon)].
  \end{split}
\end{equation}
As such, these currents are a generalization of the Carter constant for point particles to linearized gravity in the Kerr spacetime, at least in the limit of geometric optics.

We now turn to the current $\pb{\pb{s} \mathcal{D}} j^a [\var \bs g]$.
First, note that, from equations~\eqref{eqn:D} and~\eqref{eqn:RS},

\begin{equation}
  \pb{s} \mathcal{D} \pb{s} \Omega = \frac{1}{\epsilon^2} |\zeta \kappa_0 \kappa_1|^2 \pb{s} \Omega [1 + O(\epsilon)],
\end{equation}
and so

\begin{equation}
  \pb{s} \mathcal{D}_{ab}{}^{cd} \var g_{cd} = \frac{K}{2 \hbar^2} \pb{s} \Chi_{ab}{}^{cd} \var g_{cd} [1 + O(\epsilon)].
\end{equation}
Now, note that $\pb{s} \Chi_{ab}{}^{cd} \var g_{cd}$, by equations~\eqref{eqn:Chi} and~\eqref{eqn:Pi}, can be written (in the limit of geometric optics, where differential operators commute to leading order) as a product of the form

\begin{equation}
  \pb{s} \Chi_{ab}{}^{cd} \var g_{cd} = 4 \left(\pb{s, s} \tilde{\mathcal C} \overline{\pb{-s, -s} \tilde{\mathcal C}}\right)^{-1} \pb{s} \mathcal{C}_{ab}{}^{cd} \overline{\pb{-s} \mathcal{C}_{cd}{}^{ef}} \var g_{ef} [1 + O(\epsilon)],
\end{equation}
where the operator $\left(\pb{s, s} \tilde{\mathcal C} \overline{\pb{-s, -s} \tilde{\mathcal C}}\right)^{-1}$ is a nonlocal operator having the effect of multiplying each coefficient of the expansion~\eqref{eqn:decoupled_expansion} by $64/(C_{lm\omega}^2 + 144M^2 \omega^2)$.
This operator is a nonlocal inverse to $\pb{s, s} \tilde{\mathcal C} \overline{\pb{-s, -s} \tilde{\mathcal C}}$, by equation~\eqref{eqn:C_action}.
For its geometric optics limit, note that

\begin{subequations} \label{eqn:C_go}
  \begin{align}
    \pb{2, -2} \widetilde{\mathcal C}\; \overline{\pb{2} \Omega} &= \frac{1}{2\epsilon^4} (\bar{\zeta} \kappa_0 \bar{\kappa}_1)^4 \overline{\pb{2} \Omega} [1 + O(\epsilon)], \quad &\pb{-2, 2} \widetilde{\mathcal C}\; \overline{\pb{-2} \Omega} &= \frac{1}{2\epsilon^4} (\zeta \bar{\kappa}_0 \kappa_1)^4 \overline{\pb{-2} \Omega} [1 + O(\epsilon)], \\
    \pb{2, 2} \widetilde{\mathcal C}\; \overline{\pb{-2} \Omega} &= \frac{1}{2\epsilon^4} |\kappa_0|^8 \overline{\pb{-2} \Omega} [1 + O(\epsilon)], \quad &\pb{-2, -2} \widetilde{\mathcal C}\; \overline{\pb{2} \Omega} &= \frac{1}{2\epsilon^4} |\zeta \kappa_1|^8 \overline{\pb{2} \Omega} [1 + O(\epsilon)],
  \end{align}
\end{subequations}
and so

\begin{equation}
  \left(\pb{s, s} \tilde{\mathcal C} \overline{\pb{-s, -s} \tilde{\mathcal C}}\right)^{-1} \pb{s} \Omega = \frac{4 \epsilon^8}{|\zeta \kappa_0 \kappa_1|^8} \pb{s} \Omega[1 + O(\epsilon)].
\end{equation}
Moreover, we have that [from equations~\eqref{eqn:tau_dagger_go} and~\eqref{eqn:M_go}]

\begin{equation}
  \pb{s} \mathcal{C}_{ab}{}^{cd} \overline{\pb{-s} \mathcal{C}_{cd}{}^{ef}} \var g_{ef} = -\frac{a}{4 \epsilon^8} |\zeta \kappa_0 \kappa_1|^8 (\bar{e}_R e^{i\vartheta/\epsilon} + e_L e^{-i\vartheta/\epsilon}) \begin{cases}
    r_a r_b/\kappa_0^2 + O(\epsilon) & s = 2 \\
    s_a s_b/\kappa_1^2 + O(\epsilon) & s = -2
  \end{cases},
\end{equation}
from which it follows that

\begin{equation}
  \pb{s} \Chi_{ab}{}^{cd} \var g_{cd} = -4a (\bar{e}_R e^{i\vartheta/\epsilon} + e_L e^{-i\vartheta/\epsilon}) \begin{cases}
    r_a r_b/\kappa_0^2 + O(\epsilon) & s = 2 \\
    s_a s_b/\kappa_1^2 + O(\epsilon) & s = -2
  \end{cases}.
\end{equation}
The current in question is then given by

\begin{equation} \label{eqn:D_current_go}
  \langle\pb{\pb{s} \mathcal{D}} j^a [\var \boldsymbol{g}]\rangle = \frac{1}{\hbar} K \left(|e_R|^2 - |e_L|^2\right) \mathcal N^a [1 + O(\epsilon)].
\end{equation}
This therefore provides another, entirely \emph{non-local} notion of the Carter constant for linearized gravity in the Kerr spacetime.

There are, of course, other currents whose charges reduce to the Carter constant in the geometric optics limit.
Another class of currents come from the symplectic product for the master variables, instead of the metric perturbation.
One current of interest from this class is given by equation~\eqref{eqn:Omega_current}, which has a limit in geometric optics given by [from equations~\eqref{eqn:symp_form_BCJR},~\eqref{eqn:C_go}, and~\eqref{eqn:Omega_go}]

\begin{equation} \label{eqn:Omega_current_go}
  \langle\pb{\pb{s} \Omega} j^a [\var \bs{g}]\rangle = \frac{1}{\hbar^7} K^4 (|e_R|^2 - |e_L|^2) \mathcal N^a [1 + O(\epsilon)].
\end{equation}

The results of this section [equations~\eqref{eqn:C_current_go},~\eqref{eqn:D_current_go}, and~\eqref{eqn:Omega_current_go}] give the geometric optics limits for the currents that do not involve projection operators.
We now consider the two remaining currents, $\pb{\pb{2} \mathring{\mathcal C}} j^a [\var \bs g]$ and $\pb{\pb{2} \mathring{\mathcal D}} j^a [\var \bs g]$.
For simplicity, we first consider $\pb{\pb{2} \mathring{\mathcal C}} j^a [\var \bs g]$ (the exact same argument holds for $\pb{\pb{2} \mathring{\mathcal D}} j^a [\var \bs g]$).
This current is the sum of two terms, the first of which is equal to $\pb{\pb{-2} \mathcal C} j^a [\var \bs g]$, except that it contains a projection which eliminates the ingoing modes at null infinity.
Similarly, the second term is equal to $\pb{\pb{2} \mathcal C} j^a [\var \bs g]$, except it eliminates all outgoing modes.
Consider the case where $\var g_{ab}$ represents a null fluid of gravitons where the gravitons are purely outgoing at future null infinity; that is, $k^a$ is tangent to an outgoing null congruence.
The geometric optics limit in this case would be the same as that of $\pb{\pb{-2} \mathcal C} j^a [\var \bs g]$.
Similarly, if $k^a$ is an ingoing null congruence, the geometric optics limit would be the same as that of $\pb{\pb{2} \mathcal C} j^a [\var \bs g]$.
Since these geometric optics limits are equal by equation~\eqref{eqn:C_current_go}, we recover the following result:
\begin{equation}
  \left\langle\pb{\pb{2} \mathring{\mathcal C}} j^a [\var \boldsymbol{g}]\right\rangle = \frac{1}{\hbar^7} K^4 \left(|e_R|^2 - |e_L|^2\right) \mathcal N^a [1 + O(\epsilon)],
\end{equation}
when $\var g_{ab}$ represents an ingoing or outgoing null fluid of gravitons.
A similar argument gives a similar result for $\pb{\pb{2} \mathring{\mathcal D}} j^a [\var \bs g]$.
However, the geometric optics limits for $\pb{\pb{2} \mathring{\mathcal C}} j^a [\var \bs g]$ and $\pb{\pb{2} \mathring{\mathcal D}} j^a [\var \bs g]$ are only given by simple expressions when $k^a$ is either tangent to an ingoing or outgoing null congruence, but not for general geometric optics solutions $\var g_{ab}$.

We conclude this discussion with a brief review of a classification scheme for conserved currents in geometric optics that we used in~\cite{Grant:2019qyo}.
In the limit of geometric optics, one often finds that conserved currents depend on the quantities $e_R$ and $e_L$ in one of the following four ways; depending on this dependence, we call such currents \emph{energy}, \emph{zilch}, \emph{chiral}, and \emph{antichiral currents}:

\begin{equation}
  \langle j^a\rangle = Q \mathcal N^a \begin{cases}
    1 + O(\epsilon) & \textrm{energy currents} \\
    (|e_R|^2 - |e_L|^2) + O(\epsilon) & \textrm{zilch currents} \\
    e_R \bar{e}_L + O(\epsilon) & \textrm{chiral currents} \\
    \bar{e}_R e_L + O(\epsilon) & \textrm{antichiral currents}
  \end{cases}.
\end{equation}
This classification scheme is a specialization of that of~\cite{Anco:2002xn}.
For conserved currents that are $\mathbb{R}$-bilinear functionals of $(\var \Psi)_{ABCD}$ (a property which is satisfied by all currents considered in this paper), there is a relationship between $Q$ and the type of current in this classification: for energy and zilch currents,

\begin{equation}
  Q = Q_{a_1 \cdots a_n} p^{a_1} \cdots p^{a_n},
\end{equation}
where $Q_{a_1 \cdots a_n}$ is a rank $n$ Killing tensor and $n$ is odd for energy currents and even for zilch currents.
Moreover, for chiral and antichiral currents, $Q$ cannot be written in the above form.
Since we wanted to construct conserved currents which were related to the Carter constant, which is a conserved quantity arising from a rank two Killing tensor, it is unsurprising that all currents which we considered were zilch currents.

Another interesting result of this classification scheme is an odd result for the symplectic product for the master variables.
The symplectic product for linearized gravity, when applied to $\var g_{ab}$ and $\lie_\xi \var g_{ab}$, gives an energy current in geometric optics, and the associated conserved quantity is proportional to $\xi^a p_a$ (which would be proportional to the energy in the case $\xi^a = t^a$).
This current is known as the \emph{canonical energy current}.
However, using the symplectic product for the master variables, one finds that a similar current, obtained by using $\pb{\pm s} \Omega$ and $\lie_\xi \pb{\pm s} \Omega$, gives a chiral current.
In this sense, the symplectic product for the master variables cannot be used to construct a current whose geometric optics limit behaves like energy.

\section{Fluxes at null infinity and the horizon} \label{section:fluxes}

Another desirable property for a conserved current is that its flux through the horizon ($H$) and through null infinity ($\scri$) be finite.
In this section, we provide formulae for these fluxes, using results for the asymptotic falloffs in appendix~\ref{appendix:asymptotics}.
More details on the definitions of these fluxes are given in appendix~\ref{appendix:fluxes_integration}.

We begin with some notation: first, the Boyer-Lindquist coordinate system is not well suited to working at the horizon or null infinity.
Instead, one uses the ingoing and outgoing coordinate systems $(v, r, \theta, \psi)$ and $(u, r, \theta, \chi)$, defined in terms of Boyer-Lindquist coordinates and the tortoise coordinate~\eqref{eqn:tortoise} by

\begin{subequations}
  \begin{align}
    v &= t + r^*, \qquad &\psi &= \phi + \int \frac{a \ud r}{\Delta}, \\
    u &= t - r^*, \qquad &\chi &= \phi - \int \frac{a \ud r}{\Delta}.
  \end{align}
\end{subequations}
The ingoing coordinate system is relevant near the future horizon ($H^+$) and past null infinity ($\scri^-$), while the outgoing coordinate system is relevant near the past horizon ($H^-$) and future null infinity ($\scri^+$).
When dealing with a generic surface $S$, we will write $w$ and $\alpha$ instead of either $v$ and $\psi$ or $u$ and $\chi$:

\begin{equation}
  w = \begin{cases}
    v & \textrm{ at $H^+$, $\scri^-$} \\
    u & \textrm{ at $H^-$, $\scri^+$}
  \end{cases}, \qquad \alpha = \begin{cases}
    \psi & \textrm{ at $H^+$, $\scri^-$} \\
    \chi & \textrm{ at $H^-$, $\scri^+$}
  \end{cases}.
\end{equation}
This greatly simplifies definitions.
For example, we will write the flux of a current $\pb{\ldots} j^a$ through a surface $S$ as $\ud^2 \pb{\ldots} Q/\ud w \ud \Omega|_S$, which we will define more explicitly in equation~\eqref{eqn:differential_fluxes}, where the differential solid angle is defined by

\begin{equation}
  \ud \Omega \equiv \sin \theta \ud \theta \ud \alpha.
\end{equation}

We next remark that, in this paper, we compute fluxes of the conserved currents~\eqref{eqn:C_current}, \eqref{eqn:C_current_proj}, \eqref{eqn:D_current}, and~\eqref{eqn:D_current_proj} \emph{only} when acting upon the metric perturbations $\Im[\var_\pm g_{ab}]$.
We are free to do so, as these metric perturbations are related by a gauge transformation to any $l \geq 2$ metric perturbation $\var g_{ab}$.
Moreover, this specialization allows us to use equations~\eqref{eqn:C_tensor_action} and~\eqref{eqn:D_tensor_action} in order to write the fluxes in terms of the fluxes of the currents

\begin{equation} \label{eqn:mode_currents}
  \pb{\pm 2} j^a_{ll'm\omega pp'} \equiv \pb{S} j_{\textrm{EH}}^a \Big[(\var_\pm \bs{g})_{lm\omega p}, \overline{(\var_\pm \bs{g})_{l'm\omega p'}}\Big],
\end{equation}
assuming that we average over $w$ and $\alpha$.
These currents are functions of the Debye potentials $\pb{\pm 2} \psi$, instead of the metric perturbation.
In particular, they are functions of the coefficients $\pb{s} \widehat{\psi}_{lm\omega p}^{\; \textrm{in/out/down/up}}$.
In terms of the fluxes of the currents~\eqref{eqn:mode_currents}, we have that (averaging over $w$ and $\alpha$)

\begin{subequations}
  \begin{align}
    \left\langle\frac{\ud^2 \pb{\pb{s} \mathcal{C}} Q}{\ud w \ud \Omega}\right\rangle_{w, \alpha} &= \frac{i}{32} \int_{-\infty}^\infty \ud \omega \sum_{l, l' = 2}^\infty \sum_{|m| \leq \min(l, l')} \sum_{p, p' = \pm 1} pp' \pb{s} C_{lm\omega p} \overline{\pb{s} C_{l'm\omega p'}} \frac{\ud^2 \pb{s} Q_{ll'm\omega pp'}}{\ud w \ud \Omega}, \\
    \left\langle\frac{\ud^2 \pb{\pb{s} \mathcal{D}} Q}{\ud w \ud \Omega}\right\rangle_{w, \alpha} &= \frac{i}{16} \int_{-\infty}^\infty \ud \omega \sum_{l, l' = 2}^\infty \sum_{|m| \leq \min(l, l')} \sum_{p, p' = \pm 1} \pb{2} \lambda_{l'm\omega} \frac{\ud^2 \pb{s} Q_{ll'm\omega pp'}}{\ud w \ud \Omega}.
  \end{align}
\end{subequations}
As these quantities are all $\mathbb{R}$-bilinear, it is convenient to define

\begin{equation}
  \pb{s} \Upsilon^{\textrm{in}/\textrm{out}/\textrm{down}/\textrm{up}}_{ll'm\omega pp'} \equiv \pb{s} \widehat{\psi}^{\textrm{in}/\textrm{out}/\textrm{down}/\textrm{up}}_{lm\omega p} \overline{\pb{s} \widehat{\psi}^{\textrm{in}/\textrm{out}/\textrm{down}/\textrm{up}}_{l'm\omega p'}}.
\end{equation}

Moreover, the fluxes will each have a nontrivial angular dependence.
To determine this, we define, for some quantity $q[\pb{s} \psi]$, with coefficients $q_{lm\omega p} [\pb{s} \psi]$ in an expansion, the angular dependences $\pb{q} S^{\textrm{in}/\textrm{out}/\textrm{down}/\textrm{up}}_{lm\omega p} (\theta)$ by

\begin{equation} \label{eqn:S_def}
  q_{lm\omega p} (t, r, \theta, \phi) \equiv \begin{cases}
    \pb{s} \widehat{\psi}^{\textrm{in}}_{lm\omega p} e^{i(m\psi - \omega v)} \pb{q} S^{\textrm{in}}_{lm\omega p} (\theta) \Delta^{n_q^{\textrm{in}}} + \pb{s} \widehat{\psi}^{\textrm{out}}_{lm\omega p} e^{i(m\chi - \omega u)} \pb{q} S^{\textrm{out}}_{lm\omega p} (\theta) \Delta^{n_q^{\textrm{out}}} & r \to r_+ \\
    \pb{s} \widehat{\psi}^{\textrm{down}}_{lm\omega p} e^{i(m\psi - \omega v)} \pb{q} S^{\textrm{down}}_{lm\omega p} (\theta) r^{n_q^{\textrm{down}}} + \pb{s} \widehat{\psi}^{\textrm{up}}_{lm\omega p} e^{i(m\chi - \omega u)} \pb{q} S^{\textrm{up}}_{lm\omega p} (\theta) r^{n_q^{\textrm{up}}} & r \to \infty
  \end{cases},
\end{equation}
for some integers $n_q^{\textrm{in}/\textrm{out}/\textrm{down}/\textrm{up}}$.
Assuming appropriate smoothness conditions, equation~\eqref{eqn:S_def} simplifies further if we specialize to the various surfaces at which we are computing these quantities:

\begin{equation}
  \left.q_{lm\omega p} (t, r, \theta, \phi)\right|_S \sim \begin{cases}
    \pb{s} \widehat{\psi}^{\textrm{in}}_{lm\omega p} e^{i(m\psi - \omega v)} \pb{q} S^{\textrm{in}}_{lm\omega p} (\theta) \Delta^{n_q^{\textrm{in}}} & S = H^+ \\
    \pb{s} \widehat{\psi}^{\textrm{out}}_{lm\omega p} e^{i(m\chi - \omega u)} \pb{q} S^{\textrm{out}}_{lm\omega p} (\theta) \Delta^{n_q^{\textrm{out}}} & S = H^- \\
    \pb{s} \widehat{\psi}^{\textrm{down}}_{lm\omega p} e^{i(m\psi - \omega v)} \pb{q} S^{\textrm{down}}_{lm\omega p} (\theta) r^{n_q^{\textrm{down}}} & S = \scri^- \\
    \pb{s} \widehat{\psi}^{\textrm{up}}_{lm\omega p} e^{i(m\chi - \omega u)} \pb{q} S^{\textrm{up}}_{lm\omega p} (\theta) r^{n_q^{\textrm{up}}} & S = \scri^+
  \end{cases}.
\end{equation}
In other words, only ``in'' modes contribute at $H^+$, ``out'' modes at $H^-$, etc.
The various quantities $q$ which we will be considering will be components of metric perturbations and perturbed connection coefficients.
The relevant integers $n_q^{\textrm{in/out/down/up}}$ are (effectively) given in table~\ref{table:gr_asymp}.
Moreover, the various angular dependences are given by equations~\eqref{eqn:perturbed_metric_asymp} and~\eqref{eqn:perturbed_chris_asymp}, and computed in appendix~\ref{appendix:asymptotics}.

Using table~\ref{table:gr_asymp} and equations~\eqref{eqn:symplectic_flux_l} and~\eqref{eqn:symplectic_flux_n}, we find that

\begin{subequations} \label{eqn:base_fluxes_down}
  \begin{align}
    \left.\frac{\ud^2 \pb{+2} Q^{\textrm{down}}_{ll'm\omega pp'}}{\ud u \ud \Omega}\right|_{\scri^+} &= 0, \\
    \left.\frac{\ud^2 \pb{+2} Q^{\textrm{down}}_{ll'm\omega pp'}}{\ud v \ud \Omega}\right|_{\scri^-} &= -\frac{i}{64\pi} \pb{-2} \Upsilon^{\textrm{down}}_{ll'm\omega pp'} \pb{\var_+ C_{l\bar{m}\bar{m}}} S^{\textrm{down}}_{lm\omega p} \overline{\pb{\var_+ g_{\bar{m}\bar{m}}} S^{\textrm{down}}_{l'm\omega p'}} + \overline{\interchange{l, p}{l', p'}}, \\
    \left.\frac{\ud^2 \pb{+2} Q^{\textrm{down}}_{ll'm\omega pp'}}{\ud v \ud \Omega}\right|_{H^+} &= -\frac{i}{64\pi} \pb{-2} \Upsilon^{\textrm{in}}_{ll'm\omega pp'} \pb{\var_+ C_{l\bar{m}\bar{m}}} S^{\textrm{in}}_{lm\omega p} \overline{\pb{(\var_+ g)_{\bar{m}\bar{m}}} S^{\textrm{in}}_{l'm\omega p'}} + \overline{\interchange{l, p}{l', p'}}, \\
    \left.\frac{\ud^2 \pb{+2} Q^{\textrm{down}}_{ll'm\omega pp'}}{\ud u \ud \Omega}\right|_{H^-} &= \frac{i\Sigma_+}{32\pi} \pb{-2} \Upsilon^{\textrm{out}}_{ll' m\omega pp'} \left(\pb{\var_+ C_{n\bar{m}\bar{m}}} S^{\textrm{out}}_{lm\omega p} \overline{\pb{\var_+ g_{\bar{m}\bar{m}}} S^{\textrm{out}}_{l'm\omega p'}} - \pb{\var_+ C_{n(l\bar{m})}} S^{\textrm{out}}_{lm\omega p} \overline{\pb{\var_+ g_{n\bar{m}}} S^{\textrm{out}}_{l'm\omega p'}}\right) \nonumber \\
    &\hspace{1em}+ \overline{\interchange{l, p}{l', p'}},
  \end{align}
\end{subequations}
where the superscript ``down'' indicates that we have performed a projection such that $\pb{s} \widehat{\psi}^{\textrm{up}}_{lm\omega p} = 0$, and

\begin{subequations} \label{eqn:base_fluxes_up}
  \begin{align}
    \left.\frac{\ud^2 \pb{-2} Q^{\textrm{up}}_{ll'm\omega pp'}}{\ud u \ud \Omega}\right|_{\scri^+} &= \frac{i}{32\pi} \pb{2} \Upsilon^{\textrm{up}}_{ll'm\omega pp'} \pb{\var_- C_{nmm}} S^{\textrm{up}}_{lm\omega p} \overline{\pb{\var_- g_{mm}} S^{\textrm{up}}_{l'm\omega p'}} + \overline{\interchange{l, p}{l', p'}}, \\
    \left.\frac{\ud^2 \pb{-2} Q^{\textrm{up}}_{ll'm\omega pp'}}{\ud v \ud \Omega}\right|_{\scri^-} &= 0, \\
    \left.\frac{\ud^2 \pb{-2} Q^{\textrm{up}}_{ll'm\omega pp'}}{\ud v \ud \Omega}\right|_{H^+} &= -\frac{i}{64\pi} \pb{2} \Upsilon^{\textrm{in}}_{ll'm\omega pp'} \left(\pb{\var_- C_{lmm}} S^{\textrm{in}}_{lm\omega p} \overline{\pb{\var_- g_{mm}} S^{\textrm{in}}_{l'm\omega p'}} - \pb{\var_- C_{l(nm)}} S^{\textrm{in}}_{lm\omega p} \overline{\pb{\var_- g_{lm}} S^{\textrm{in}}_{l'm\omega p'}}\right) \nonumber \\
    &\hspace{1em}+ \overline{\interchange{l, p}{l', p'}}, \\
    \left.\frac{\ud^2 \pb{-2} Q^{\textrm{up}}_{ll'm\omega pp'}}{\ud u \ud \Omega}\right|_{H^-} &= \frac{i\Sigma_+}{32\pi} \pb{2} \Upsilon^{\textrm{out}}_{ll'm\omega pp'} \pb{\var_- C_{nmm}} S^{\textrm{out}}_{lm\omega p} \overline{\pb{\var_- g_{mm}} S^{\textrm{out}}_{l'm\omega p'}} + \overline{\interchange{l, p}{l', p'}},
  \end{align}
\end{subequations}
and the superscript ``up'' denotes the fact that we have performed a projection to set $\pb{s} \widehat{\psi}^{\textrm{down}}_{lm\omega p} = 0$.
If these projections are not performed, then the respective fluxes \emph{diverge}, as is evident from table~\ref{table:gr_asymp} and equation~\eqref{eqn:symplectic_components}.
Since the fluxes of $\pb{\pb{s} \mathcal{C}} j^a$ and $\pb{\pb{s} \mathcal{D}} j^a$ can be written in terms of those of $\pb{s} j_{ll'm\omega pp'}^a$, there are issues with these currents as well.

These divergences motivated the introduction of the projection operators in section~\ref{section:projection}.
With these projection operators, we have sacrificed locality (which we had already sacrificed in $\pb{\pb{s} \mathcal{D}} j^a$) in order to obtain finite fluxes.
As mentioned at the end of section~\ref{section:geometric_optics_examples}, the geometric optics limits are similar to those of the currents $\pb{\pb{s} \mathcal{C}} j^a$ and $\pb{\pb{s} \mathcal{D}} j^a$.
We also have that

\begin{subequations} \label{eqn:projection_fluxes}
  \begin{align}
    \left\langle\frac{\ud^2 \pb{\pb{2} \mathring{\mathcal C}} Q}{\ud w \ud \Omega}\right\rangle_{w, \alpha} &= \frac{i}{32} \int_{-\infty}^\infty \ud \omega \sum_{l, l' = 2}^\infty \sum_{|m| \leq \min(l, l')} \sum_{p, p' = \pm 1} \nonumber \\
                            &\hspace{3.5em}\times pp' \left\{\pb{2} C_{lm\omega p} \overline{\pb{2} C_{l'm\omega p'}} \frac{\ud^2 \pb{2} Q^{\textrm{down}}_{ll'm\omega pp'}}{\ud w \ud \Omega} + \pb{-2} C_{lm\omega p} \overline{\pb{-2} C_{l'm\omega p'}} \frac{\ud^2 \pb{-2} Q^{\textrm{up}}_{ll'm\omega pp'}}{\ud w \ud \Omega}\right\}, \\
    \left\langle\frac{\ud^2 \pb{\pb{2} \mathring{\mathcal D}} Q}{\ud w \ud \Omega}\right\rangle_{w, \alpha} &= \frac{i}{16} \int_{-\infty}^\infty \ud \omega \sum_{l, l' = 2}^\infty \sum_{|m| \leq \min(l, l')} \sum_{p, p' = \pm 1} \pb{2} \lambda_{l'm\omega} \left\{\frac{\ud^2 \pb{2} Q^{\textrm{down}}_{ll'm\omega pp'}}{\ud w \ud \Omega} + \frac{\ud^2 \pb{-2} Q^{\textrm{up}}_{ll'm\omega pp'}}{\ud w \ud \Omega}\right\}.
  \end{align}
\end{subequations}
Using equations~\eqref{eqn:base_fluxes_down},~\eqref{eqn:base_fluxes_up}, and~\eqref{eqn:projection_fluxes}, we have completely determined the fluxes of the charges $\pb{\pb{2} \mathring{\mathcal C}} j^a$ and $\pb{\pb{2} \mathring{\mathcal D}} j^a$.

Using the symplectic product for linearized gravity, we have not been able to construct a \emph{local} current with finite fluxes which reduces to the Carter constant in geometric optics.
However, we can do so using the symplectic product we defined in equation~\eqref{eqn:symp_form_BCJR} for the master variables.
We find that the fluxes for $\pb{\pb{s} \Omega} j^a$, averaged over $w$ and $\alpha$, are given by an expansion of the form

\begin{equation} \label{eqn:Omega_flux}
  \left\langle\frac{\ud^2 \pb{\pb{s} \Omega} Q}{\ud w \ud \Omega}\right\rangle_{w, \alpha} \equiv \int_{-\infty}^\infty \ud \omega\; \sum_{l, l' = 2}^\infty\; \sum_{|m| < l, l'} \sum_{p, p' = \pm 1} \frac{\ud^2 \pb{\pb{s} \Omega} Q_{ll'm\omega pp'}}{\ud w \ud \Omega},
\end{equation}
where

\begin{subequations}
  \begin{align}
    &\begin{aligned}
      \left.\frac{\ud^2 \pb{\pb{s} \Omega} Q_{ll'm\omega pp'}}{\ud u \ud \Omega}\right|_{\scri^+} = \frac{\omega}{32\pi} \bigg\{&C_{l'm\omega} \pb{s} \Theta_{lm\omega} \pb{s} \Theta_{l'm\omega} \Big[\pb{s} \widehat{\psi}^{\textrm{up}}_{lm\omega p} \overline{\pb{-s} \widehat{\psi}^{\textrm{up}}_{l'm\omega p'}} + \overline{\interchange{l, p, s}{l', p', -s}}\Big] \\
      &+ \pb{s} C_{l'm\omega p'} \pb{-s} \Theta_{lm\omega} \pb{-s} \Theta_{l'm\omega} \Big[\pb{-s} \widehat{\psi}^{\textrm{up}}_{lm\omega p} \overline{\pb{s} \widehat{\psi}^{\textrm{up}}_{l'm\omega p'}} + \overline{\interchange{l, p, s}{l', p', -s}}\Big]\bigg\},
    \end{aligned} \\
    &\begin{aligned}
      \left.\frac{\ud^2 \pb{\pb{s} \Omega} Q_{ll'm\omega pp'}}{\ud v \ud \Omega}\right|_{\scri^-} = -\frac{\omega}{32\pi} \bigg\{&C_{l'm\omega} \pb{s} \Theta_{lm\omega} \pb{s} \Theta_{l'm\omega} \Big[\pb{s} \widehat{\psi}^{\textrm{down}}_{lm\omega p} \overline{\pb{-s} \widehat{\psi}^{\textrm{down}}_{l'm\omega p'}} + \overline{\interchange{l, p, s}{l', p', -s}}\Big] \\
      &+ \pb{s} C_{l'm\omega p'} \pb{-s} \Theta_{lm\omega} \pb{-s} \Theta_{l'm\omega} \Big[\pb{-s} \widehat{\psi}^{\textrm{down}}_{lm\omega p} \overline{\pb{s} \widehat{\psi}^{\textrm{down}}_{l'm\omega p'}} + \overline{\interchange{l, p, s}{l', p', -s}}\Big]\bigg\},
    \end{aligned}
  \end{align}
\end{subequations}
and

\begin{subequations}
  \begin{align}
    &\begin{aligned}
      \left.\frac{\ud^2 \pb{\pb{s} \Omega} Q_{ll'm\omega pp'}}{\ud v \ud \Omega}\right|_{H^+} = -\frac{Mr_+ k_{m\omega}}{16\pi} \bigg\{&C_{l'm\omega} \pb{s} \Theta_{lm\omega} \pb{s} \Theta_{l'm\omega} \Big[\pb{s} \kappa_{m\omega} \pb{s} \widehat{\psi}^{\textrm{in}}_{lm\omega p} \overline{\pb{-s} \widehat{\psi}^{\textrm{in}}_{l'm\omega p'}} \\
      &\hspace{10.5em}+ \overline{\interchange{l, p, s}{l', p', -s}}\Big] \\
      &+ \pb{s} C_{l'm\omega p'} \pb{-s} \Theta_{lm\omega} \pb{-s} \Theta_{l'm\omega} \Big[\pb{-s} \kappa_{m\omega} \pb{-s} \widehat{\psi}^{\textrm{in}}_{lm\omega p} \overline{\pb{s} \widehat{\psi}^{\textrm{in}}_{l'm\omega p'}} \\
      &\hspace{12.5em}+ \overline{\interchange{l, p, s}{l', p', -s}}\Big]\bigg\},
    \end{aligned} \\
    &\begin{aligned}
      \left.\frac{\ud^2 \pb{\pb{s} \Omega} Q_{ll'm\omega pp'}}{\ud u \ud \Omega}\right|_{H^-} = \frac{Mr_+ k_{m\omega}}{16\pi} \bigg\{&C_{l'm\omega} \pb{s} \Theta_{lm\omega} \pb{s} \Theta_{l'm\omega} \Big[\pb{s} \kappa_{m\omega} \pb{s} \widehat{\psi}^{\textrm{out}}_{lm\omega p} \overline{\pb{-s} \widehat{\psi}^{\textrm{out}}_{l'm\omega p'}} \\
      &\hspace{10.5em}+ \overline{\interchange{l, p, s}{l', p', -s}}\Big] \\
      &+ \pb{s} C_{l'm\omega p'} \pb{-s} \Theta_{lm\omega} \pb{-s} \Theta_{l'm\omega} \Big[\pb{-s} \kappa_{m\omega} \pb{-s} \widehat{\psi}^{\textrm{out}}_{lm\omega p} \overline{\pb{s} \widehat{\psi}^{\textrm{out}}_{l'm\omega p'}} \\
      &\hspace{12.5em}+ \overline{\interchange{l, p, s}{l', p', -s}}\Big]\bigg\},
    \end{aligned}
  \end{align}
\end{subequations}
where

\begin{equation}
  \pb{s} \kappa_{m\omega} = 1 - \frac{is(r_+ - M)}{2Mr_+ k_{m\omega}}.
\end{equation}

\section{Discussion} \label{section:discussion}

\begin{table}[t!]
  \centering
  \caption{\label{table:summary} Summary of the properties of the conserved currents considered in this paper.
    For convenience, we give the equation numbers (within section~\ref{section:definitions}) in which these currents are defined.
    We then give the limit of the corresponding charges in geometric optics, where $K$ is the Carter constant of a graviton (see section~\ref{section:geometric_optics} for the definitions of the polarization coefficients $e_R$ and $e_L$, as well as the justification of the factors of $\hbar$).
    The next column indicates whether the fluxes of these currents through future and past null infinity ($\scri^\pm$) and the future and past horizons ($H^\pm$) are finite.
    We finally indicate which of these currents are local functionals of the metric perturbation.}
  \renewcommand{\arraystretch}{1.5}
  \begin{tabular}{|*{8}{c|}} \hline
    & Definition & Geometric optics limit & \multicolumn{4}{c|}{Finite fluxes?} & \\
    Current & (equation) & of charge (per graviton) & $\scri^+$ & $\scri^-$ & $H^+$ & $H^-$ & Local? \\\hline
    $\pb{\pb{2} \mathcal{C}} j^a [\var \bs g]$ & \multirow{2}{*}{\eqref{eqn:C_current}} & \multirow{2}{*}{$K^4 (|e_R|^2 - |e_L|^2)/\hbar^7$} & $\times$ & \checkmark & \checkmark & \checkmark & \checkmark \\
    $\pb{\pb{-2} \mathcal{C}} j^a [\var \bs g]$ & & & \checkmark & $\times$ & \checkmark & \checkmark & \checkmark \\\hline
    $\pb{\pb{2} \mathring{\mathcal C}} j^a [\var \bs g]$ & \eqref{eqn:C_current_proj} & $K^4 (|e_R|^2 - |e_L|^2)/\hbar^7\;$~\footnote{\label{foot:tab} \!\!This result only holds, if the null fluid of gravitons is either completely ingoing or outgoing at null infinity; see the discussion near the end of section~\ref{section:geometric_optics_examples} for more details.} & \checkmark & \checkmark & \checkmark & \checkmark & $\times$ \\\hline
    $\pb{\pb{2} \mathcal{D}} j^a [\var \bs g]$ & \multirow{2}{*}{\eqref{eqn:D_current}} & \multirow{2}{*}{$K (|e_R|^2 - |e_L|^2)/\hbar$} & $\times$ & \checkmark & \checkmark & \checkmark & $\times$ \\
    $\pb{\pb{-2} \mathcal{D}} j^a [\var \bs g]$ & & & \checkmark & $\times$ & \checkmark & \checkmark & $\times$ \\\hline
    $\pb{\pb{2} \mathring{\mathcal D}} j^a [\var \bs g]$ & \eqref{eqn:D_current_proj} & $K (|e_R|^2 - |e_L|^2)/\hbar\;$~\textsuperscript{\ref{foot:tab}} & \checkmark & \checkmark & \checkmark & \checkmark & $\times$\\\hline
    \begin{tabular}{c}
      $\pb{\pb{2} \Omega} j^a [\var \bs g]$ \\
      $\pb{\pb{-2} \Omega} j^a [\var \bs g]$ \\
    \end{tabular} & \eqref{eqn:Omega_current} & $K^4 (|e_R|^2 - |e_L|^2)/\hbar^7$ & \checkmark & \checkmark & \checkmark & \checkmark & \checkmark \\\hline
  \end{tabular}
\end{table}

In this paper, we have constructed a class of conserved currents for linearized gravity whose conserved charges reduce to the sum of the Carter constants (to some positive power) for a null fluid of gravitons in the geometric optics limit.
These conserved currents are constructed from symplectic products of two solutions constructed via the method of symmetry operators.
Moreover, some of these currents yield finite fluxes at the horizon and null infinity, although most that are finite at null infinity are not local.
A full summary of their properties is given in table~\ref{table:summary}.
Note that only the currents $\pb{\pb{s} \Omega} j^a$ are both local and possess finite fluxes.

That some of these currents possess diverging fluxes at null infinity is not ideal.
It may be possible to find a symmetry operator, differing from those that appear in this paper by a gauge transformation, that is both local and maps to a solution with a non-divergent symplectic product.
In the absence of a clear example of such a symmetry operator, we have instead decided to consider nonlocal symmetry operators which are easier to define.
We have also shown that there exists a symplectic product for the master variables (instead of the metric perturbation) which yields finite fluxes.
This symplectic product can also be used to construct a current which gives (positive powers of) the Carter constant in the limit of geometric optics.
However, note that this is not the physical symplectic product for linearized gravity.

One motivation for seeking conserved currents is the hope to derive, for the dynamical system of a point particle coupled to linearized gravity in the Kerr spacetime, a ``unified conservation law'' that would generalize the conservation of the Carter constant for a point particle by itself.
The local currents considered in this paper could be relevant for such a conservation law, but the potential relevance of the nonlocal currents is less obvious.
We plan to further explore these currents, particularly their applications, in future work.

\section*{Acknowledgments}

We thank Lars Andersson and Kartik Prabhu for helpful conversations.
We acknowledge the support of NSF Grants PHY-1404105 and PHY-1707800 to Cornell University.

\appendix

\section{Integration along the horizon and null infinity} \label{appendix:fluxes_integration}

The flux of a current $\pb{\ldots} j^a$ through a surface $S$ of constant $r$ (such as the horizon or null infinity) is defined by

\begin{equation} \label{eqn:differential_fluxes}
  \left.\frac{\ud^2 \pb{\ldots} Q}{\ud w \ud \Omega}\right|_S \equiv \lim_{\to S} (r^2 + a^2) \pb{\ldots} j^a N_a,
\end{equation}
where $N_a$ is the surface normal, and the factor of $r^2 + a^2$ comes from the fact that the determinant of the induced metric on surfaces of constant $r$ is $(r^2 + a^2) \sin \theta$.
The surface normals are proportional to $(\ud r)_a$,

\begin{equation}
  N_a \propto (\ud r)_a = n_a - \frac{\Delta}{2\Sigma} l_a,
\end{equation}
and the usual scaling freedom is fixed by requiring\footnote{Note that, if one were integrating these currents on a finite portion of these surfaces, the normalization of $N_a$ would not matter.
  However, for equation~\eqref{eqn:differential_fluxes} to hold---that is, when integrating over an infinitesimal portion $\ud w$, for $w = u$ or $v$, we must normalize $N_a$ appropriately.} that either $N^a \nabla_a u = 1$ (for $H^-$ and $\scri^+$) or $N^a \nabla_a v = 1$ (for $H^+$ and $\scri^-$).
It turns out, however, that these requirements are the same, and fix the normalization such that

\begin{equation}
  N_a = \frac{1}{r^2 + a^2} \left(\Sigma n_a - \frac{\Delta}{2} l_a\right).
\end{equation}
As such, we find that

\begin{subequations}
  \begin{align}
    \left.\frac{\ud^2 Q}{\ud v \ud \Omega}\right|_{H^+} &= \lim_{r \to r_+, v \textrm{ fixed}} \Sigma \left(j_n - \frac{\Delta}{2\Sigma} j_l\right), \\
    \left.\frac{\ud^2 Q}{\ud u \ud \Omega}\right|_{H^-} &= \lim_{r \to r_+, u \textrm{ fixed}} \Sigma \left(j_n - \frac{\Delta}{2\Sigma} j_l\right), \\
    \left.\frac{\ud^2 Q}{\ud v \ud \Omega}\right|_{\scri^-} &= \lim_{r \to \infty, v \textrm{ fixed}} r^2 \left(j_n - \frac{1}{2} j_l\right), \\
    \left.\frac{\ud^2 Q}{\ud u \ud \Omega}\right|_{\scri^+} &= \lim_{r \to \infty, v \textrm{ fixed}} r^2 \left(j_n - \frac{1}{2} j_l\right).
  \end{align}
\end{subequations}

From this discussion, for the calculations in section~\ref{section:fluxes}, we need the components of symplectic products along $l_a$ and $n_a$:

\begin{subequations} \label{eqn:symplectic_components}
  \begin{align}
    \pb{S} j^{\textrm{EH}}_l \left[\var_+ \boldsymbol{g}, \overline{\var_+ \boldsymbol{g}}\right] &= -\frac{1}{16\pi} \Im\left[(\var_+ C)_{l\bar{m}\bar{m}} \overline{(\var_+ g)^{\bar{m}\bar{m}}}\right], \label{eqn:symplectic_flux_l} \\
    \pb{S} j^{\textrm{EH}}_n \left[\var_+ \boldsymbol{g}, \overline{\var_+ \boldsymbol{g}}\right] &= -\frac{1}{16\pi} \Im\left[(\var_+ C)_{n\bar{m}\bar{m}} \overline{(\var_+ g)^{\bar{m}\bar{m}}} - (\var_+ C)_{n(l\bar{m})} \overline{(\var_+ g)^{(n\bar{m})}}\right], \label{eqn:symplectic_flux_n}
  \end{align}
\end{subequations}
where $l$, $n$, $m$, and $\bar{m}$ subscripts denote contraction on an index with the corresponding null tetrad vector, and where the non-zero perturbed connection coefficients are

\begin{subequations}
  \begin{align}
    (\var_+ C)_{l\bar{m}\bar{m}} &= -\frac{1}{2} [D + 2(\epsilon - \bar{\epsilon}) - \rho] (\var_+ g)_{\bar{m}\bar{m}}, \\
    (\var_+ C)_{n(l\bar{m})} &= -\frac{1}{4} (D + 2\epsilon + \rho) (\var_+ g)_{(n\bar{m})} - \frac{1}{2} \tau (\var_+ g)_{\bar{m}\bar{m}}, \\
    (\var_+ C)_{n\bar{m}\bar{m}} &= -\frac{1}{4} (\npdelta + 2\bar{\alpha}) (\var_+ g)_{(n\bar{m})} - \frac{1}{2} [\npDelta + 2(\gamma - \bar{\gamma}) - 2\mu] (\var_+ g)_{\bar{m}\bar{m}}.
  \end{align}
\end{subequations}
One can obtain the analogous expressions for $\var_-$ by performing a $'$ transformation.
For the symplectic product defined using the master variables, we find that

\begin{subequations}
  \begin{align}
    \pb{S} j^{\textrm{BCJR}}_l \left[\var_1 \pb{s} \Omega, \var_1 \pb{-s} \Omega; \var_2 \pb{s} \Omega, \var_2 \pb{-s} \Omega\right] &= \var_1 \pb{s} \Omega (D - s\Gamma_l) \var_2 \pb{-s} \Omega + \var_1 \pb{-s} \Omega (D + s\Gamma_l) \var_2 \pb{s} \Omega - \interchange{1}{2}, \\
    \pb{S} j^{\textrm{BCJR}}_n \left[\var_1 \pb{s} \Omega, \var_1 \pb{-s} \Omega; \var_2 \pb{s} \Omega, \var_2 \pb{-s} \Omega\right] &= \var_1 \pb{s} \Omega (\npDelta - s\Gamma_n) \var_2 \pb{-s} \Omega + \var_1 \pb{-s} \Omega (\npDelta + s\Gamma_n) \var_2 \pb{s} \Omega - \interchange{1}{2}.
  \end{align}
\end{subequations}

\section{Asymptotic behavior} \label{appendix:asymptotics}

In order to determine fluxes at null infinity and the horizon, we also need to know the asymptotic behavior of the quantities that appear in equation~\eqref{eqn:symplectic_components} and its $'$ transform.
These are given in table~\ref{table:gr_asymp}.
To determine these falloff rates, we write the quantities that appear in~\eqref{eqn:symplectic_components} and its $'$ transform in terms of differential operators acting upon the Debye potential, using the operators defined in equation~\eqref{eqn:D_L_operators}: the perturbed metric satisfies

\begin{subequations} \label{eqn:var_g_exp}
  \begin{align}
    (\var_+ g)_{(n\bar{m})} &= -\frac{1}{\sqrt{2} \bar{\zeta}} \left[\left(\mathscr{D}_0 + \frac{1}{\zeta} - \frac{2}{\bar{\zeta}}\right) \left(\mathscr{L}_2^+ - \frac{3ia \sin \theta}{\zeta}\right) + \left(\mathscr{L}_2^+ + \frac{ia \sin \theta}{\zeta} + \frac{2ia \sin \theta}{\bar{\zeta}}\right) \left(\mathscr{D}_0 - \frac{3}{\zeta}\right)\right] \pb{-2} \psi, \\
    (\var_+ g)_{\bar{m}\bar{m}} &= -\left(\mathscr{D}_0 + \frac{1}{\zeta}\right) \left(\mathscr{D}_0 - \frac{3}{\zeta}\right) \pb{-2} \psi, \\
    (\var_- g)_{(lm)} &= \frac{\zeta^2}{2 \sqrt{2} \bar{\zeta} \Delta} \left[\left(\mathscr{L}_2 + \frac{ia \sin \theta}{\zeta} + \frac{2ia \sin \theta}{\bar{\zeta}}\right) \left(\mathscr{D}_0^+ - \frac{3}{\zeta}\right) + \left(\mathscr{D}_0^+ + \frac{1}{\zeta} - \frac{2}{\bar{\zeta}}\right) \left(\mathscr{L}_2 - \frac{3ia \sin \theta}{\zeta}\right)\right] \Delta^2 \pb{2} \psi, \\
    (\var_- g)_{mm} &= \frac{\zeta^2}{4\bar{\zeta}^2} \left(\mathscr{D}_0^+ + \frac{1}{\zeta}\right) \left(\mathscr{D}_0^+ - \frac{3}{\zeta}\right) \Delta^2 \pb{2} \psi,
  \end{align}
\end{subequations}
whereas the relevant perturbed connection coefficients are given by

\begin{subequations} \label{eqn:var_gamma_exp}
  \begin{align}
    (\var_+ C)_{l\bar{m}\bar{m}} &= -\frac{1}{2} \left(\mathscr{D}_0 + \frac{1}{\zeta}\right) (\var_+ g)_{\bar{m}\bar{m}}, \\
    (\var_+ C)_{n(l\bar{m})} &= -\frac{1}{4} \left(\mathscr{D}_0 - \frac{1}{\zeta}\right) (\var_+ g)_{(n\bar{m})} + \frac{ia \sin \theta}{2\sqrt{2} \Sigma} (\delta_+ g)_{\bar{m}\bar{m}}, \\
    (\var_+ C)_{n\bar{m}\bar{m}} &= -\frac{1}{4\sqrt{2} \bar{\zeta}} \left(\mathscr{L}_{-1}^+ - \frac{2ia \sin \theta}{\bar{\zeta}}\right) (\var_+ g)_{(n\bar{m})} + \frac{\Delta}{4 \Sigma} \left(\mathscr{D}_0^+ - \frac{2}{\zeta} - \frac{2}{\bar{\zeta}}\right) (\var_+ g)_{\bar{m}\bar{m}}, \\
    (\var_- C)_{nmm} &= \frac{\Delta}{4 \Sigma} \left(\mathscr{D}_0^+ - \frac{1}{\zeta} + \frac{2}{\bar{\zeta}}\right) (\var_- g)_{mm}, \\
    (\var_- C)_{l(nm)} &= \frac{1}{8 \Sigma} \left(\mathscr{D}_0^+ - \frac{3}{\zeta}\right) \Delta (\var_- g)_{(lm)} + \frac{ia \sin \theta}{2\sqrt{2} \zeta^2} (\var_- g)_{mm}, \\
    (\var_- C)_{lmm} &= -\frac{1}{4\sqrt{2} \zeta} \mathscr{L}_{-1} (\var_- g)_{(lm)} - \frac{1}{2} \left(\mathscr{D}_0 - \frac{2}{\zeta}\right) (\var_- g)_{mm}.
  \end{align}
\end{subequations}

In order to compute the asymptotic behavior of these quantities, one needs to determine the asymptotic behavior of derivatives of the master variables.
However, applying the na\"ive approach, which uses the asymptotic expansions given by equations~\eqref{eqn:radial_falloffs_in_out} and~\eqref{eqn:radial_falloffs_down_up}, along with

\begin{equation} \label{eqn:naive}
  \begin{split}
    &\left.\begin{aligned}
        \mathscr{D}_{0(\pm m)(\pm \omega)} f(r) e^{\pm i\omega r^*} &= \frac{\ud f}{\ud r} e^{\pm i\omega r^*} \\
        \mathscr{D}_{0(\pm m)(\pm \omega)} f(r) e^{\mp i\omega r^*} &= \left[\frac{\ud f}{\ud r} \mp 2i\omega f(r)\right] e^{\mp i\omega r^*} \\
      \end{aligned}\right\} r^* \to \infty, \\
    &\left.\begin{aligned}
        \mathscr{D}_{0(\pm m)(\pm \omega)} f(r) e^{\pm ik_{m\omega} r^*} &= \frac{\ud f}{\ud r} e^{\pm i k_{m\omega} r^*} \\
        \mathscr{D}_{0(\pm m)(\pm \omega)} f(r) e^{\mp ik_{m\omega} r^*} &= \left[\frac{\ud f}{\ud r} \mp \frac{4Mr^+}{\Delta} ik_{m\omega} f(r)\right] e^{\mp ik_{m\omega} r^*} \\
      \end{aligned}\right\} r^* \to -\infty,
  \end{split}
\end{equation}
results in cancellations in the leading-order behavior.
Instead, we use the radial Teukolsky-Starobinsky identity~\eqref{eqn:radial_starobinsky}, which provides a differential equation that is independent of the radial Teukolsky equation~\eqref{eqn:radial_teukolsky}.
Using the radial Teukolsky equation, one can reduce the radial Teukolsky-Starobinsky identity to the following expression for derivatives of $\pb{s} \widehat{\Omega}_{lm\omega p} (r)$~\cite{chandrasekhar1983mathematical}:

\begin{equation} \label{eqn:deriv_relation}
  \mathscr{D}_{0(\mp m)(\mp \omega)} \Delta^{(2 \pm 2)/2} \pb{\pm 2} \widehat{\Omega}_{lm\omega p} \equiv \pb{\pm 2} \Xi_{lm\omega p} \Delta^{(2 \pm 2)/2} \pb{\pm 2} \widehat{\Omega}_{lm\omega p} + \pb{\pm 2} \Pi_{lm\omega p} \Delta^{(2 \mp 2)/2} \pb{\mp 2} \widehat{\Omega}_{lm\omega p},
\end{equation}
where this equation defines the coefficients $\pb{\pm s} \Xi_{lm\omega p}$ and $\pb{\pm s} \Pi_{lm\omega p}$.
These equations also clearly hold for $\pb{s} \widehat{\psi}_{lm\omega p} (r)$.

\begin{table}[!t]
  \centering
  \caption{\label{table:gr_asymp} Asymptotic behavior of the solutions for linearized gravity.}
  \begin{tabular}{|l|c c|c c|} \hline
    & \multicolumn{2}{c|}{Ingoing [$e^{i(m\psi - \omega v)} \times$]} & \multicolumn{2}{c|}{Outgoing [$e^{i(m\chi - \omega u)} \times$]} \\
    & $r \to r_+$ & $r \to \infty$ & $r \to r_+$ & $r \to \infty$ \\\hline
    $(\var_+ g_{lm\omega p})_{n\bar{m}}$ & $\Delta$ & $1/r^2$ & $1$ & $r$ \\
    $(\var_+ g_{lm\omega p})_{\bar{m}\bar{m}}$ & $1$ & $1/r$ & $1$ & $1$ \\
    $(\var_- g_{lm\omega p})_{lm}$ & $1/\Delta$ & $r$ & $1$ & $1/r^2$ \\
    $(\var_- g_{lm\omega p})_{mm}$ & $1$ & $1$ & $1$ & $1/r$ \\
    $(\var_+ C_{lm\omega p})_{l\bar{m}\bar{m}}$ & $1/\Delta$ & $1/r$ & $1$ &  $1/r$ \\
    $(\var_+ C_{lm\omega p})_{n(l\bar{m})}$ & $1$ & $1/r^2$ & $1$ & $1/r^2$ \\
    $(\var_+ C_{lm\omega p})_{n\bar{m}\bar{m}}$ & $\Delta$ & $1/r^2$ & $1$ & $1$ \\
    $(\var_- C_{lm\omega p})_{nmm}$ & $\Delta$ & $1/r$ & $1$ & $1/r$ \\
    $(\var_- C_{lm\omega p})_{l(nm)}$ & $1$ & $1/r^2$ & $1$ & $1/r^2$ \\
    $(\var_- C_{lm\omega p})_{lmm}$ & $1/\Delta$ & $1$ & $1$ & $1/r^2$ \\\hline
  \end{tabular}
\end{table}

Plugging equation~\eqref{eqn:deriv_relation} [for $\pb{s} \psi_{lm\omega p} (r)$] into equations~\eqref{eqn:var_g_exp} and~\eqref{eqn:var_gamma_exp}, and then taking the limits $r \to \infty$ and $r \to r_+$, yields the asymptotic forms given in table~\ref{table:gr_asymp}.
Using this same calculation, we can determine the angular dependences of the quantities in~\eqref{eqn:var_g_exp} and~\eqref{eqn:var_gamma_exp}.
Defining, for $s \geq 0$,

\begin{equation}
  \pb{\pm s} \eta^+_{lm\omega} = \pm 2i(2s - 1) \omega r_+ - \pb{2} \lambda_{lm\omega}, \qquad \pb{\pm s} \eta^\infty_{lm\omega} = \pm 2(2s - 1) \omega a \cos \theta + \pb{2} \lambda_{lm\omega},
\end{equation}
they are given by

\begin{subequations} \label{eqn:perturbed_metric_asymp}
  \begin{align}
    &\pb{\var_+ g_{n\bar{m}}} S_{lm\omega p}^{\textrm{in}} = \frac{4ik_{m\omega} \sqrt{Mr_+} \pb{-2} \kappa_{m\omega}}{\zeta_+} \mathscr{L}_{2(-m)(-\omega)} \pb{-2} \Theta_{lm\omega}, \\
    &\mathrlap{\pb{\var_+ g_{n\bar{m}}} S_{lm\omega p}^{\textrm{out}} = \frac{\pb{-2} \eta^+_{lm\omega} \zeta_+ + 8 Mr_+ ik_{m\omega} \pb{-1} \kappa_{m\omega}}{4 (Mr_+)^{3/2} ik_{m\omega} \pb{-1} \kappa_{m\omega} \zeta_+^2} \mathscr{L}_{2(-m)(-\omega)} \pb{-2} \Theta_{lm\omega},} \\
    &\pb{\var_+ g_{n\bar{m}}} S_{lm\omega p}^{\textrm{down}} = 2\sqrt{2} i\omega \mathscr{L}_{2(-m)(-\omega)} \pb{-2} \Theta_{lm\omega}, &\pb{\var_+ g_{n\bar{m}}} S_{lm\omega p}^{\textrm{up}} = -\sqrt{2} \mathscr{L}_{2(-m)(-\omega)} \pb{-2} \Theta_{lm\omega}, \\
    &\pb{\var_+ g_{\bar{m}\bar{m}}} S_{lm\omega p}^{\textrm{in}} = 4 (2Mr_+)^{3/2} k_{m\omega}^2 \pb{-2} \kappa_{m\omega} \pb{-1} \kappa_{m\omega} \pb{-2} \Theta_{lm\omega}, \\
    &\mathrlap{\pb{\var_+ g_{\bar{m}\bar{m}}} S_{lm\omega p}^{\textrm{out}} = -\frac{24 Mr_+ i\omega k_{m\omega} \pb{-1} \kappa_{m\omega} \zeta_+ + [i\zeta_+ (2 - \pb{-1} \eta^+_{lm\omega}) + 8Mr_+ k_{m\omega}] \pb{-2} \eta^+_{lm\omega}}{4 ik_{m\omega}^2 (2Mr_+)^{5/2} \pb{-1} \kappa_{m\omega} \zeta_+} \pb{-2} \Theta_{lm\omega},} \\
    &\pb{\var_+ g_{\bar{m}\bar{m}}} S_{lm\omega p}^{\textrm{down}} = 4 \omega^2 \pb{-2} \Theta_{lm\omega}, &\pb{\var_+ g_{\bar{m}\bar{m}}} S_{lm\omega p}^{\textrm{up}} = \frac{i \pb{2} \eta^\infty_{lm\omega}}{\omega} \pb{-2} \Theta_{lm\omega}, \\
    &\pb{\var_- g_{lm}} S_{lm\omega p}^{\textrm{in}} = \frac{\pb{2} \eta^+_{lm\omega} \zeta_+ - 8Mr_+ ik_{m\omega} \pb{1} \kappa_{m\omega}}{8 (Mr_+)^{3/2} ik_{m\omega} \pb{1} \kappa_{m\omega}} \mathscr{L}_{2m\omega} \pb{2} \Theta_{lm\omega}, \\
    &\pb{\var_- g_{lm}} S_{lm\omega p}^{\textrm{out}} = 2 \sqrt{Mr_+} ik_{m\omega} \pb{2} \kappa_{m\omega} \zeta_+ \mathscr{L}_{2m\omega} \pb{2} \Theta_{lm\omega}, \\
    &\pb{\var_- g_{lm}} S_{lm\omega p}^{\textrm{down}} = \frac{\mathscr{L}_{2m\omega}}{\sqrt{2}} \pb{2} \Theta_{lm\omega}, &\pb{\var_- g_{lm}} S_{lm\omega p}^{\textrm{up}} = \sqrt{2} i \omega \mathscr{L}_{2m\omega} \pb{2} \Theta_{lm\omega}, \\
    &\mathrlap{\pb{\var_- g_{mm}} S_{lm\omega p}^{\textrm{in}} = \frac{24 Mr_+ i\omega k_{m\omega} \pb{1} \kappa_{m\omega} \zeta_+ + [i\zeta_+ (2 - \pb{1} \eta^+_{lm\omega}) - 8Mr_+ k_{m\omega}] \pb{2} \eta^+_{lm\omega}}{16 ik_{m\omega}^2 (2Mr_+)^{5/2} \pb{1} \kappa_{m\omega} \zeta_+} \pb{2} \Theta_{lm\omega},} \\
    &\pb{\var_- g_{mm}} S_{lm\omega p}^{\textrm{out}} = -(2Mr_+)^{3/2} k_{m\omega}^2 \pb{2} \kappa_{m\omega} \pb{1} \kappa_{m\omega} \pb{2} \Theta_{lm\omega}, \\
    &\pb{\var_- g_{mm}} S_{lm\omega p}^{\textrm{down}} = \frac{i \pb{-2} \eta^\infty_{lm\omega}}{4 \omega} \pb{2} \Theta_{lm\omega}, &\pb{\var_- g_{mm}} S_{lm\omega p}^{\textrm{up}} = -\omega^2 \pb{2} \Theta_{lm\omega},
  \end{align}
\end{subequations}
and

\begin{subequations} \label{eqn:perturbed_chris_asymp}
  \begin{align}
    &\mathrlap{\pb{\var_+ C_{l\bar{m}\bar{m}}} S_{lm\omega p}^{\textrm{in}} = 4(2Mr_+)^{5/2} ik_{m\omega}^3 \pb{-2} \kappa_{m\omega} \pb{-1} \kappa_{m\omega} \pb{-2} \Theta_{lm\omega},} \\
    &\mathrlap{\begin{aligned}
        \pb{\var_+ C_{l\bar{m}\bar{m}}} S_{lm\omega p}^{\textrm{out}} &= \Big\{4Mr_+ ik_{m\omega} \pb{1} \kappa_{m\omega} [24 Mr_+ i\omega k_{m\omega} \pb{-1} \kappa_{m\omega} + i(2 - \pb{-1} \eta^+_{lm\omega}) \pb{-2} \eta^+_{lm\omega}] \\
        &\hspace{2em}- \zeta_+ \{8Mr_+ i\omega k_{m\omega} [3 \pb{-1} \kappa_{m\omega} (2 - \pb{1} \eta^+_{lm\omega}) - 4 \pb{-2} \eta^+_{lm\omega}] \\
        &\hspace{5em}+ i \pb{-2} \eta^+_{lm\omega} [|\pb{-1} \eta^+_{lm\omega}|^2 + 4(\pb{2} \lambda_{lm\omega} + 1)]\}\Big\} \frac{\pb{-2} \Theta_{lm\omega}}{16 k_{m\omega}^3 (2Mr_+)^{7/2} |\pb{-1} \kappa_{m\omega}|^2 \zeta_+},
      \end{aligned}} \\
    &\pb{\var_+ C_{l\bar{m}\bar{m}}} S_{lm\omega p}^{\textrm{down}} = 4 i \omega^3 \pb{-2} \Theta_{lm\omega}, &\pb{\var_+ C_{l\bar{m}\bar{m}}} S_{lm\omega p}^{\textrm{up}} = -\frac{i \pb{2} \eta^\infty_{lm\omega}}{2 \omega} \pb{-2} \Theta_{lm\omega}, \\
    &\mathrlap{\pb{\var_+ C_{n(l\bar{m})}} S_{lm\omega p}^{\textrm{in}} = -4 (Mr_+)^{3/2} k_{m\omega}^2 \pb{-2} \kappa_{m\omega} \pb{-1} \kappa_{m\omega} \zeta_+^{-2} (\zeta_+ \mathscr{L}_{2(-m)(-\omega)} - ia \sin \theta) \pb{-2} \Theta_{lm\omega},} \\
    &\mathrlap{\begin{aligned}
        \pb{\var_+ C_{n(l\bar{m})}} S_{lm\omega p}^{\textrm{out}} &= \zeta_+^{-3} \Big\{[24Mr_+ i\omega k_{m\omega} \pb{-1} \kappa_{m\omega} + i(2 - \pb{-1} \eta^+_{lm\omega}) \pb{-2} \eta^+_{lm\omega}] \zeta_+ (\zeta_+ \mathscr{L}_{2(-m)(-\omega)} - ia \sin \theta) \\
        &\hspace{4em}+ 8Mr_+ k_{m\omega} \pb{-2} \eta^+_{lm\omega} (2\zeta_+ \mathscr{L}_{2(-m)(-\omega)} - ia \sin \theta) \\
        &\hspace{4em}+ 6 (4Mr_+)^2 ik_{m\omega}^2 \pb{-1} \kappa_{m\omega} \mathscr{L}_{2(-m)(-\omega)}\Big\} \frac{\pb{-2} \Theta_{lm\omega}}{64 (Mr_+)^{5/2} ik_{m\omega}^2 \pb{-1} \kappa_{m\omega}},
      \end{aligned}} \\
    &\mathrlap{\pb{\var_+ C_{n(l\bar{m})}} S_{lm\omega p}^{\textrm{down}} = -\sqrt{2} \omega^2 \mathscr{L}_{2(-m)(-\omega)} \pb{-2} \Theta_{lm\omega},} \\
    &\mathrlap{\begin{aligned}
        \pb{\var_+ C_{n(l\bar{m})}} S_{lm\omega p}^{\textrm{up}} &= -\Big\{[4 \omega^2 a^2 (2 \cos^2 \theta - 3) - 12 i \omega (M + iam) + \pb{2} \lambda_{lm\omega} (\pb{2} \lambda_{lm\omega} + 2)] \mathscr{L}_{2(-m)(-\omega)} \\
        &\hspace{3em}+ 4a\omega \sin \theta (12 a\omega \cos \theta + \pb{2} \eta^\infty_{lm\omega})\Big\} \frac{\pb{-2} \Theta_{lm\omega}}{8\sqrt{2} \omega^2},
      \end{aligned}} \\
    &\mathrlap{\begin{aligned}
        \pb{\var_+ C_{n\bar{m}\bar{m}}} S_{lm\omega p}^{\textrm{in}} &= \sqrt{Mr_+} ik_{m\omega} \pb{-2} \kappa_{m\omega} \zeta_+^{-4} \Big\{\zeta_+^2 [(2 - \pb{-1} \eta^+_{lm\omega}) - \mathscr{L}_{(-1)(-m)(-\omega)} \mathscr{L}_{2(-m)(-\omega)}] \\
        &\hspace{12em}+ 16Mr_+ ik_{m\omega} \pb{-3/2} \kappa_{m\omega} \zeta_+ + 2a^2 \sin^2 \theta \\
        &\hspace{12em}+ ia \sin \theta \zeta_+ \mathscr{L}_{2(-m)(-\omega)}\Big\} \frac{\pb{-2} \Theta_{lm\omega}}{\sqrt{2}},
      \end{aligned}} \\
    &\mathrlap{\begin{aligned}
        \pb{\var_+ C_{n\bar{m}\bar{m}}} S_{lm\omega p}^{\textrm{out}} &= \zeta_+^{-4} \Big\{\pb{-2} \eta^+_{lm\omega} [\zeta_+ (ia \sin \theta - \zeta_+ \mathscr{L}_{(-1)(-m)(-\omega)}) \mathscr{L}_{2(-m)(-\omega)} - 8Mr_+ ik_{m\omega} \zeta_+ + 2a^2 \sin^2 \theta] \\
        &\hspace{3.5em}- 8Mr_+ ik_{m\omega} \pb{-1} \kappa_{m\omega} (\zeta_+ \mathscr{L}_{(-1)(-m)(-\omega)} - ia \sin \theta) \mathscr{L}_{2(-m)(-\omega)} \\
        &\hspace{3.5em}+ [24Mr_+ \omega k_{m\omega} \pb{-1} \kappa_{m\omega} + (2 - \pb{-1} \eta^+_{lm\omega}) \pb{-2} \eta^+_{lm\omega}] \zeta_+^2\Big\} \frac{\pb{-2} \Theta_{lm\omega}}{(8Mr_+)^{3/2} ik_{m\omega} \pb{-1} \kappa_{m\omega}}, \\
      \end{aligned}} \\
    &\pb{\var_+ C_{n\bar{m}\bar{m}}} S_{lm\omega p}^{\textrm{down}} = -5 \omega^2 \pb{-2} \Theta_{lm\omega}, &\pb{\var_+ C_{n\bar{m}\bar{m}}} S_{lm\omega p}^{\textrm{up}} = -\frac{2 \pb{2} \eta^\infty_{lm\omega} - \mathscr{L}_{(-1)(-m)(-\omega)} \mathscr{L}_{2(-m)(-\omega)}}{4} \pb{-2} \Theta_{lm\omega}, \\
    &\mathrlap{\begin{aligned}
        \pb{\var_- C_{nmm}} S_{lm\omega p}^{\textrm{in}} &= \Big\{4Mr_+ ik_{m\omega} \pb{-1} \kappa_{m\omega} [24Mr_+ i\omega k_{m\omega} \pb{1} \kappa_{m\omega} + i(2 - \pb{1} \eta^+_{lm\omega}) \pb{2} \eta^+_{lm\omega}] \\
        &\hspace{2em}+ \zeta_+ \{8Mr_+ i\omega k_{m\omega} [3 \pb{1} \kappa_{m\omega} (2 - \pb{-1} \eta^+_{lm\omega}) - 4 \pb{2} \eta^+_{lm\omega}] \\
        &\hspace{5em}+ i \pb{2} \eta^+_{lm\omega} [|\pb{1} \eta^+_{lm\omega}|^2 + 4(\pb{2} \lambda_{lm\omega} + 1)]\}\Big\} \frac{\pb{2} \Theta_{lm\omega}}{k_{m\omega}^3 (8Mr_+)^{7/2} |\pb{1} \kappa_{m\omega}|^2 \zeta_+^3},
      \end{aligned}} \\
    &\mathrlap{\pb{\var_- C_{nmm}} S_{lm\omega p}^{\textrm{out}} = -\frac{(2Mr_+)^{5/2} ik_{m\omega}^3 \pb{2} \kappa_{m\omega} \pb{1} \kappa_{m\omega}}{2\zeta_+^2} \pb{2} \Theta_{lm\omega},} \\
    &\pb{\var_- C_{nmm}} S_{lm\omega p}^{\textrm{down}} = \frac{i \pb{-2} \eta^\infty_{lm\omega}}{16 \omega} \pb{2} \Theta_{lm\omega}, &\pb{\var_- C_{nmm}} S_{lm\omega p}^{\textrm{up}} = -\frac{i \omega^3}{2} \pb{2} \Theta_{lm\omega}, \\
    &\mathrlap{\begin{aligned}
        \pb{\var_- C_{l(nm)}} S_{lm\omega p}^{\textrm{in}} &= \zeta_+^{-3} \Big\{[24Mr_+ i\omega k_{m\omega} \pb{1} \kappa_{m\omega} + i \pb{2} \eta^+_{lm\omega} (2 - \pb{1} \eta^+_{lm\omega})] \zeta_+ (\zeta_+ \mathscr{L}_{2m\omega} + ia \sin \theta) \\
        &\hspace{4em}- 8Mr_+ k_{m\omega} \pb{2} \eta^+_{lm\omega} (2 \zeta_+ \mathscr{L}_{2m\omega} + ia \sin \theta) \\
        &\hspace{4em}+ 6(4Mr_+)^2 ik_{m\omega}^2 \pb{1} \kappa_{m\omega} \mathscr{L}_{2m\omega}\Big\} \frac{\pb{2} \Theta_{lm\omega}}{256 (Mr_+)^{5/2} ik_{m\omega}^2 \pb{1} \kappa_{m\omega}},
      \end{aligned}} \\
    &\mathrlap{\pb{\var_- C_{l(nm)}} S_{lm\omega p}^{\textrm{out}} = -(Mr_+)^{3/2} k_{m\omega}^2 \pb{2} \kappa_{m\omega} \pb{1} \kappa_{m\omega} \zeta_+^{-2} (\zeta_+ \mathscr{L}_{2m\omega} + ia \sin \theta) \pb{2} \Theta_{lm\omega},} \\
    &\mathrlap{\begin{aligned}
        \pb{\var_- C_{l(nm)}} S_{lm\omega p}^{\textrm{down}} &= -\Big\{[4 \omega^2 a^2 (2\cos^2 \theta - 3) + 12 i\omega (M - iam) + \pb{2} \lambda_{lm\omega} (\pb{2} \lambda_{lm\omega} + 2)] \mathscr{L}_{2m\omega} \\
        &\hspace{3em}+ 4 a\omega \sin \theta (12a\omega \cos \theta + \pb{-2} \eta^\infty_{lm\omega})\Big\} \frac{\pb{2} \Theta_{lm\omega}}{32\sqrt{2} \omega^2},
      \end{aligned}} \\
    &\pb{\var_- C_{l(nm)}} S_{lm\omega p}^{\textrm{up}} = -\frac{\omega^2 \mathscr{L}_{2m\omega}}{2\sqrt{2}} \pb{2} \Theta_{lm\omega}, \\
    &\mathrlap{\begin{aligned}
        \pb{\var_- C_{lmm}} S_{lm\omega p}^{\textrm{in}} &= \zeta^{-2} \Big\{\pb{2} \eta^+_{lm\omega} [2a^2 \sin^2 \theta - \zeta_+ (\zeta_+ \mathscr{L}_{(-1)m\omega} + 3ia \sin \theta) \mathscr{L}_{2m\omega} - 8Mr_+ ik_{m\omega} \zeta_+] \\
        &\hspace{3.5em}+ 8Mr_+ ik_{m\omega} \pb{1} \kappa_{m\omega} (\zeta_+ \mathscr{L}_{(-1)m\omega} + 3ia \sin \theta) \mathscr{L}_{2m\omega} \\
        &\hspace{3.5em}- [24Mr_+ \omega k_{m\omega} \pb{1} \kappa_{m\omega} + \pb{2} \eta^+_{lm\omega} (2 - \pb{1} \eta^+_{lm\omega})] \zeta_+^2\Big\} \frac{\pb{2} \Theta_{lm\omega}}{2 (8Mr_+)^{3/2} ik_{m\omega} \pb{1} \kappa_{m\omega}},
      \end{aligned}} \\
    &\mathrlap{\begin{aligned}
        \pb{\var_- C_{lmm}} S_{lm\omega p}^{\textrm{out}} &= \sqrt{Mr_+} ik_{m\omega} \pb{2} \kappa_{m\omega} \zeta_+^{-2} \Big\{\zeta_+ (8Mr_+ ik_{m\omega} \pb{2} \kappa_{m\omega} - 3ia \sin \theta \mathscr{L}_{2m\omega}) + 2a^2 \sin^2 \theta \\
          &\hspace{11.5em}-\zeta_+^2 [\mathscr{L}_{(-1)m\omega} \mathscr{L}_{2m\omega} + (2 - \pb{1} \eta^+_{lm\omega})]\Big\} \frac{\pb{2} \Theta_{lm\omega}}{2\sqrt{2}},
        \end{aligned}} \\
    &\mathrlap{\pb{\var_- C_{lmm}} S_{lm\omega p}^{\textrm{down}} = -\frac{2 \pb{-2} \eta^\infty_{lm\omega} + \mathscr{L}_{(-1)m\omega} \mathscr{L}_{2m\omega}}{8} \pb{2} \Theta_{lm\omega},} &\pb{\var_- C_{lmm}} S_{lm\omega p}^{\textrm{up}} = -\frac{3 \omega^2}{2} \pb{2} \Theta_{lm\omega}.
  \end{align}
\end{subequations}

\bibliographystyle{unsrt}
\bibliography{adwg}

\begin{thebibliography}{10}

\bibitem{PhysRev.174.1559}
Brandon Carter.
\newblock Global structure of the kerr family of gravitational fields.
\newblock {\em Phys. Rev.}, 174:1559--1571, Oct 1968.

\bibitem{1970CMaPh..18..265W}
M.~{Walker} and R.~{Penrose}.
\newblock {On quadratic first integrals of the geodesic equations for type
  $\{22\}$ spacetimes}.
\newblock {\em Communications in Mathematical Physics}, 18:265--274, December
  1970.

\bibitem{PhysRevD.16.3395}
Brandon Carter.
\newblock Killing tensor quantum numbers and conserved currents in curved
  space.
\newblock {\em Phys. Rev. D}, 16:3395--3414, Dec 1977.

\bibitem{PhysRevD.19.1093}
B.~Carter and R.~G. McLenaghan.
\newblock Generalized total angular momentum operator for the dirac equation in
  curved space-time.
\newblock {\em Phys. Rev. D}, 19:1093--1097, Feb 1979.

\bibitem{Andersson:2015xla}
Lars Andersson, Thomas B{\"a}ckdahl, and Pieter Blue.
\newblock {Spin geometry and conservation laws in the Kerr spacetime}.
\newblock In Lydia Bieri and Shing-Tung Yau, editors, {\em Surveys in
  Differential Geometry}, volume~20, chapter~8, pages 183--226. International
  Press, August 2015.

\bibitem{Grant:2019qyo}
Alexander~M. Grant and \'Eanna~\'E. Flanagan.
\newblock {Conserved currents for electromagnetic fields in the Kerr
  spacetime}.
\newblock {\em Class. Quant. Grav.}, 37(18):185021, 2020.

\bibitem{AmaroSeoane:2012km}
Pau Amaro-Seoane et~al.
\newblock {eLISA/NGO: Astrophysics and cosmology in the gravitational-wave
  millihertz regime}.
\newblock {\em GW Notes}, 6:4--110, 2013.

\bibitem{Wardell:2015kea}
Barry Wardell and Achamveedu Gopakumar.
\newblock {Self-force: Computational Strategies}.
\newblock {\em Fund. Theor. Phys.}, 179:487--522, 2015.

\bibitem{Hughes:1999bq}
Scott~A. Hughes.
\newblock {The Evolution of circular, nonequatorial orbits of Kerr black holes
  due to gravitational wave emission}.
\newblock {\em Phys. Rev.}, D61(8):084004, 2000.
\newblock [Erratum: Phys. Rev.D90,no.10,109904(2014)].

\bibitem{Mino:2003yg}
Yasushi Mino.
\newblock {Perturbative approach to an orbital evolution around a supermassive
  black hole}.
\newblock {\em Phys. Rev.}, D67:084027, 2003.

\bibitem{Isoyama:2018sib}
Soichiro Isoyama, Ryuichi Fujita, Hiroyuki Nakano, Norichika Sago, and Takahiro
  Tanaka.
\newblock {``Flux-balance formulae'' for extreme mass-ratio inspirals}.
\newblock {\em PTEP}, 2019(1):013E01, 2019.

\bibitem{Sago01052006}
Norichika Sago, Takahiro Tanaka, Wataru Hikida, Katsuhiko Ganz, and Hiroyuki
  Nakano.
\newblock Adiabatic evolution of orbital parameters in kerr spacetime.
\newblock {\em Progress of Theoretical Physics}, 115(5):873--907, 2006.

\bibitem{Grant:2015xqa}
Alexander Grant and Eanna~E. Flanagan.
\newblock {Non-conservation of Carter in black hole spacetimes}.
\newblock {\em Class. Quant. Grav.}, 32(15):157001, 2015.

\bibitem{PhysRevLett.41.203}
Robert~M. Wald.
\newblock Construction of solutions of gravitational, electromagnetic, or other
  perturbation equations from solutions of decoupled equations.
\newblock {\em Phys. Rev. Lett.}, 41:203--206, Jul 1978.

\bibitem{Chrzanowski:1975wv}
P.~L. Chrzanowski.
\newblock {Vector Potential and Metric Perturbations of a Rotating Black Hole}.
\newblock {\em Phys. Rev.}, D11:2042--2062, 1975.

\bibitem{Aksteiner:2016mol}
Steffen Aksteiner and Thomas Bäckdahl.
\newblock {Symmetries of linearized gravity from adjoint operators}.
\newblock {\em J. Math. Phys.}, 60(8):082501, 2019.

\bibitem{penrose1987spinors}
R.~Penrose and W.~Rindler.
\newblock {\em Spinors and Space-Time: Volume 1, Two-Spinor Calculus and
  Relativistic Fields}.
\newblock Cambridge Monographs on Mathematical Physics. Cambridge University
  Press, 1987.

\bibitem{penrose1988spinors}
R.~Penrose and W.~Rindler.
\newblock {\em Spinors and Space-Time: Volume 2, Spinor and Twistor Methods in
  Space-Time Geometry}.
\newblock Cambridge Monographs on Mathematical Physics. Cambridge University
  Press, 1988.

\bibitem{Backdahl:2015yua}
Thomas B{\"a}ckdahl and Juan A.~Valiente Kroon.
\newblock {A formalism for the calculus of variations with spinors}.
\newblock {\em J. Math. Phys.}, 57:022502, 2016.

\bibitem{Stewart:1974uz}
J.M. Stewart and M.~Walker.
\newblock {Perturbations of spacetimes in general relativity}.
\newblock {\em Proc.Roy.Soc.Lond.}, A341:49--74, 1974.

\bibitem{1973ApJ...185..635T}
S.~A. {Teukolsky}.
\newblock {Perturbations of a Rotating Black Hole. I. Fundamental Equations for
  Gravitational, Electromagnetic, and Neutrino-Field Perturbations}.
\newblock {\em \apj}, 185:635--648, October 1973.

\bibitem{1973JMP....14.1453W}
R.~M. {Wald}.
\newblock {On perturbations of a Kerr black hole.}
\newblock {\em Journal of Mathematical Physics}, 14:1453--1461, 1973.

\bibitem{Newman:1961qr}
Ezra Newman and Roger Penrose.
\newblock {An Approach to gravitational radiation by a method of spin
  coefficients}.
\newblock {\em J. Math. Phys.}, 3:566--578, 1962.

\bibitem{Geroch:1973am}
Robert~P. Geroch, A.~Held, and R.~Penrose.
\newblock {A space-time calculus based on pairs of null directions}.
\newblock {\em J. Math. Phys.}, 14:874--881, 1973.

\bibitem{1974ApJ...193..443T}
S.~A. {Teukolsky} and W.~H. {Press}.
\newblock {Perturbations of a rotating black hole. III - Interaction of the
  hole with gravitational and electromagnetic radiation}.
\newblock {\em \apj}, 193:443--461, October 1974.

\bibitem{chandrasekhar1983mathematical}
S.~Chandrasekhar.
\newblock {\em The mathematical theory of black holes}.
\newblock International series of monographs on physics. Clarendon Press, 1983.

\bibitem{1973ApJ...185..649P}
W.~H. {Press} and S.~A. {Teukolsky}.
\newblock {Perturbations of a Rotating Black Hole. II. Dynamical Stability of
  the Kerr Metric}.
\newblock {\em \apj}, 185:649--674, October 1973.

\bibitem{1982JPhA...15.3737G}
D.~V. {Gal'tsov}.
\newblock {Radiation reaction in the Kerr gravitational field}.
\newblock {\em Journal of Physics A Mathematical General}, 15:3737--3749,
  December 1982.

\bibitem{1992RSPSA.439..103K}
E.~G. {Kalnins}, R.~G. {McLenaghan}, and G.~C. {Williams}.
\newblock {Symmetry Operators for Maxwell's Equations on Curved Space-Time}.
\newblock {\em Royal Society of London Proceedings Series A}, 439:103--113,
  October 1992.

\bibitem{PhysRevD.10.1070}
Jeffrey~M. Cohen and Lawrence~S. Kegeles.
\newblock Electromagnetic fields in curved spaces: A constructive procedure.
\newblock {\em Phys. Rev. D}, 10:1070--1084, Aug 1974.

\bibitem{PhysRevD.19.1641}
Lawrence~S. Kegeles and Jeffrey~M. Cohen.
\newblock Constructive procedure for perturbations of spacetimes.
\newblock {\em Phys. Rev. D}, 19:1641--1664, Mar 1979.

\bibitem{PhysRevD.14.317}
Paul~L. Chrzanowski, Richard~A. Matzner, Vernon~D. Sandberg, and Michael~P.
  Ryan.
\newblock Zero-mass plane waves in nonzero gravitational backgrounds.
\newblock {\em Phys. Rev. D}, 14:317--326, Jul 1976.

\bibitem{Aksteiner:2016pjt}
Steffen Aksteiner, Lars Andersson, and Thomas Bäckdahl.
\newblock {New identities for linearized gravity on the Kerr spacetime}.
\newblock {\em Phys. Rev. D}, 99(4):044043, 2019.

\bibitem{Stewart:1978tm}
John~M. Stewart.
\newblock {{Hertz-Bromwich-Debye-Whittaker-Penrose} Potentials in General
  Relativity}.
\newblock {\em Proc. Roy. Soc. Lond.}, A367:527--538, 1979.

\bibitem{bardeen}
J.~Bardeen.
\newblock {A Re-examination of the Teukolsky-Starobinsky Identities}.
\newblock {\em unpublished}, 2007.

\bibitem{Burnett57}
Gregory~A. Burnett and Robert~M. Wald.
\newblock A conserved current for perturbations of einstein-maxwell
  space-times.
\newblock {\em Proceedings of the Royal Society of London A: Mathematical,
  Physical and Engineering Sciences}, 430(1878):57--67, 1990.

\bibitem{Toth:2018qrx}
G.\'abor~Zsolt T\'oth.
\newblock {Noether currents for the Teukolsky master equation}.
\newblock {\em Class. Quant. Grav.}, 35(18):185009, 2018.

\bibitem{Bini:2002jx}
Donato Bini, Christian Cherubini, Robert~T Jantzen, and Remo~J. Ruffini.
\newblock {Teukolsky master equation: De Rham wave equation for the
  gravitational and electromagnetic fields in vacuum}.
\newblock {\em Prog. Theor. Phys.}, 107:967--992, 2002.

\bibitem{mtw}
C.W. Misner, K.S. Thorne, and J.A. Wheeler.
\newblock {\em Gravitation}.
\newblock W. H. Freeman, 1973.

\bibitem{Isaacson:1968zza}
Richard~A. Isaacson.
\newblock {Gravitational Radiation in the Limit of High Frequency. II.
  Nonlinear Terms and the Ef fective Stress Tensor}.
\newblock {\em Phys. Rev.}, 166:1272--1279, 1968.

\bibitem{Burnett:1989gp}
G.~A. Burnett.
\newblock {The High Frequency Limit in General Relativity}.
\newblock {\em J. Math. Phys.}, 30:90--96, 1989.

\bibitem{Anco:2002xn}
Stephen~C. Anco and Juha Pohjanpelto.
\newblock {Conserved currents of massless fields of spin $s > 0$}.
\newblock {\em Proc. Roy. Soc. Lond.}, A459:1215--1240, 2003.

\end{thebibliography}

\end{document}